\documentclass[twocolumn]{emulateapj}

\usepackage{epsfig}
\usepackage{longtable}
\usepackage{comment}
\usepackage{color}
\usepackage[normalem]{ulem}
\usepackage[colorlinks=true, pdfstartview=FitV, linkcolor=blue,citecolor=blue, urlcolor=blue]{hyperref}
\usepackage{amsmath,amssymb}
\usepackage{array}
\usepackage{accents}
\newcommand{\cxc}{{\em Chandra}}

\shorttitle{Fraction of cool-cores in X-ray and SZ samples}
\shortauthors{Andrade-Santos et al.}

\begin{document}


%%\title{PLCKG345.40-39.34: A {\it Planck} detected double cluster}
%%\title{supermassive Black Hole Binaries, Sloshing, and Cold Fronts}
\title{The fraction of cool-core clusters in X-ray vs. SZ samples \\ using {\em Chandra} observations}


\author{
Felipe Andrade-Santos$^1$, 
Christine Jones$^1$, 
William R. Forman$^1$,
Lorenzo Lovisari$^1$,
Alexey Vikhlinin$^1$, 
Reinout J. van Weeren$^1$,
Stephen S. Murray$^1$, 
Monique Arnaud$^2$,
Gabriel W. Pratt$^2$,
Jessica D\'emocl\`es$^{2}$,
Ralph Kraft$^1$,
Pasquale Mazzotta$^{1,3}$,
Hans B\"ohringer$^4$,
Gayoung Chon$^4$,
Simona Giacintucci$^5$,
Tracy E. Clarke$^6$,
Stefano Borgani$^{7,8}$,
Larry David$^1$,
Marian Douspis$^9$,
Etienne Pointecouteau$^{10,11}$,
H{\scriptsize \AA}kon Dahle$^{12}$,
Shea Brown$^{13}$,
Nabila Aghanim$^{14}$,
Elena Rasia$^8$
}
\affil{
$^1$Harvard-Smithsonian Center for Astrophysics, 60 Garden Street, Cambridge, MA 02138, USA \\
$^2$Laboratoire AIM, CEA/IRFU, CNRS/INSU, Universit\'e Paris Diderot, CEA-Saclay, 91191 Gif-Sur-Yvette, France\\
$^3$Dipartimento di Fisica, Universit\`a di Roma Tor Vergata, via della Ricerca Scientifica 1, 00133 Roma, Italy\\
$^4$Max Planck Institute for Extraterrestrial Physics, Giessenbachstrasse, 85748 Garching, Germany\\
$^5$Naval Research Laboratory, 4555 Overlook Avenue SW, Code 7213, Washington, DC 20375, USA\\
$^6$U.S. Naval Research Laboratory, Code 7213, 4555 Overlook Ave. SW., Washington, DC  20375\\
$^7$Dipartimento di Fisica dell' Universit\`a di Trieste, Sezione di Astronomia, via Tiepolo 11, I-34131 Trieste, Italy\\
$^8$INAF, Osservatorio Astronomico di Trieste, via Tiepolo 11, I-34131, Trieste, Italy\\
$^9$Institut d'Astrophysique Spatiale, CNRS (UMR8617) Universit\'e Paris-Sud 11, B\^atiment 121, Orsay, France\\
$^{10}$CNRS; IRAP; 9 Av. colonel Roche, BP 44346, F-31028 Toulouse cedex 4, France\\
$^{11}$Universit\'e de Toulouse; UPS-OMP; IRAP; Toulouse, France\\
$^{12}$Institute of Theoretical Astrophysics, University of Oslo, P.O. Box 1029, Blindern, NO-0315 Oslo, Norway\\
$^{13}$Department of Physics \& Astronomy, The University of Iowa, Iowa City, IA, 52245\\
$^{14}$Institut d'Astrophysique Spatiale, CNRS, Univ. Paris-Sud, Universit\'e Paris-Saclay, B\^at. 121, 91405 Orsay Cedex, France
}



%%%%%%%%%%%%%%%%%%%%%%%%%%%%%%%%%%%%%%%%%%%%%%%%%%%%%%%%%%%%%%%%%%%%%%%%%%%%%%%%%%
%%%%%%%%%%%%%%%%%%%%%%%%%%%%%%%%%%%%%%%%%%%%%%%%%%%%%%%%%%%%%%%%%%%%%%%%%%%%%%%%%%
%%
%%                                  ABSTRACT
%%
%%%%%%%%%%%%%%%%%%%%%%%%%%%%%%%%%%%%%%%%%%%%%%%%%%%%%%%%%%%%%%%%%%%%%%%%%%%%%%%%%%
%%%%%%%%%%%%%%%%%%%%%%%%%%%%%%%%%%%%%%%%%%%%%%%%%%%%%%%%%%%%%%%%%%%%%%%%%%%%%%%%%%

\begin{abstract}

We derive and compare the fractions of cool-core clusters in the {\em Planck}
Early Sunyaev-Zel'dovich sample of 164 clusters with $z \leq 0.35$  
and in a flux-limited X-ray sample of 100 clusters with $z \leq 0.30$, 
using {\em Chandra} observations. We use four metrics to identify
cool-core clusters: 1) the concentration
parameter: the ratio of the integrated emissivity profile within 0.15
$r_{500}$ to that within $r_{500}$, and 2) the ratio of the integrated
emissivity profile within 40 kpc to
that within 400 kpc, 3) the cuspiness of the gas density profile: the
negative of the logarithmic derivative of the gas density with respect to the
radius, measured at 0.04 $r_{500}$, and 4) the central gas density,
measured at  0.01 $r_{500}$. We find that the sample of X-ray selected
clusters, as characterized by each of these metrics, contains a 
significantly larger fraction of cool-core clusters compared to the
sample of SZ selected clusters (44$\pm$7\% vs. 28$\pm$4\% using the concentration
parameter in the 0.15--1.0 $r_{500}$ range, 61$\pm$8\% vs. 36$\pm$5\% using the concentration
parameter in the 40--400 kpc range, 64$\pm$8\% vs. 38$\pm$5\%
using the cuspiness,
and 53$\pm$7\% vs. 39$\pm$5\% using the central gas
density). Qualitatively, cool-core clusters are more X-ray luminous at
fixed mass. Hence, our X-ray 
flux-limited sample, compared to the approximately mass-limited SZ
sample, is over-represented with cool-core clusters. 
We describe a simple quantitative model that 
uses the excess luminosity of cool-core clusters compared to non-cool-core clusters 
at fixed mass to successfully predict the observed fraction of
cool-core clusters in X-ray selected samples.
%Thus, the effect of cool-cores on the X-ray selection is clearly
%understood and can be accounted for in cosmological studies. 

\end{abstract}



\section{Introduction}

Clusters of galaxies are the largest gravitationally bound structures
in the Universe. In the standard $\Lambda$CDM cosmology,
massive halos dominated by
dark matter assemble by the accretion of smaller groups and clusters \citep[e.g.,][]{1982Forman,2011Allen,2012Kravtsov}. 
Under the influence of gravity, diffuse matter and smaller collapsed halos fall into
larger halos and, occasionally, halos of comparable mass merge with
one another. X-ray observations of substructures in clusters of galaxies 
\citep[see, for instance,][]{1984Jones,1979Jones,1999Jones,1995Mohr,1996Buote,2005Jeltema,2010Bohringer,2010Lagana,
2012Andrade-Santos,2013Andrade-Santos} and measurements of the growth of structure
\citep{2009bVik,2010Mantz,2011Allen,2013Benson,2014Planck,2016Planck}
demonstrate that clusters are still in the process of formation.

Early X-ray observations of galaxy clusters revealed that a
significant fraction present a bright and dense core, whose central
cooling time is much shorter than the Hubble time. These observations
led to the development of the cooling flow model \citep{1977Fabian,1984Fabian,1994Fabian,2012Fabian}.
Analyzing deep {\em Chandra} observations of Hydra-A, \citet{2001David} showed
that the spectral fits yielded significantly smaller mass
deposition rates than expected.
Using XMM-{\em Newton}, \citet{2003Peterson}
presented high-resolution X-ray spectra of 14 putative cooling-flow
clusters that exhibit a severe deficit of very cool emission relative
to the predictions of the isobaric cooling-flow model. 
However, as predicted by the cooling-flow model, 
a temperature drop in the center of many clusters is
observed, typically reaching one third of the peak temperature
\citep[e.g.,][]{2003Peterson,2005Vik}. Clusters presenting such a temperature drop 
in their cores are referred to as cool-core (CC) clusters \citep{2001Molendi}.

Using a very large, high-resolution cosmological N-body
simulation (Millennium-XXL), \citet{2012Angulo} showed that cosmological conclusions
based on galaxy cluster surveys depend critically on the selection
biases, which include the wavelength used
for identification of clusters of galaxies. \citet{2011PlanckXMMvalid,2013PlanckXMMvalid}
presented the first observational
evidence of different morphological properties in X-ray vs. SZ
selected samples: SZ selected clusters have a less 
peaked density distribution and are less X-ray luminous at a given mass than
X-ray selected clusters. \citet{2013Wen} presented a method to diagnose 
substructure and dynamical state for 2092 optical galaxy clusters. 
They found that the fraction of relaxed clusters is 28\% in the full
sample, while the fraction increases to 46\% for the subsample matched
with ROSAT detections, indicating that the wavelength used for
detecting clusters plays a significant role in the dynamical state of the population that is
selected. \citet{2013McDonald} showed that CC clusters in his SZ
sample represent 40$\pm$10\% of the cluster population at low
redshift. \citet{2014Sommer} constructed near-complete samples
based on X-ray and SZ catalogs. They found that roughly 70$\pm$10\% of
the clusters in the X-ray selection have no radio halos (indicating
they are relaxed), whereas the fraction in the {\em Planck} 
selection is only 30$\pm$10\%, in agreement with findings from \citet{2013Wen,2013McDonald}.

More recently, \citet{2016Rossetti} compared the dynamical state of the 132 clusters with the
highest signal to noise ratio from the {\em Planck} Sunyaev-Zel'dovich (SZ) catalogue,
to that of three X-ray-selected samples 
(HIFLUGCS\footnote{HIFLUGCS -- The HIghest X-ray FLUx Galaxy Cluster Sample \citep{2002Reiprich}},
MACS\footnote{MACS -- The MAssive Cluster Survey \citep{2001Ebeling}}, and
REXCESS\footnote{REXCESS -- The REpresentative XMM-Newton ClustEr Structure Survey \citep{2007Bohringer}}).
They showed that the fractions of relaxed clusters in the the X-ray
samples are significantly larger than that in the Planck sample, and
interpreted this result as an indication of a cool core bias 
\citep{2011Eckert} affecting X-ray selection.

Recently, \citet{2017Rossetti} compared the fraction of cool core clusters in a
{\rm Planck} cosmology SZ sample \citep[PSZ1,][]{2014Planck} to that in the MACS X-ray sample.
Using the concentration parameter which measures the ratio of the
integrated surface brightness in two fixed physical apertures, 
as defined by \citet{2008Santos}, 
they showed that the cool core fraction is significantly higher
in the MACS X-ray selected sample than in the Planck cosmology SZ sample (59$\pm$5\% in their X-ray sample
vs. 29$\pm$4\% in their SZ sample). This result 
agrees with that presented by \citet{2017Andrade-Santos}, which is fully
described in this paper.
The X-ray sample presented in this
paper spans a higher mass range compared to the mass
range in the X-ray sample presented by \citet{2017Rossetti}. 
Therefore, we also make use of three parameters
which are computed at radii that scale with total mass. We note that the work presented by \citet{2017Rossetti} and the
work presented here were developed in parallel \citep{2016Jones,2017Andrade-Santos}.

In this paper we compare the nature of the cores for 164 {\em Planck}
ESZ clusters at $z < 0.35$ to the 100 highest flux X-ray clusters at Galactic latitudes
$|b| > 20^\circ$ and $0.025 < z < 0.30$. The X-ray sample is extended to 100
clusters from the sample of 52 clusters
presented by \citet{2004Voevodkin}, by lowering the flux limit to $f_{\rm X} > 7.5 \times
10^{-12} \rm ~erg ~ s^{-1} ~ cm^{-2}$ in the 0.5 -- 2.0 keV band.

Throughout this paper, we assume a standard $\Lambda$CDM cosmology
with $H_0= ~ 70 \rm ~km~s^{-1}~Mpc^{-1}$, $\Omega_{\rm M}=0.3$, and
$\Omega_{\rm \Lambda}=0.7$. All uncertainties are quoted at the $1\sigma$ level.

%%%%%%%%%%%%%%%%%%%%%%%%%%%%%%%%%%%%%%%%%%%%%%%%%%%%%%%%%%%%%%%%%%%%%%%%%%%%%%%%%%
%%%%%%%%%%%%%%%%%%%%%%%%%%%%%%%%%%%%%%%%%%%%%%%%%%%%%%%%%%%%%%%%%%%%%%%%%%%%%%%%%%
%%
%%                                  SAMPLES
%%
%%%%%%%%%%%%%%%%%%%%%%%%%%%%%%%%%%%%%%%%%%%%%%%%%%%%%%%%%%%%%%%%%%%%%%%%%%%%%%%%%%
%%%%%%%%%%%%%%%%%%%%%%%%%%%%%%%%%%%%%%%%%%%%%%%%%%%%%%%%%%%%%%%%%%%%%%%%%%%%%%%%%%

\section{SZ and X-ray selected clusters}

The main goal of this work is to compare the fraction of CC clusters
in X-ray and SZ selected samples. In this section, we describe the two
samples. 

\subsection{SZ sample}\label{sec:esz}

The first catalog of 189 SZ clusters detected by
the {\em Planck} mission was released
in early 2011 \citep{2011PlanckCol}.
A \cxc ~XVP (X-ray Visionary Program -- PI: Jones) and HRC
Guaranteed Time Observations (PI: Murray) combined
to form the {\em
Chandra-Planck} Legacy Program for Massive Clusters of 
Galaxies\footnote{\scriptsize \href{http://hea-www.cfa.harvard.edu/CHANDRA_PLANCK_CLUSTERS/}{hea-www.cfa.harvard.edu/CHANDRA\_PLANCK\_CLUSTERS/}}.
%\footnote{\scriptsize \href{http://hea-www.cfa.harvard.edu/CHANDRA_PLANCK_CLUSTERS/}{http://goo.gl/3yHJt9}}, 
For each of the 164 ESZ
{\em Planck} clusters at $z \le 0.35$, we obtained 
\cxc ~exposures sufficient to collect at least 10,000 source counts.

\begin{figure}[!t]
\centerline{
\includegraphics[width=0.47\textwidth]{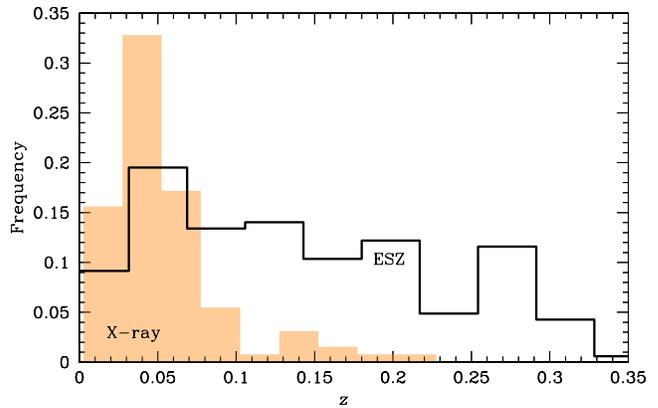}
}
\caption{\small{
Distribution of cluster redshifts in the ESZ and X-ray selected
samples. The ESZ sample extends to higher redshifts. The solid
histogram shows the X-ray flux limited sample while the solid line
corresponds to the ESZ cluster sample.
}}\label{fig:histz}
\end{figure} 

\subsection{X-ray sample}\label{sec:xray}

\citet{2004Voevodkin} compiled a sample of the X-ray brightest
clusters in the local universe by selecting the highest flux clusters detected in the 
ROSAT All-Sky survey at $|b| > 20^\circ$ and 
$z > 0.025$ -- using the HIFLUGCS catalog \citep{2002Reiprich} as
reference. The sample used here is an extension of the   
\citet{2004Voevodkin} sample, where the flux limit in the 0.5 -- 2.0 keV band was
lowered to $f_{\rm X} > 7.5 \times 10^{-12} \rm
~erg ~ s^{-1} ~ cm^{-2}$. This sample contains 100 clusters and has an
effective redshift depth of $z < 0.3$.
All have {\em Chandra} observations. Of the 100 X-ray selected clusters, 49 are also 
in the ESZ sample, and 47 are in the HIFLUGCS catalog.

\begin{figure}[!t]
\centerline{
\includegraphics[width=0.47\textwidth]{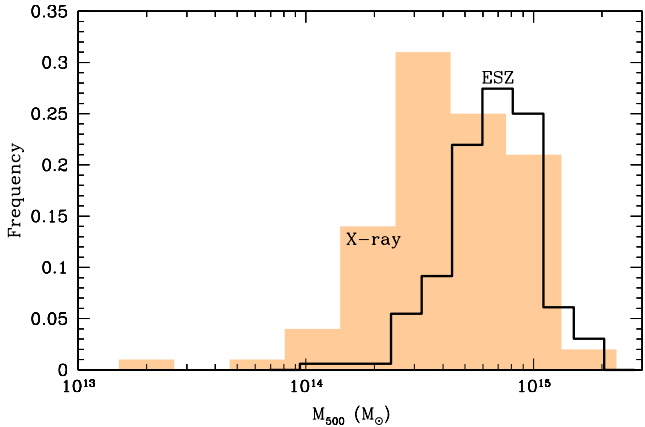}
}
\caption{\small{
Distribution of cluster masses within $r_{500}$ for the ESZ and X-ray selected
samples (see Sections \ref{sec:esz} and \ref{sec:xray}). The X-ray sample extends to lower masses than
does the ESZ sample.
The solid histogram shows the X-ray flux limited sample, while the solid line
corresponds to the ESZ cluster sample.
}}\label{fig:histm}
\end{figure}

\subsection{Comparisons between the X-ray and ESZ selected clusters}

Figure \ref{fig:histz} presents the redshift distribution in
both samples. The {\em Planck} detected clusters are clearly more
broadly distributed in redshift than are the X-ray clusters. This is due to the nature of the selection:
for resolved clusters the Sunyaev-Zel'dovich signal is independent of the redshift of
the cluster because it is the CMB that is distorted (the CMB photons
originated  at the epoch of recombination -- from a constant redshift
of $z \sim 1000$), while the X-ray selected clusters constitute
a flux-limited sample, which strongly favors the X-ray brighter, lower redshift clusters. 
Figure \ref{fig:histm} presents the mass
distribution of both samples. The X-ray sample spans a much larger
mass range, extending to lower masses than the {\em Planck}
ESZ sample. The difference between the lowest observed mass in
the X-ray and ESZ samples is caused by different detection
thresholds. Note that the highest observed mass is the same
for both samples (see Figure \ref{fig:histm}).

\subsection{Subclusters}

A small fraction ($\sim$ 10 -- 20\%) of the clusters in both the X-ray and SZ
samples present X-ray bright subclusters. In our analyses we exclude
the secondary subclusters. Only the principal cluster component is used in
the comparisons between the X-ray and SZ samples. However, we present
the metrics for all cluster components in Tables
\ref{tab:ESZmetrics} and \ref{tab:Xraymetrics}.

%%%%%%%%%%%%%%%%%%%%%%%%%%%%%%%%%%%%%%%%%%%%%%%%%%%%%%%%%%%%%%%%%%%%%%%%%%%%%%%%%%
%%%%%%%%%%%%%%%%%%%%%%%%%%%%%%%%%%%%%%%%%%%%%%%%%%%%%%%%%%%%%%%%%%%%%%%%%%%%%%%%%%
%%
%%                                  DATA REDUCTION
%%
%%%%%%%%%%%%%%%%%%%%%%%%%%%%%%%%%%%%%%%%%%%%%%%%%%%%%%%%%%%%%%%%%%%%%%%%%%%%%%%%%%
%%%%%%%%%%%%%%%%%%%%%%%%%%%%%%%%%%%%%%%%%%%%%%%%%%%%%%%%%%%%%%%%%%%%%%%%%%%%%%%%%%

\section{Data Reduction}
\label{sec:chandra} 

Our {\em Chandra} data reduction followed the process described in
\citet{2005Vik}. We applied the calibration files \texttt{CALDB 4.7.2}. 
The data reduction included corrections for the time dependence of the charge
transfer inefficiency and gain, and also a check for periods
of high background, which were then omitted. Standard blank sky background files
and readout artifacts were subtracted. We also detected compact X-ray sources in the 0.7--2.0 keV and 2.0--7.0 keV
bands using CIAO wavdetect and then masked these sources before performing the spectral and spatial
analyses of the cluster emission. For each cluster, we used all available {\em
Chandra} observations within 2 Mpc of the cluster center with all CCDs (ACIS-I and ACIS-S). 

\begin{figure*}[!t]
\centerline{
\includegraphics[width=0.48\textwidth]{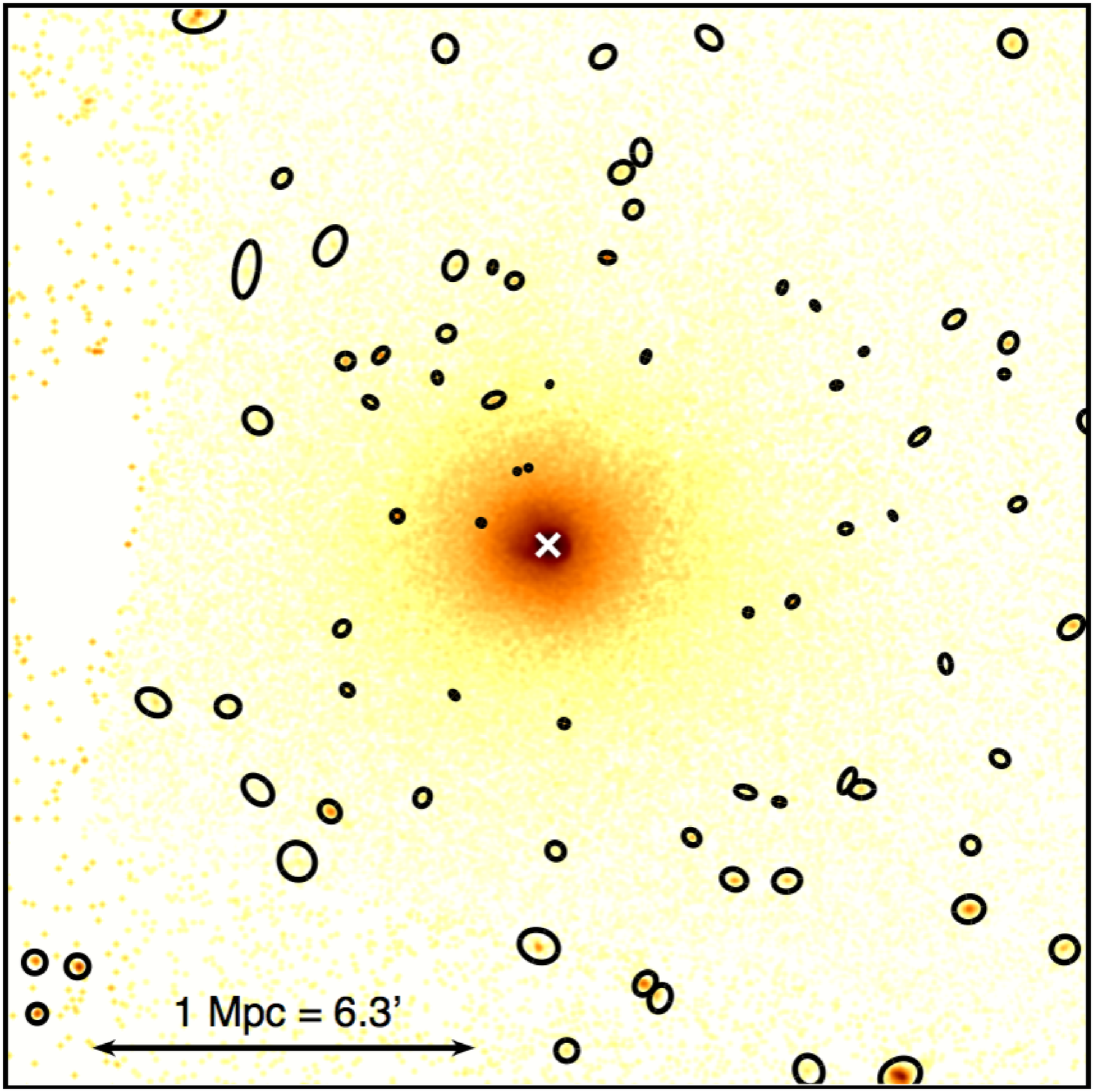}
\includegraphics[width=0.47\textwidth]{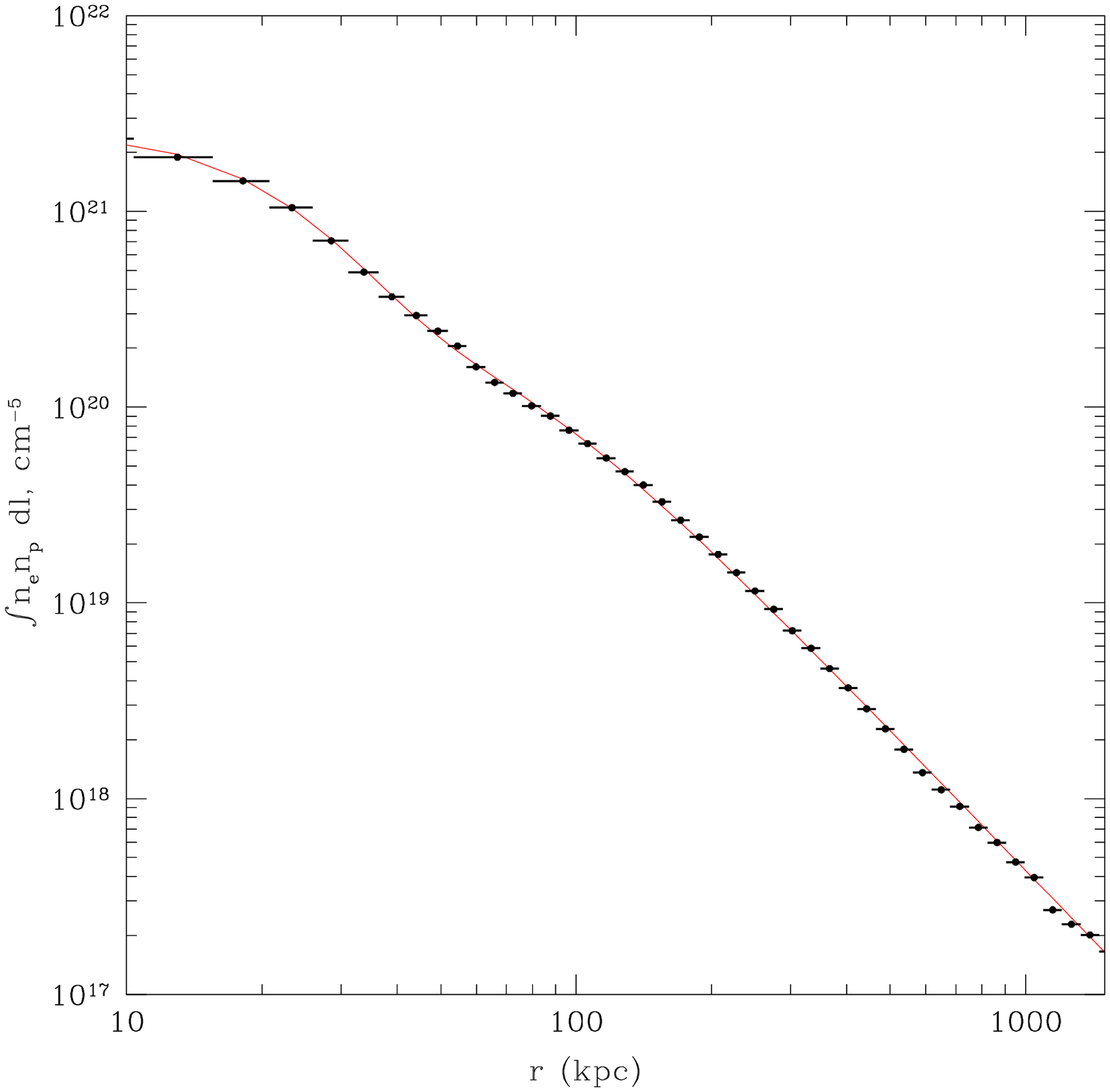}
}
\centerline{
\includegraphics[width=0.47\textwidth]{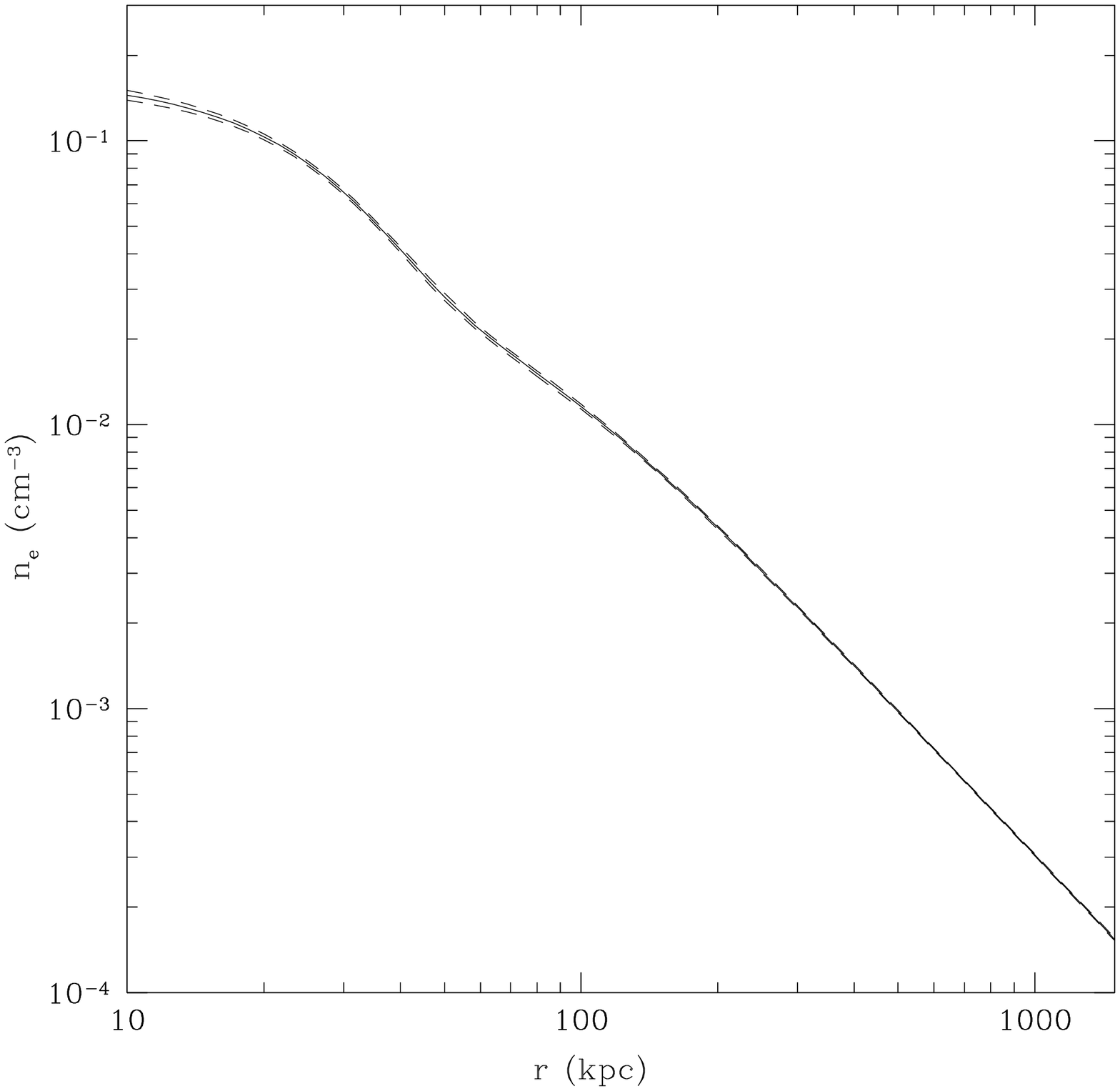}
\includegraphics[width=0.47\textwidth]{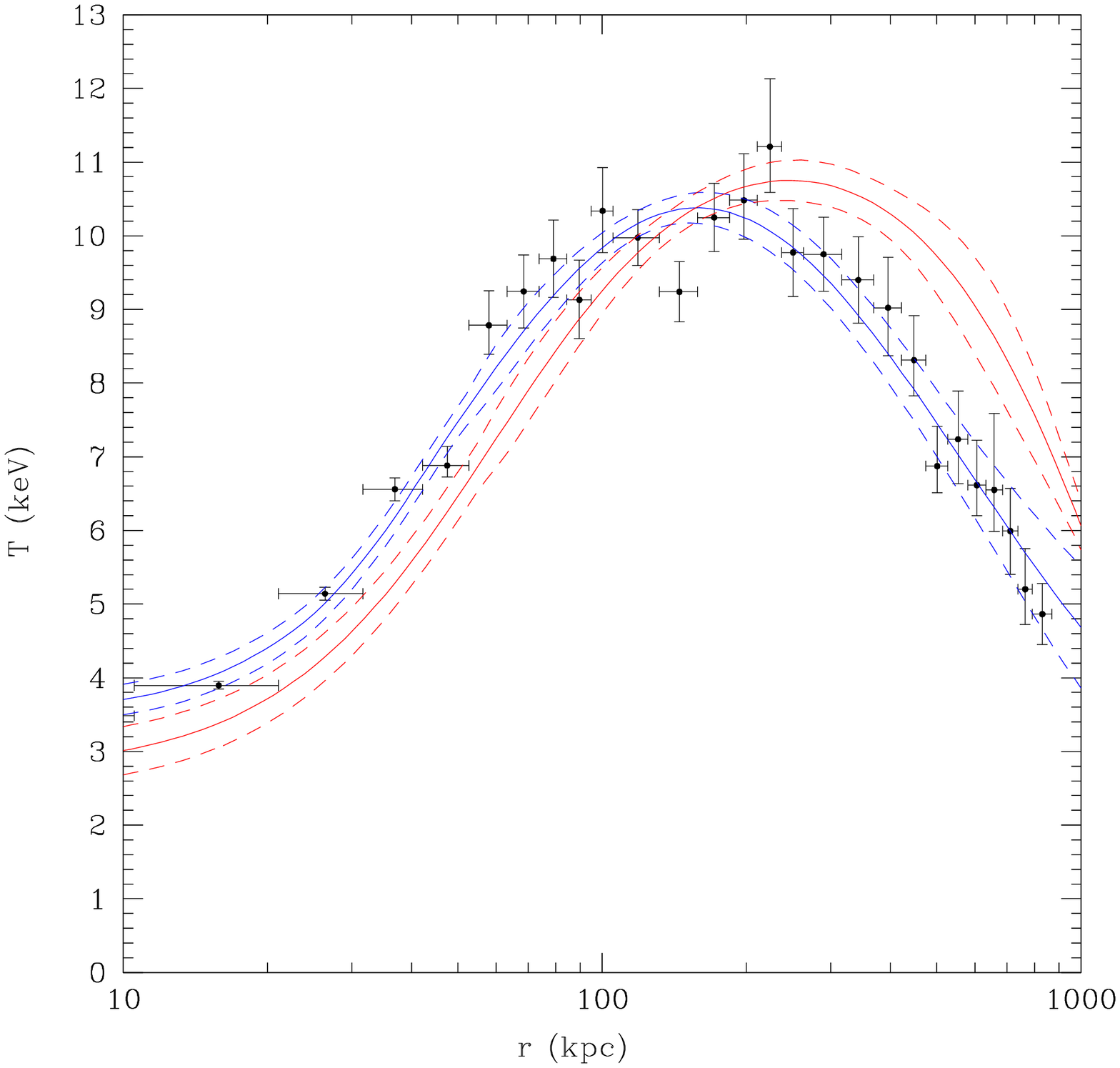}
}
\caption{\small{Example of the X-ray image and the projected emissivity, gas density,
and temperature profiles for a cool-core cluster (A2204). Top left panel
shows the 0.5-2.0 keV, background-subtracted, exposure map corrected
ACIS-I image. The total filtered {\em Chandra} exposure is 117
ks. Black ellipses correspond to the masked X-ray point
sources and the cross corresponds to the cluster center. Top right panel shows the projected emissivity profile. The
solid line shows the emission measure integral of the best fit
to the emissivity profile given by Equation (\ref{eq:nenp}). 
Bottom left panel shows the gas density profile. The
solid line shows the density profile obtained from the emissivity
profile given by Equation
(\ref{eq:nenp}).
Bottom right panel shows the gas temperature profile. The red and blue lines
show the de-projected and projected temperature profiles,
respectively (Equations \ref{eq:tprof} and \ref{eq:tspec}). The dashed lines in the gas and temperature profiles show the 68\%
confidence range. This is an example of a cluster with deep {\em Chandra} exposure,
which allow us to extract the temperature profile in many radial bins.
}}\label{fig:emm_dens}
\end{figure*}

\begin{figure*}[!t]
\centerline{
\includegraphics[width=0.48\textwidth]{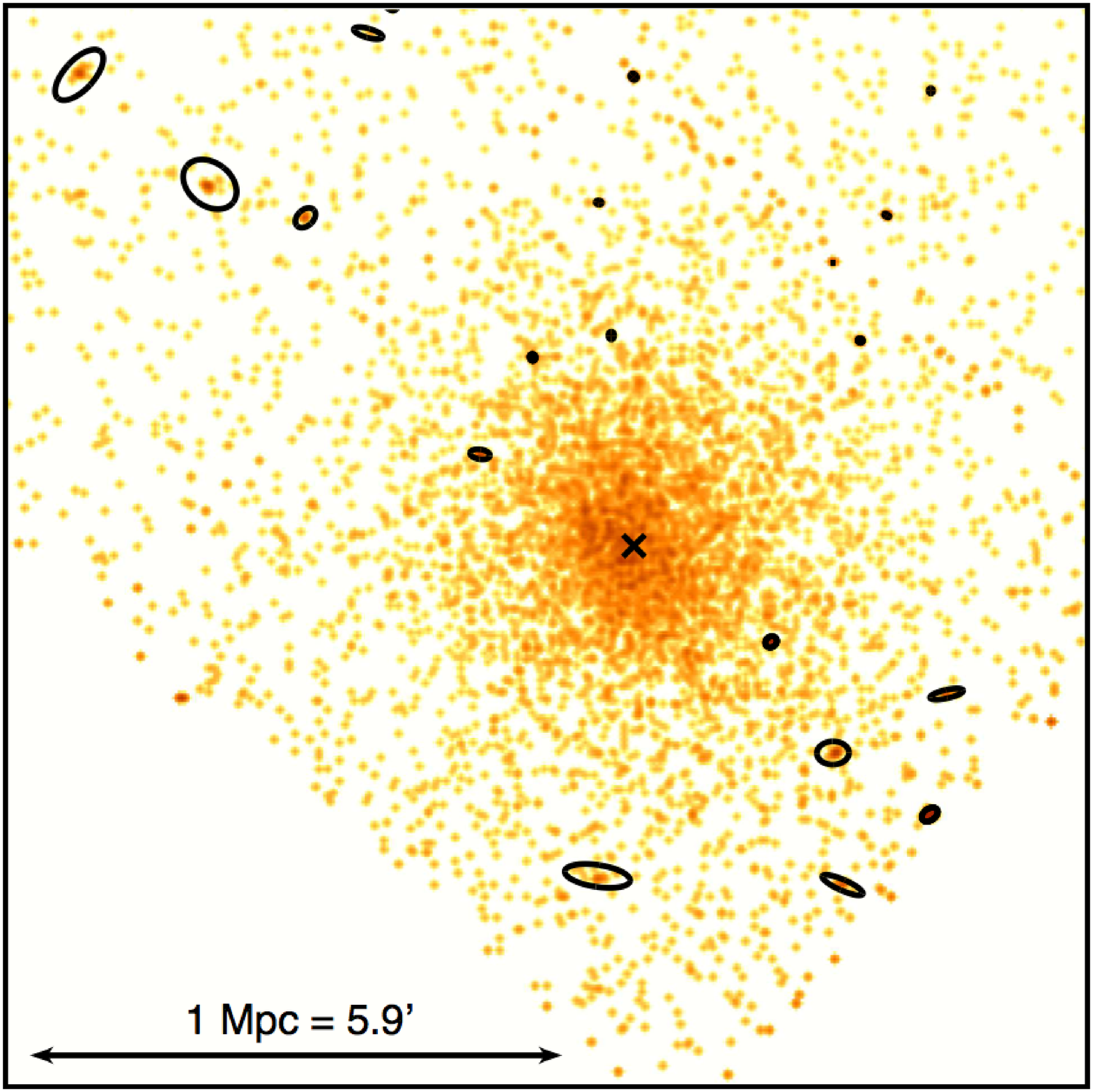}
\includegraphics[width=0.47\textwidth]{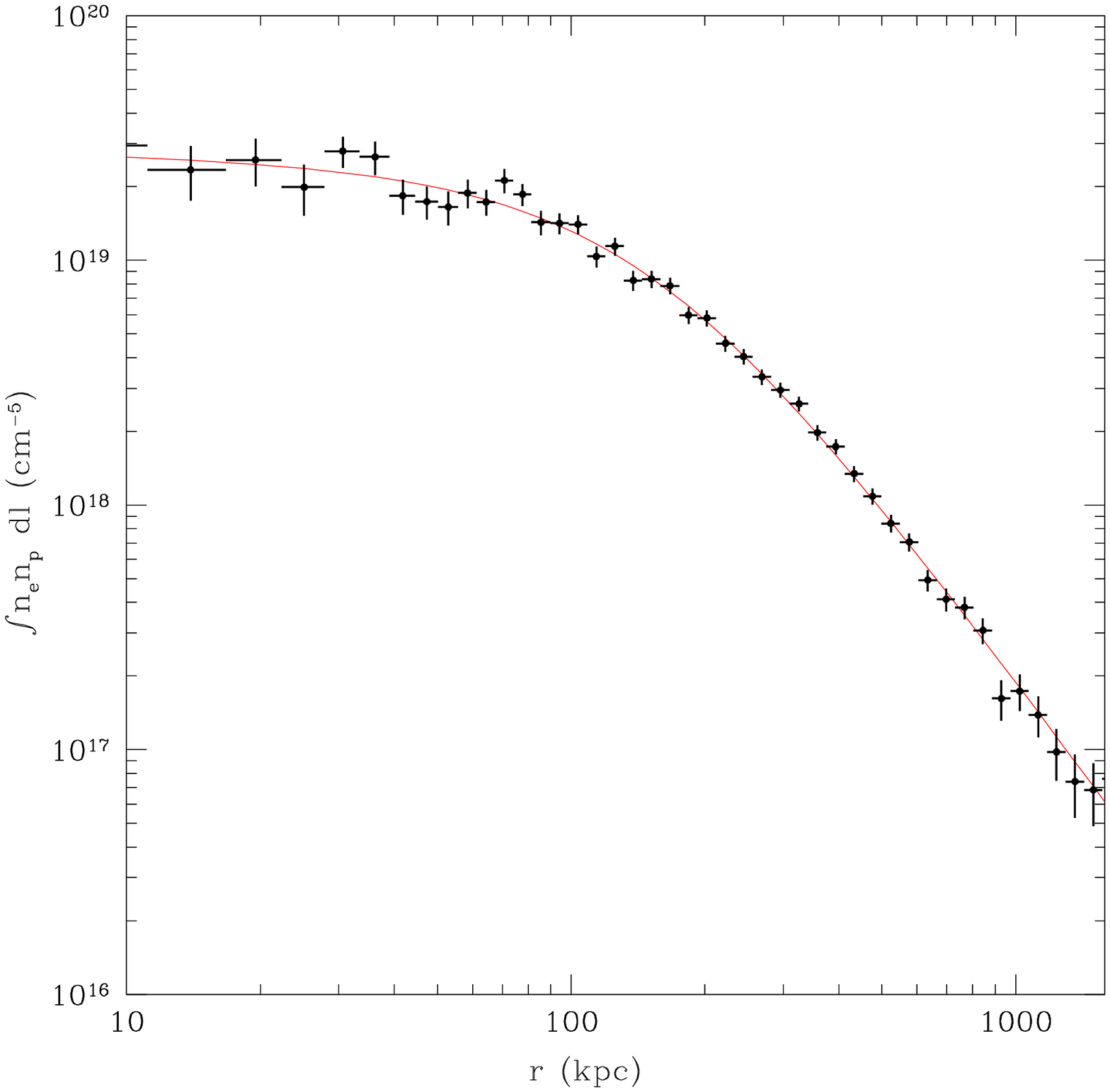}
}
\centerline{
\includegraphics[width=0.47\textwidth]{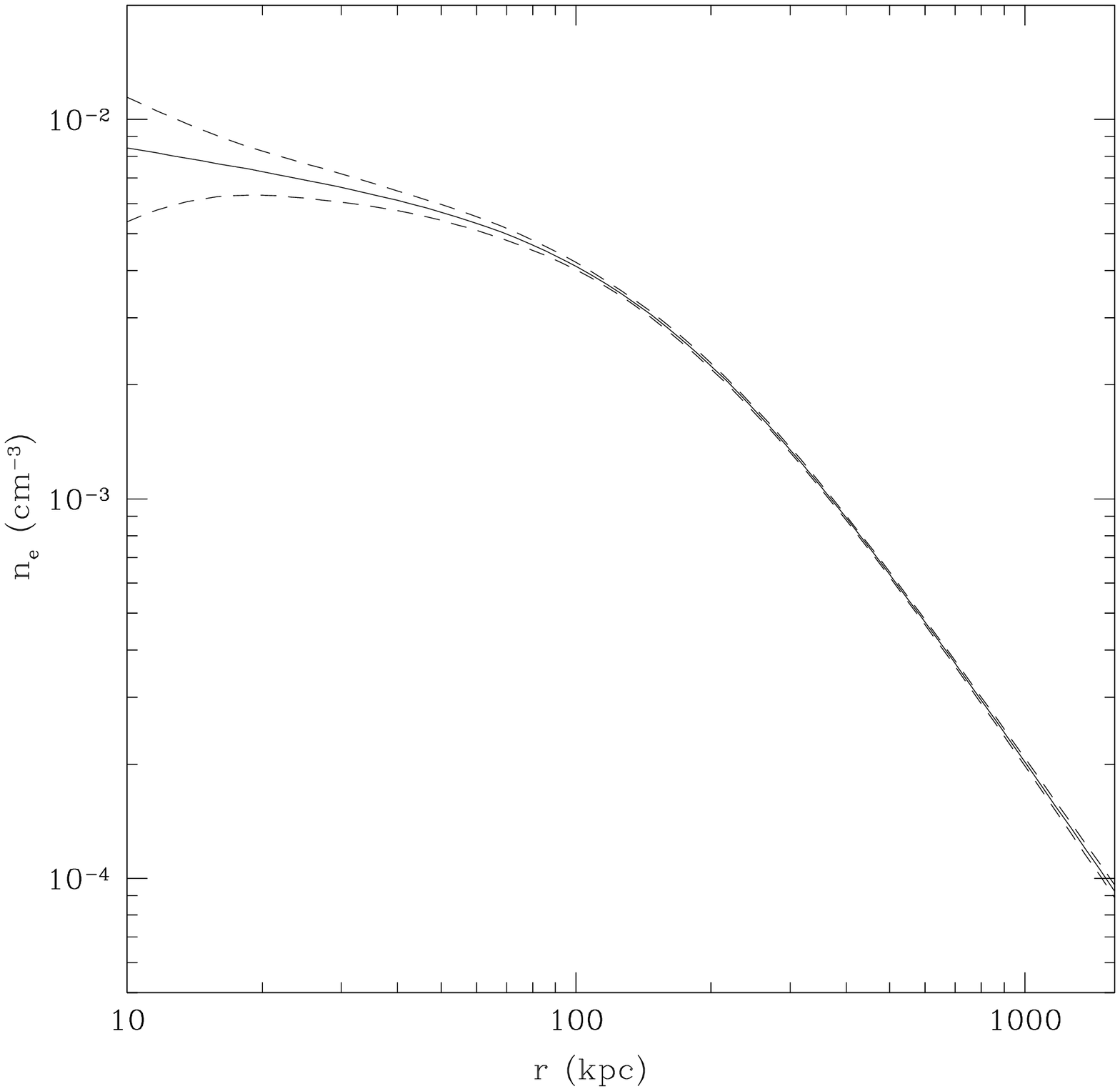}
\includegraphics[width=0.47\textwidth]{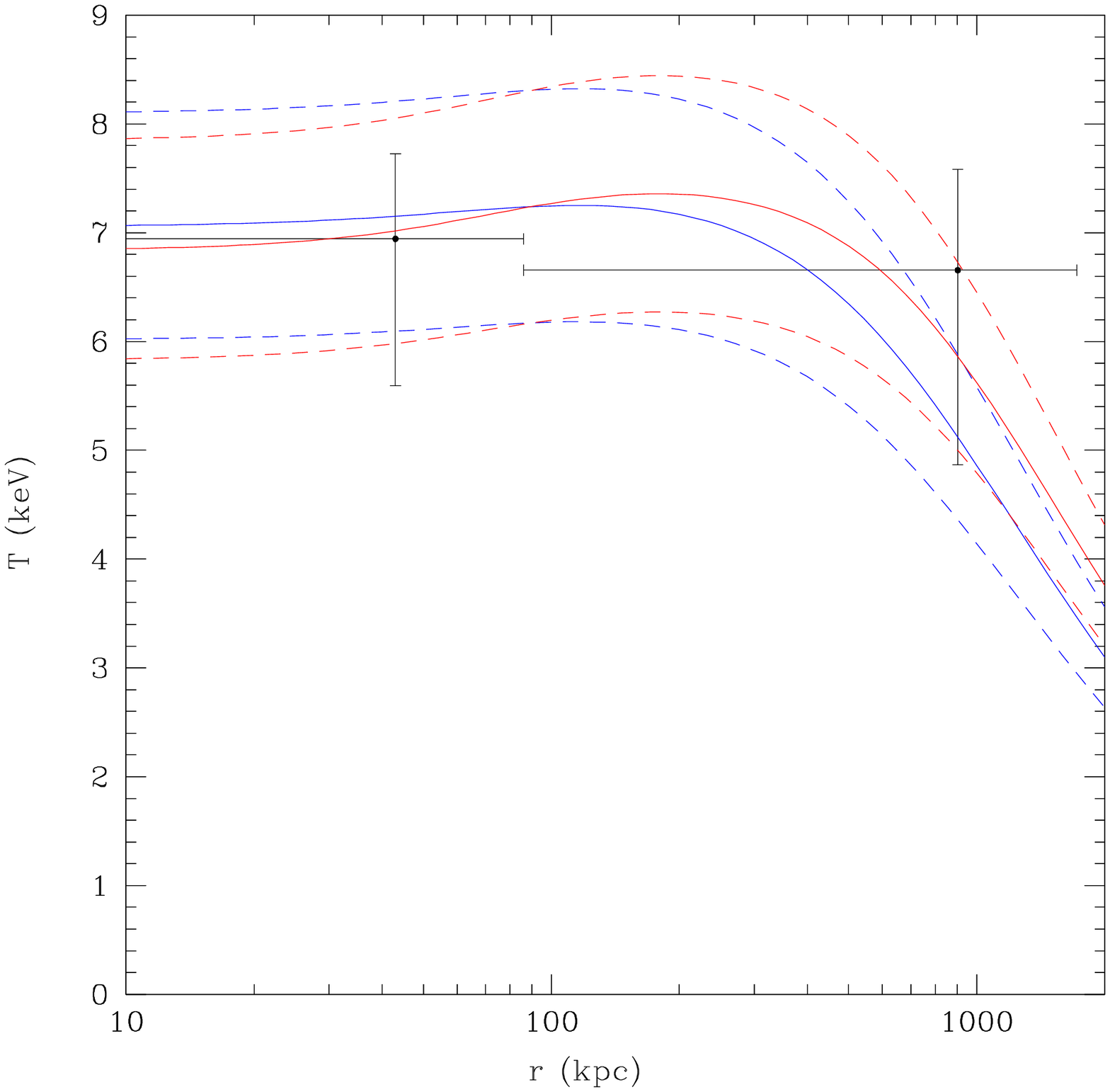}
}
\caption{\small{
Similar panels as in Figure \ref{fig:emm_dens}, for the non-cool-core
cluster PLCKESZ G000.44-41.83. This is an example of a cluster with 
moderate data quality, which illustrates the necessity of metrics to
determine the nature of the cluster core in the absence of a detailed
temperature profile, like that shown in Figure \ref{fig:emm_dens}. 
The total filtered {\em Chandra} exposure is 14 ks.
}}\label{fig:emm_dens_G000}
\end{figure*}

%%%%%%%%%%%%%%%%%%%%%%%%%%%%%%%%%%%%%%%%%%%%%%%%%%%%%%%%%%%%%%%%%%%%%%%%%%%%%%%%%%
%%%%%%%%%%%%%%%%%%%%%%%%%%%%%%%%%%%%%%%%%%%%%%%%%%%%%%%%%%%%%%%%%%%%%%%%%%%%%%%%%%
%%
%%                               EMM PROFILES
%%
%%%%%%%%%%%%%%%%%%%%%%%%%%%%%%%%%%%%%%%%%%%%%%%%%%%%%%%%%%%%%%%%%%%%%%%%%%%%%%%%%%
%%%%%%%%%%%%%%%%%%%%%%%%%%%%%%%%%%%%%%%%%%%%%%%%%%%%%%%%%%%%%%%%%%%%%%%%%%%%%%%%%%

\section{Emission measure profiles}

We refer to \citet{2006Vik} for a detailed description of
the procedures we used to compute the emission measure profile for
each cluster.
We outline here only the main aspects of the method. 

We measured the surface brightness profiles in the 0.7--2.0 keV energy band, 
which maximizes the signal to noise ratio in
{\em Chandra} observations for typical cluster gas temperatures. We used the X-ray halo peak
as the cluster center.
The readout artifacts and blank-field background \citep[see section 2.3.3 of][]{2006Vik} 
were subtracted from the X-ray images, and the results were then exposure-corrected, using exposure maps
computed assuming an absorbed optically-thin thermal plasma with $kT = 5.0$
keV, abundance = 0.3 solar, with the Galactic column density
and including corrections for bad pixels and CCD gaps, which do not 
take into account spatial variations of the effective area. 
We subtracted any small uniform component corresponding to 
soft X-ray foreground adjustments, if required (determined by fitting 
a thermal model in a region of the detector field distant from the
cluster center, taking into account the expected thermal contribution from the cluster).

Following these steps, we extracted the surface brightness in narrow 
concentric annuli ($r_{\rm out}/r_{\rm in} = 1.05$) centered on the
X-ray halo peak and computed the {\em Chandra} area-averaged effective area for each
annulus (see \citet{2005Vik}, for details on calculating the effective area). 
To compute the emission measure and temperature profiles, we assumed spherical 
symmetry. The spherical assumption is expected to introduce only small 
deviations  in the emission measure profile \citep{2003Piffaretti}.
Using the modeled de-projected temperature (see Section \ref{sec:temp}), effective area, and 
metallicity as a function of radius, we converted the {\em Chandra} count 
rate in the 0.7--2.0 keV band into the emission integral, 
${\rm EI} =  \int n_{\rm e} n_{\rm p} dV$, within each cylindrical
shell. Tables \ref{tab:ESZmetrics} and \ref{tab:Xraymetrics} list
the maximum cluster radius where the emission integral is computed ($r_{\rm max}$)
for each cluster. Seven clusters in the ESZ sample have 
$r_{\rm max} < r_{500}$, and in the X-ray sample, nine clusters
have this condition (four of them are also in the ESZ sample). These
numbers represent only 4\% and 9\% of the clusters in the ESZ and X-ray
samples, respectively.

We fit the emission measure profile assuming the gas density
profile follows that given by \citet{2006Vik}: 

\begin{eqnarray}
n_{\rm e}n_{\rm p} &=& n_0^2
\frac{ (r/r_{\rm c})^{-\alpha}}{(1+r^2/r_{\rm c}^2)^{3\beta-\alpha/2}}
\frac{1}{(1+r^\gamma/r_{\rm s}^\gamma)^{\epsilon/\gamma}} \nonumber \\
&+& \frac{n^2_{02}}{(1+r^2/r_{\rm c2}^2)^{3\beta_2}},\label{eq:nenp}
\end{eqnarray} 
where the parameters $n_0$ and $n_{02}$ determine the normalizations of
both additive components. $\alpha$, $\beta$, $\beta_2$, and $\epsilon$ are indexes controlling the slope of the
curve at characteristic radii given by the parameters $r_{\rm c}$,
$r_{\rm c2}$, and $r_{\rm s}$. $\gamma$ controls the width of the
transition region given by $r_{\rm s}$. Although the relation given
by Equation \ref{eq:nenp} is based on a
classic  $\beta$-model \citep{1976Cavaliere}, it is modified to account
for a central power-law type cusp and a steeper emission measure slope at large radii. In addition,
a second $\beta$-model is included, to better characterize the cluster core.
For further details on this equation, we refer the reader to
\citet{2006Vik}. In
the fit to the emissivity profile, all parameters are free to vary.
For a typical metallicity of
0.3 $Z_\odot$, the reference values from \citet{1989AndersGrevesse}
yield $n_{\rm e}/n_{\rm p} = 1.1995$. Examples of projected emissivity
and gas density
profiles are presented in Figures \ref{fig:emm_dens} and \ref{fig:emm_dens_G000}.

%%%%%%%%%%%%%%%%%%%%%%%%%%%%%%%%%%%%%%%%%%%%%%%%%%%%%%%%%%%%%%%%%%%%%%%%%%%%%%%%%%
%%%%%%%%%%%%%%%%%%%%%%%%%%%%%%%%%%%%%%%%%%%%%%%%%%%%%%%%%%%%%%%%%%%%%%%%%%%%%%%%%%
%%
%%                                  MASS ESTIMATES
%%
%%%%%%%%%%%%%%%%%%%%%%%%%%%%%%%%%%%%%%%%%%%%%%%%%%%%%%%%%%%%%%%%%%%%%%%%%%%%%%%%%%
%%%%%%%%%%%%%%%%%%%%%%%%%%%%%%%%%%%%%%%%%%%%%%%%%%%%%%%%%%%%%%%%%%%%%%%%%%%%%%%%%%

\section{Total cluster mass estimates}

Using the gas mass and temperature, we estimated
the total cluster mass from the $Y_X$--$M$ scaling relation of \citet{2009Vik},
\begin{eqnarray}
M_{\rm 500,Y_X} = E^{-2/5}(z)A_{\rm YM}\left(\frac{Y_{\rm X}}{3\times10^{14}M_\odot
  {\rm keV}}\right)^{B_{\rm YM}}, \nonumber \\
\label{e_yx_m}
\end{eqnarray}
where $Y_{\rm X} = M_{\rm gas,500} \times kT_{\rm X}$, $M_{\rm
  gas,500}$ is computed using the best fit parameters of Equation (\ref{eq:nenp}), and $T_{\rm X}$ is the measured
temperature in the (0.15--1) $\times ~
r_{500}$ range. 
$A_{\rm YM}=5.77\times10^{14}h^{1/2}  M_\odot$ and $B_{\rm YM}=0.57$ \citep{2012Maughan}. Here,
$M_{\rm Y_X,500}$ is the total mass within $r_{500}$,
and $E(z)=[\Omega_{\rm M}(1+z)^3 + (1-\Omega_{\rm M}-\Omega_\Lambda)(1+z)^2 +
\Omega_\Lambda]^{1/2}$ is the function describing the evolution of the Hubble
parameter. 

Using Equation (\ref{e_yx_m}), $r_{500}$  is computed by
solving 
\begin{eqnarray}
M_{\rm 500,Y_X} \equiv 500 \rho_c (4\pi/3) r_{500}^{3},\label{m500_def}
\end{eqnarray}
where $\rho_c$ is the critical density of the Universe at the cluster
redshift. In practice, Equation (\ref{e_yx_m}) is evaluated at 
a given radius, whose result is compared
to the evaluation of Equation (\ref{m500_def}) at the same radius. This
process is repeated in an iterative procedure, until the fractional mass difference
is less than 1\%. 

%%%%%%%%%%%%%%%%%%%%%%%%%%%%%%%%%%%%%%%%%%%%%%%%%%%%%%%%%%%%%%%%%%%%%%%%%%%%%%%%%%
%%%%%%%%%%%%%%%%%%%%%%%%%%%%%%%%%%%%%%%%%%%%%%%%%%%%%%%%%%%%%%%%%%%%%%%%%%%%%%%%%%
%%
%%                                  METRICS
%%
%%%%%%%%%%%%%%%%%%%%%%%%%%%%%%%%%%%%%%%%%%%%%%%%%%%%%%%%%%%%%%%%%%%%%%%%%%%%%%%%%%
%%%%%%%%%%%%%%%%%%%%%%%%%%%%%%%%%%%%%%%%%%%%%%%%%%%%%%%%%%%%%%%%%%%%%%%%%%%%%%%%%%

\section{Cool-Core Metrics}

It is not possible to measure the temperature profile 
to determine the presence of a cool-core for all clusters in our SZ
and X-ray samples because of the X-ray data quality. Figures \ref{fig:emm_dens}
and \ref{fig:emm_dens_G000} illustrate the difference in data quality in the sample. 
Instead, we apply a more robust approach of using four metrics
described below, to characterize the presence of cool cores.

\begin{figure*}[!t]
\centerline{
\includegraphics[width=0.47\textwidth]{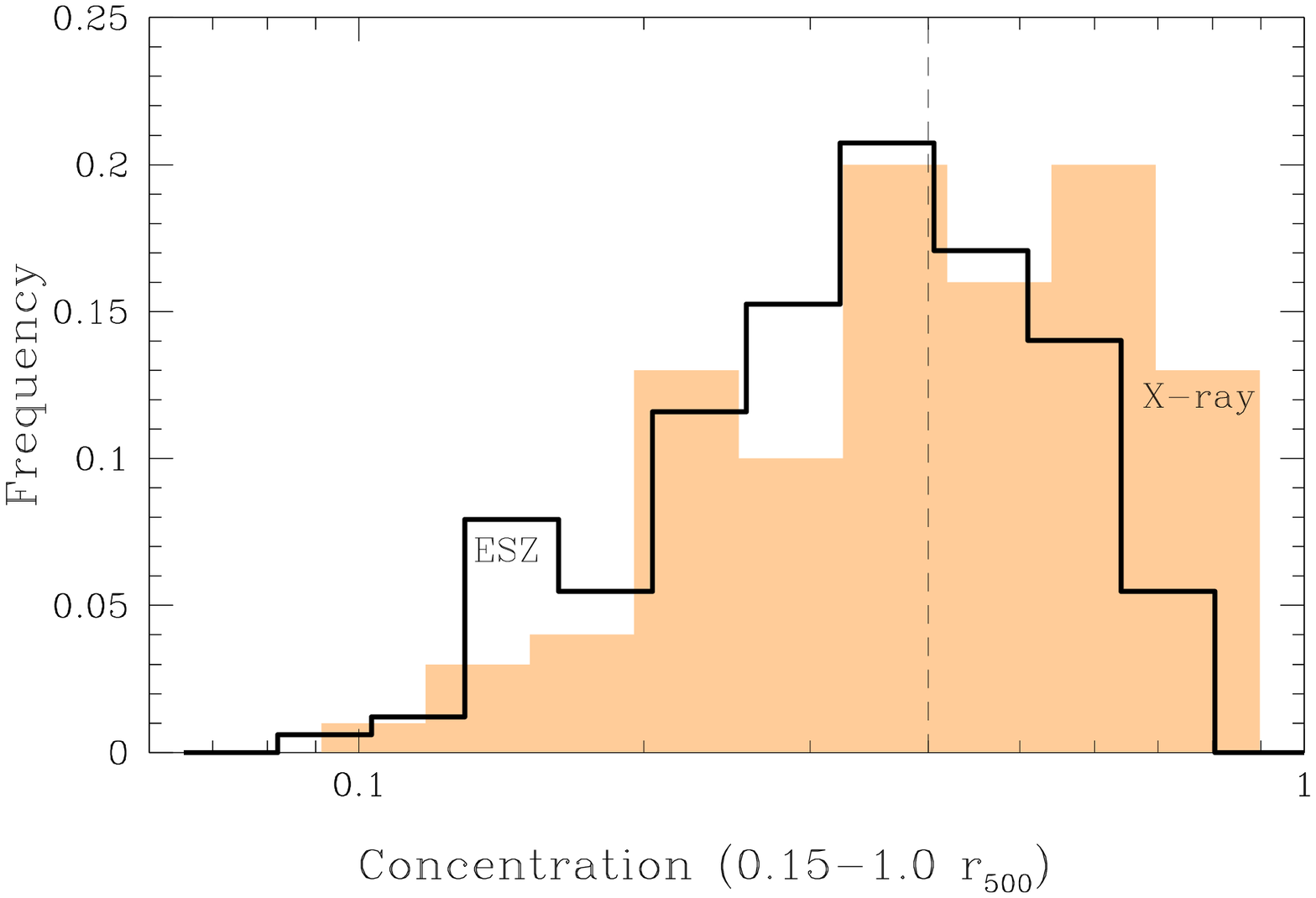}
\includegraphics[width=0.47\textwidth]{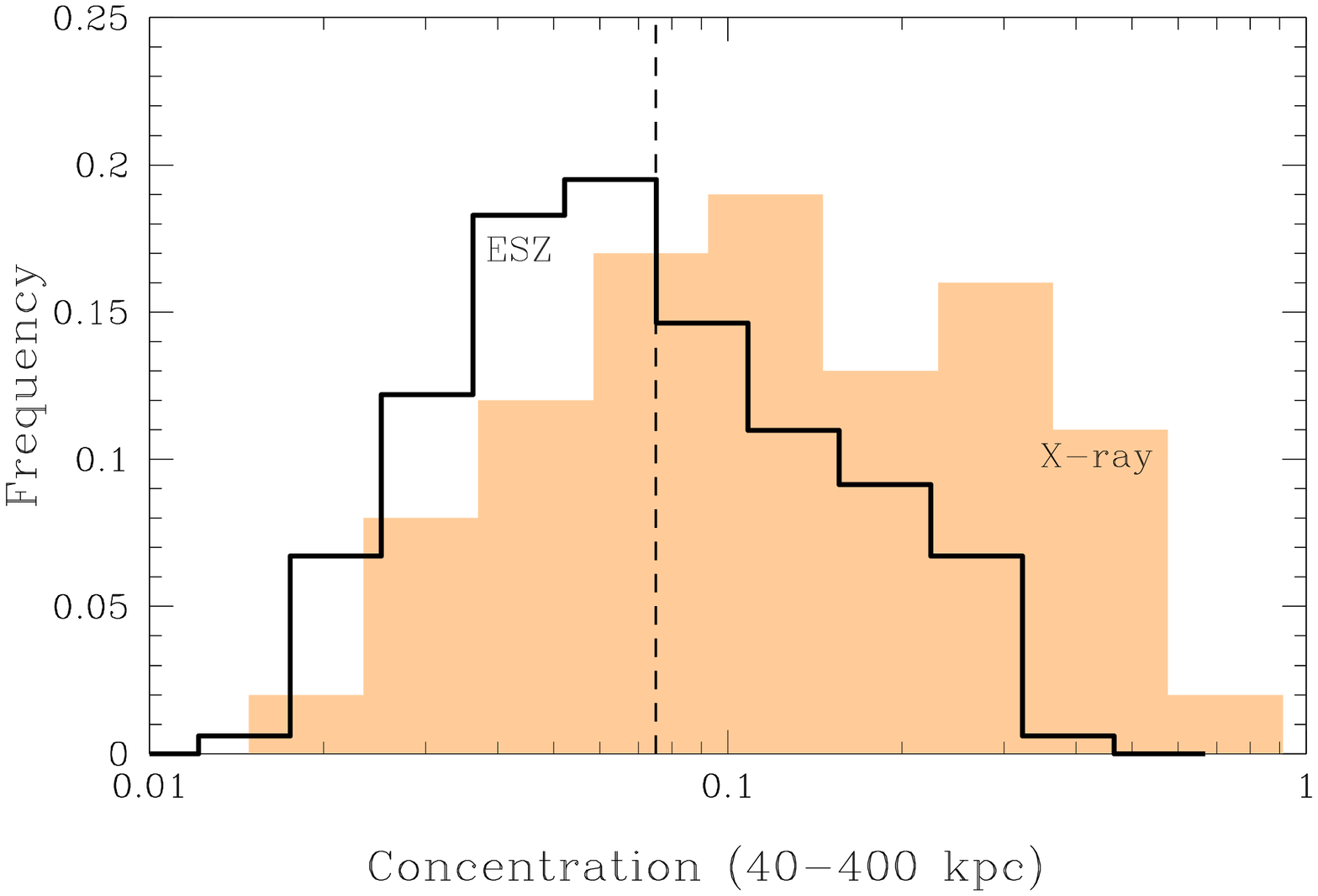}
}
\vspace{0.3cm}
\centerline{
\includegraphics[width=0.47\textwidth]{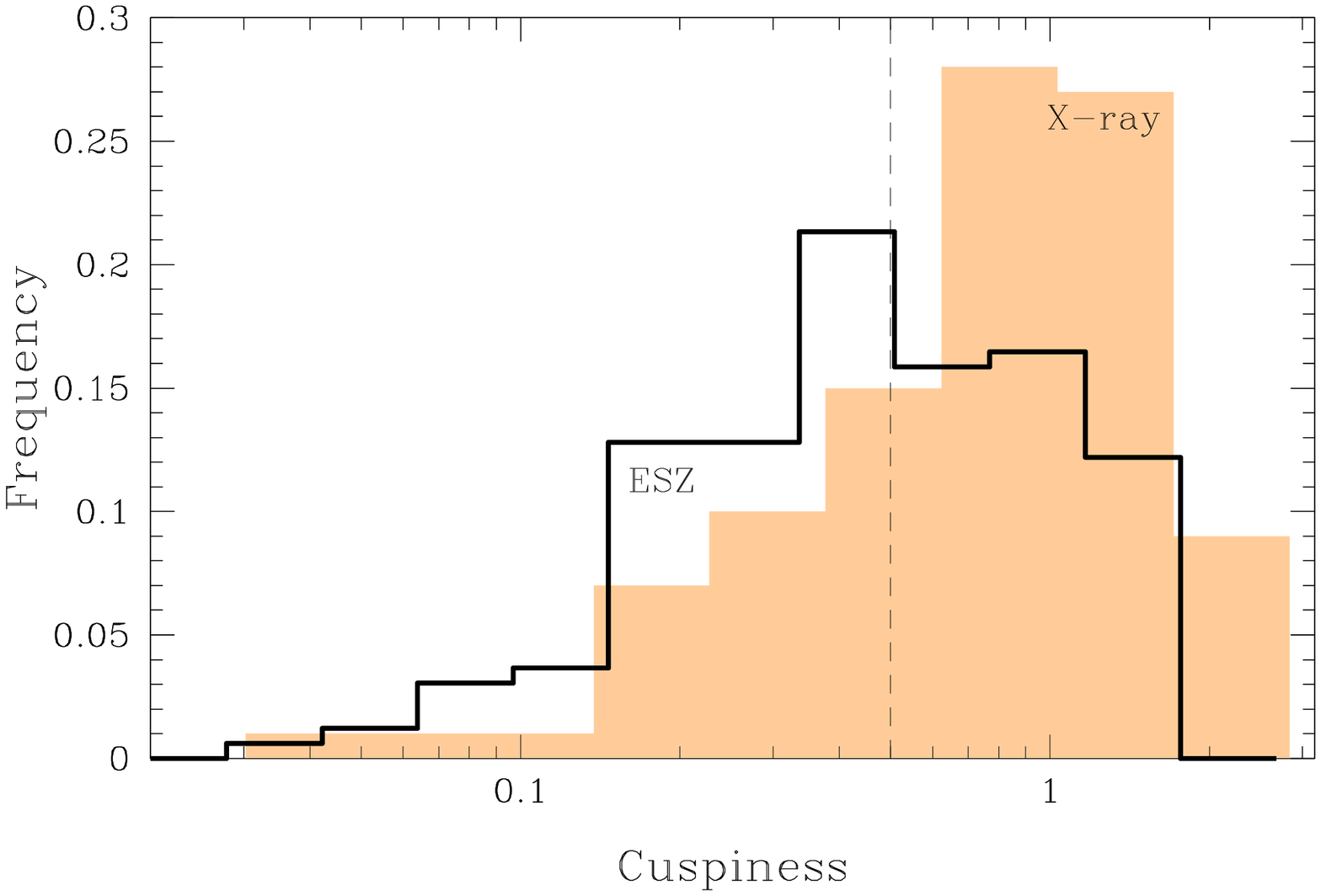}
\includegraphics[width=0.47\textwidth]{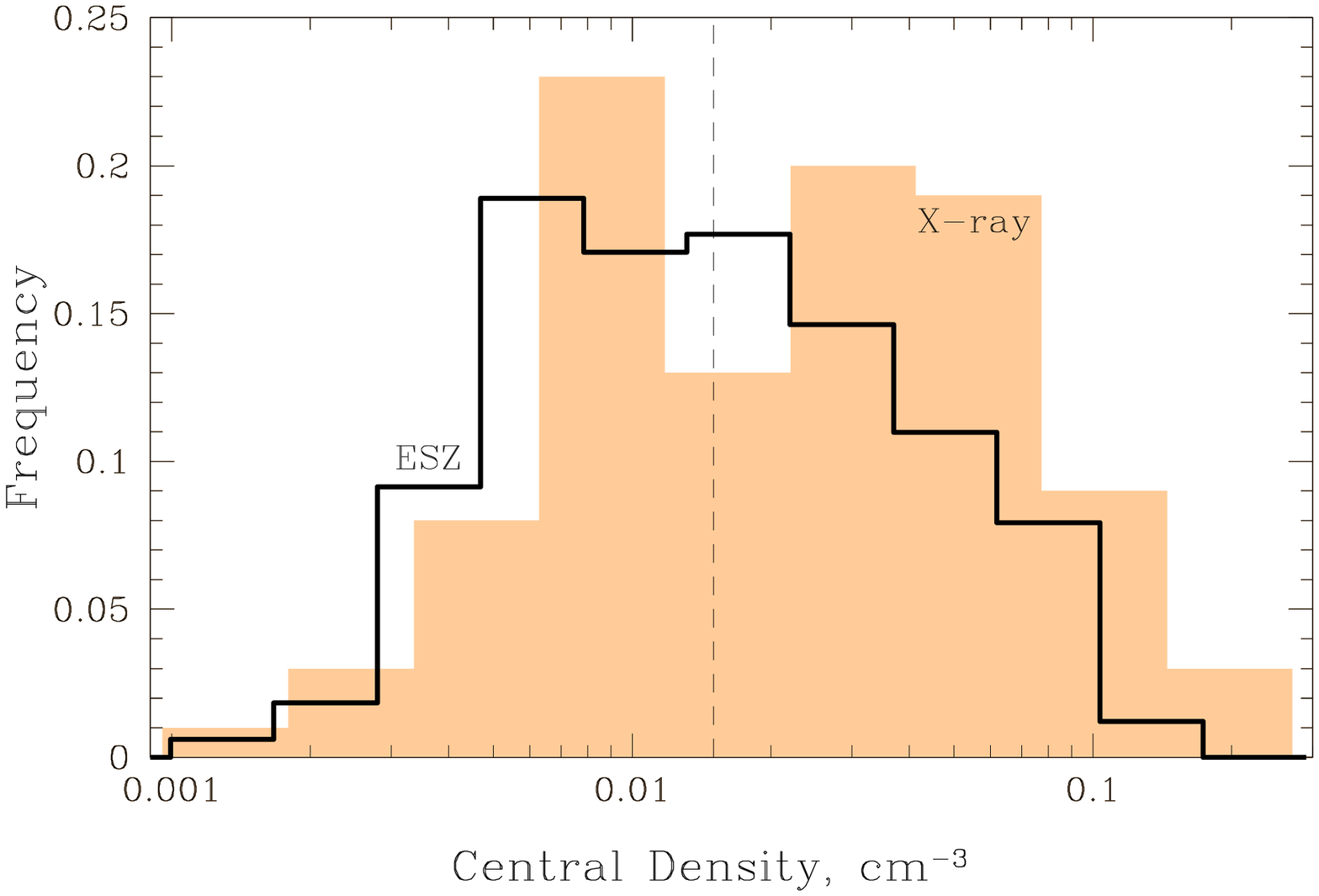}
}
\caption{\small{
Top Left: Distribution of the concentration parameter in the 0.15--1.0
$r_{500}$ range ($C_{\rm SB}$ -- see Equation \ref{eq:concentration}) for the 164 {\em Planck}
ESZ (solid line) and the 100 X-ray
selected clusters (orange shaded). Top Right:
Distribution of the concentration parameter in the 40--400
kpc range ($C_{\rm SB4}$ -- see Equation \ref{eq:concentration40_400})
for both samples. Bottom left: Distribution of the cuspiness ($\delta$ -- see Equation
\ref{eq:cuspiness}) for both samples.
Bottom right: Distribution of the central density ($n_{\rm core}$) for both samples.
The dashed vertical line in each panel corresponds to
the break value used to segregate clusters into CC and NCC subsamples
(0.4 for the concentration parameter in the 0.15--1.0 $r_{500}$ range,  
0.075 for the concentration parameter in the 40--400 kpc range, 
0.5 for the cuspiness, and 
$1.5 \times 10^{-2} \rm ~cm^{-3}$ for the central gas density). 
%The solid histogram shows the X-ray flux limited sample while the solid line corresponds to the ESZ cluster sample.
}}\label{fig:hist_metrics}
\end{figure*}

\subsection{Concentration parameter in the 40--400 kpc range}

The presence of cooler gas in the cores of clusters usually
implies a larger gas density in the core, compared to the gas 
density outside the core, to maintain the pressure
balance. This increased gas density produces an X-ray bright core, since the
X-ray emissivity is roughly proportional to the square of the gas density. 
Evaluating the X-ray brightness in the cluster core compared to the
brightness within a given larger radius is a powerful method to determine if the
cluster contains a cool-core. This metric is referred to as the concentration
parameter and was originally introduced by \citet{2008Santos}:
\begin{eqnarray}
C_{\rm SB4} = \frac{\Sigma(<40 {\rm kpc})}{\Sigma(<400 {\rm kpc})},\label{eq:concentration40_400}
\end{eqnarray} 
where $\Sigma(<r)$ is the integrated projected emissivity within a
circle of radius $r$.

\begin{table*}
\caption{Cool-Core Metrics\\{\scriptsize \rm Systematic uncertainty is
    computed by varying the break value by $\pm$10\%.}}
\centering
\begin{tabular}{l c c c c}
\hline\hline
Metric & K-S $p$-value & Break Value & CC fraction X-ray (\%) & CC fraction ESZ (\%) \\ % inserts table %heading
\hline
$C_{\rm SB}$: Concentration (0.15--1.0 $r_{500}$)  & $3.1 \times 10^{-2}$ & 0.4 & 44 $\pm$ 7 (sys $_{-6}^{+6}$) & 28 $\pm$ 4 (sys $_{-6}^{+10}$) \\
$C_{\rm SB4}$: Concentration (40--400 kpc)         & $2.9 \times 10^{-4}$ & 0.075 & 61 $\pm$ 8 (sys $_{-5}^{+3}$) & 36 $\pm$ 5 (sys $_{-4}^{+4}$) \\
$\delta$: Cuspiness                           & $1.1 \times 10^{-4}$ & 0.5 & 64 $\pm$ 8 (sys $_{-4}^{+3}$) & 38 $\pm$ 5 (sys $_{-4}^{+4}$) \\
$n_{\rm core}$: Central Density                     & $1.9 \times 10^{-2}$ & $1.5 \times 10^{-2} \rm ~cm^{-3}$ & 53 $\pm$ 7 (sys $_{-2}^{+3}$) & 39 $\pm$ 5 (sys $_{-3}^{+3}$) \\ 
\hline
\end{tabular}
\label{table:metrics}
\end{table*}

\subsection{Concentration parameter in the 0.15--1.0 $r_{500}$ range}

Here we also use a modification of the original definition \citep{2008Santos}, which is
scaled by the characteristic radius $r_{500}$ as \citep{2012Maughan}:
\begin{eqnarray}
C_{\rm SB} = \frac{\Sigma(<0.15r_{500})}{\Sigma(<r_{500})}.\label{eq:concentration}
\end{eqnarray} 

\subsection{Cuspiness of the gas density profile}

A third related metric is the cuspiness of the gas density profile computed
at a fixed scaled radius of $0.04 r_{500}$ \citep{2007Vik}:
\begin{eqnarray}
\delta = - \frac{{\rm d log~}n(r)}{{\rm d log~}r}\rvert_{r=0.04r_{500}},\label{eq:cuspiness}
\end{eqnarray} 
where $n(r)$ is the gas density at a distance $r$ from the cluster
center.

\subsection{Central gas density}

A fourth useful quantity that indicates if a cluster presents a cool
core is the central gas density \citep{2010Hudson}. Here we calculate the
central density at 0.01 $r_{500}$ from the core (which will be called
$n_{\rm core}$), since the equation used to
fit the density profile may diverge at $r = 0$ (if $\alpha > 0$ in
Equation (\ref{eq:nenp})).

%To compute the concentration parameters, the cuspiness of the density
%profile, and the central density, we compute the emission measure
%profile of each cluster.

%%%%%%%%%%%%%%%%%%%%%%%%%%%%%%%%%%%%%%%%%%%%%%%%%%%%%%%%%%%%%%%%%%%%%%%%%%%%%%%%%%
%%%%%%%%%%%%%%%%%%%%%%%%%%%%%%%%%%%%%%%%%%%%%%%%%%%%%%%%%%%%%%%%%%%%%%%%%%%%%%%%%%
%%
%%                                  TEMPERATURE
%%
%%%%%%%%%%%%%%%%%%%%%%%%%%%%%%%%%%%%%%%%%%%%%%%%%%%%%%%%%%%%%%%%%%%%%%%%%%%%%%%%%%
%%%%%%%%%%%%%%%%%%%%%%%%%%%%%%%%%%%%%%%%%%%%%%%%%%%%%%%%%%%%%%%%%%%%%%%%%%%%%%%%%%

\begin{comment}
\begin{deluxetable*}{lcc}
\tablecaption{Spearman Rank Test} 
\tablewidth{0pt} 
\tablehead{ 
\colhead{Relation} &
\colhead{Correlation (X-ray)} & 
\colhead{Correlation (ESZ)} 
}
\startdata 
$C_{SB}$(0.15--1.0 $r_{500}$) vs. $C_{SB}$(40--400 kpc) & 0.87 & 0.84  \\
$C_{SB}$(0.15--1.0 $r_{500}$) vs. Cuspiness & 0.75 & 0.69 \\
$C_{SB}$(0.15--1.0 $r_{500}$) vs. Central Density & 0.90 & 0.87 \\
$C_{SB}$(40--400 kpc) vs. Cuspiness & 0.93 & 0.90 \\
$C_{SB}$(40--400 kpc) vs. Central Density & 0.93 & 0.91 \\
Cuspiness vs. Central Density & 0.92 & 0.88  
\enddata
\tablecomments{Columns list metric relations and their Spearman correlations for both the
SZ and X-ray samples. Spearman $p$-values for all correlations were
lower than $1.5 \times 10^{-19}$, indicating extremely low probability
that the metrics are not correlated.} 
\label{tab:corr}
\end{deluxetable*}
\end{comment}

\section{Temperature Profiles}\label{sec:temp}

In this paper, we present the temperature profiles for only two
clusters, although we have temperature profiles for all clusters in
our samples, which vary in the number of fitted parameters
according to the quality of the data. We provide the fitted parameters
for all clusters in Andrade-Santos et al. (2017). The two temperature profiles
presented in this paper are examples of CC and NCC clusters as well as
clusters with very different data quality (see Figures
\ref{fig:emm_dens} and \ref{fig:emm_dens_G000}). 
In this Section, we present the analytic
equations used to obtain the profiles, referring the reader to papers
where the full description of the calculations 
are presented \citep[see][]{2006Vik,2015Andrade-Santos,2016Andrade-Santos}.   
 
\citet{2006Vik} give a 3D temperature
profile that describes the general
features of the temperature profile of clusters:
\begin{equation}\label{eq:tprof}
  T_{\rm 3D}(r) =  T_0\times \frac{x+T_{\rm min}/T_0}{x+1}
  \times \frac{(r/r_{\rm t})^{-a}}{(1+(r/r_{\rm t})^b)^{c/b}}, 
\end{equation}
where $x=(r/r_{\rm cool})^{a_{\rm cool}}$. $r_{\rm t}$ and $r_{\rm cool}$
are the transition and cool core radii, respectively. $T_{\rm min}$ is the central
temperature, and $a$, $b$, $c$, and $a_{\rm cool}$ are indexes that
determine the slopes of the temperature profile in different radial
ranges. 

We derive the deprojected 3D temperature by projecting a model to
compare to the projected measured temperature. The 3D
temperature model, $T_{3D}$, is weighted by the density squared according to
the spectroscopic-like temperature \citep[][presented a formula to project
the temperature which matches the spectroscopically measured temperature within a few percent]{2004Mazzotta}:
\begin{equation}
T_{\rm 2D}=T_{\rm spec} \equiv \frac{\int n_{\rm e}^2 T_{\rm
    3D}^{1/4} dz}{\int n_{\rm e}^2 T_{\rm 3D}^{-3/4} dz},
\label{eq:tspec}
\end{equation}
to give values of $T_{2D}$ for comparison with the measured
values. $n_{\rm e}$ is the electron density, given by Equation
\ref{eq:nenp}, and  $T_{\rm 3D}$ is the deprojected 3D temperature,
given by Equation \ref{eq:tprof}.

%Using Equations (\ref{eq:nenp}, \ref{eq:tprof}, and \ref{eq:tspec}),
%we can determine the parameters of Equation (\ref{eq:tprof}) that give
%the best $T_{\rm 2D}$ match to the observed temperature profile of
%each cluster in our sample. 

\begin{figure*}[!t]
\centerline{
\includegraphics[width=0.33\textwidth]{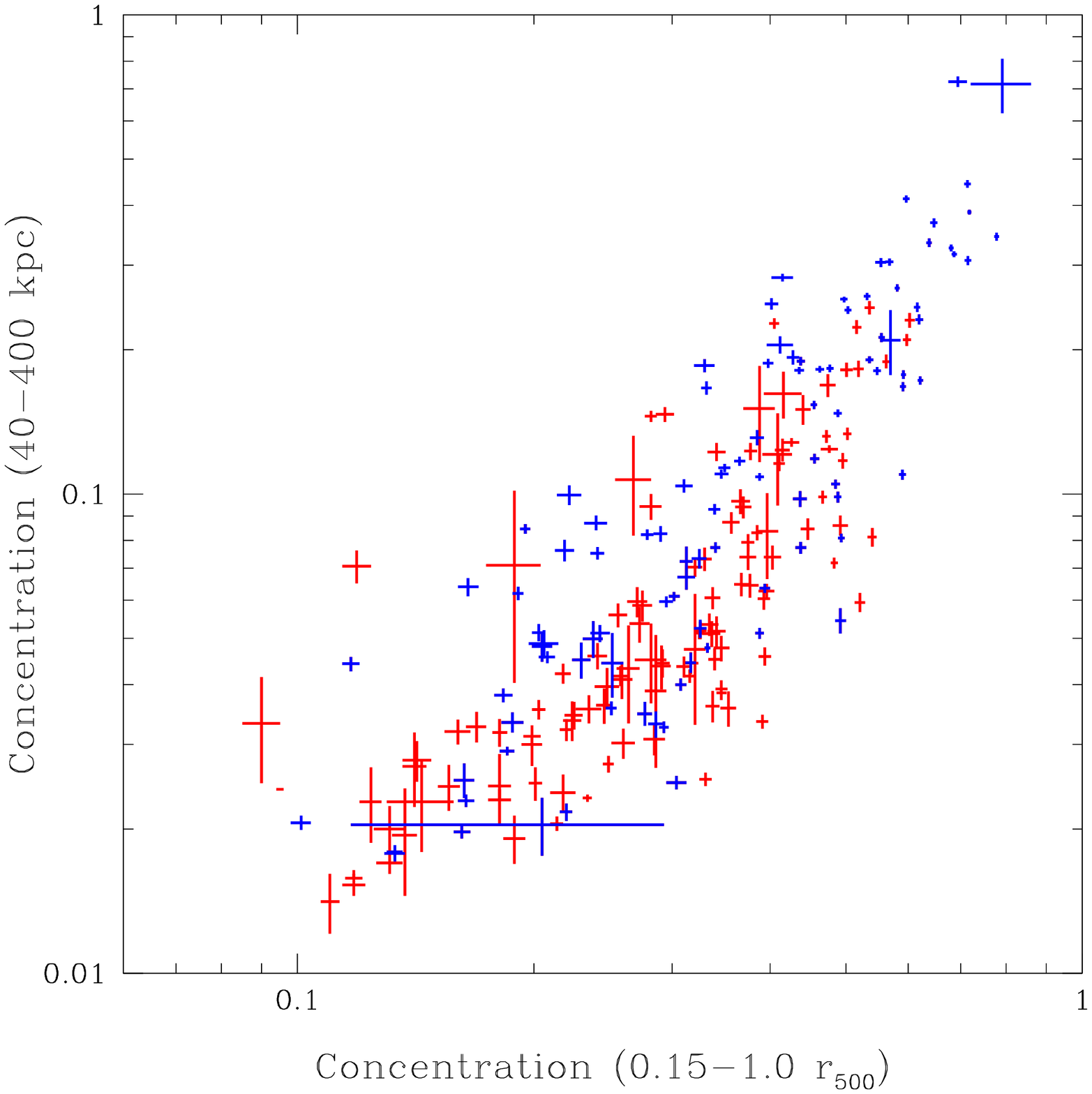}
\includegraphics[width=0.33\textwidth]{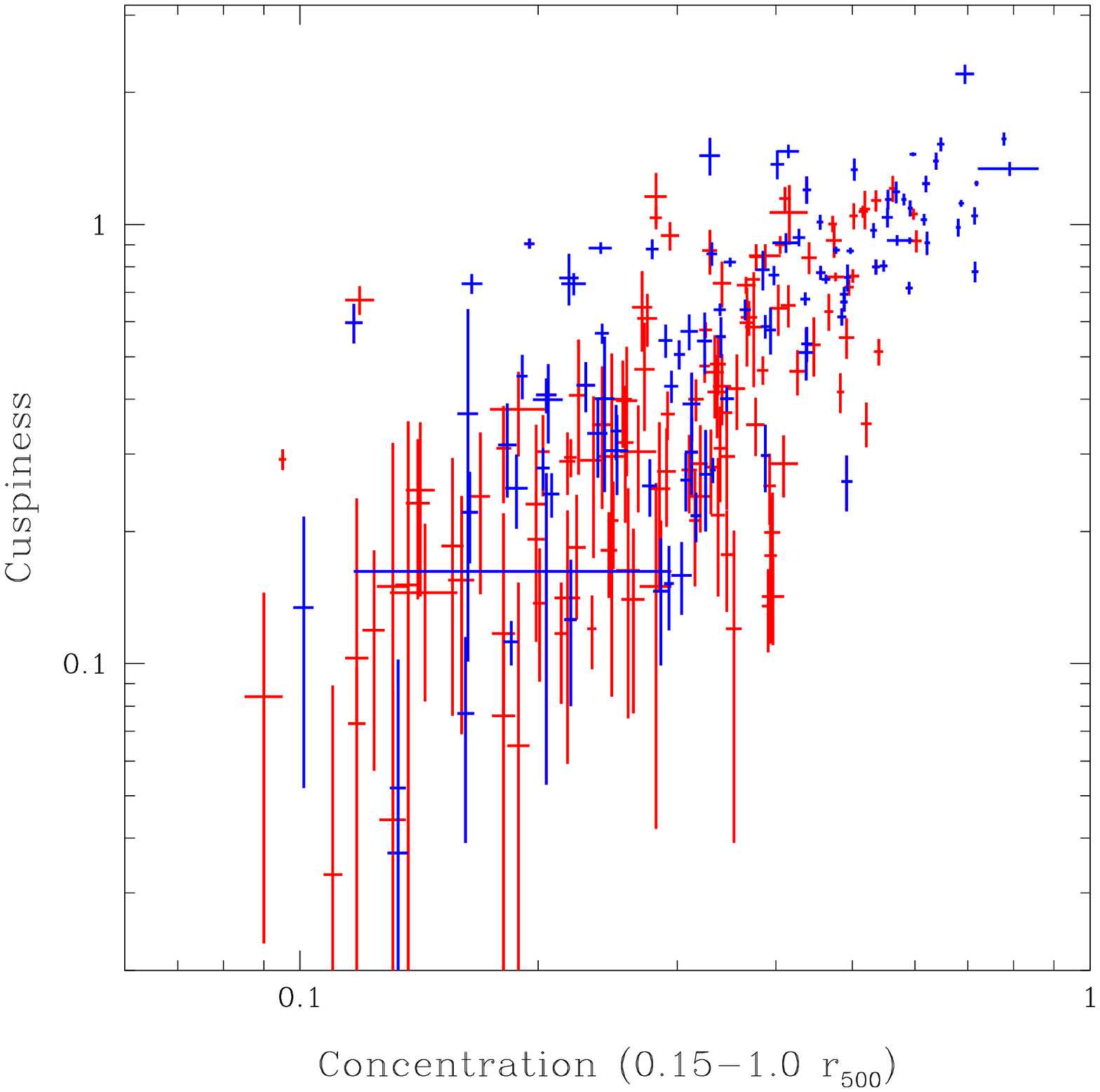}
\includegraphics[width=0.33\textwidth]{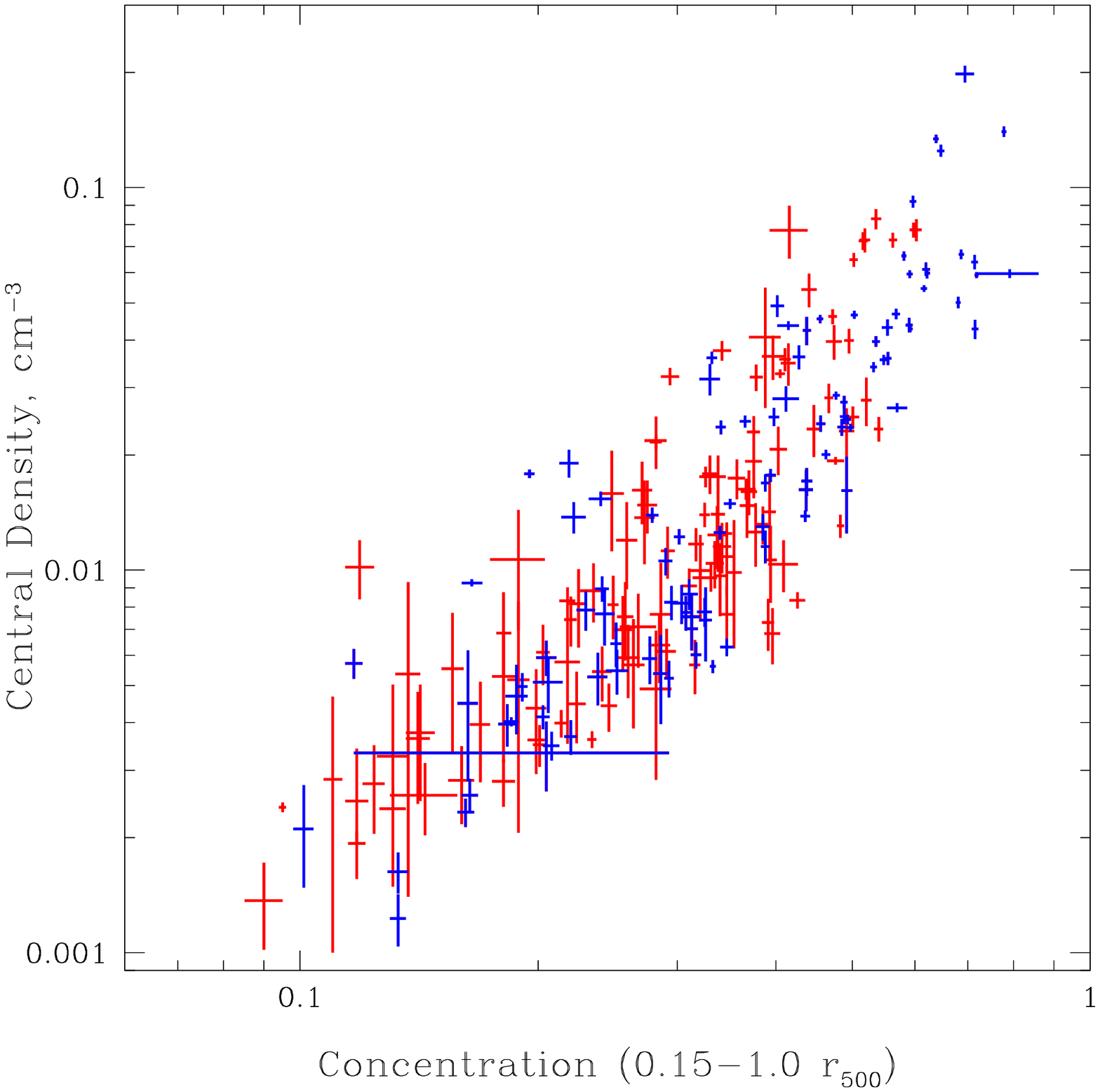}
}
\vspace{0.3cm}
\centerline{
\includegraphics[width=0.33\textwidth]{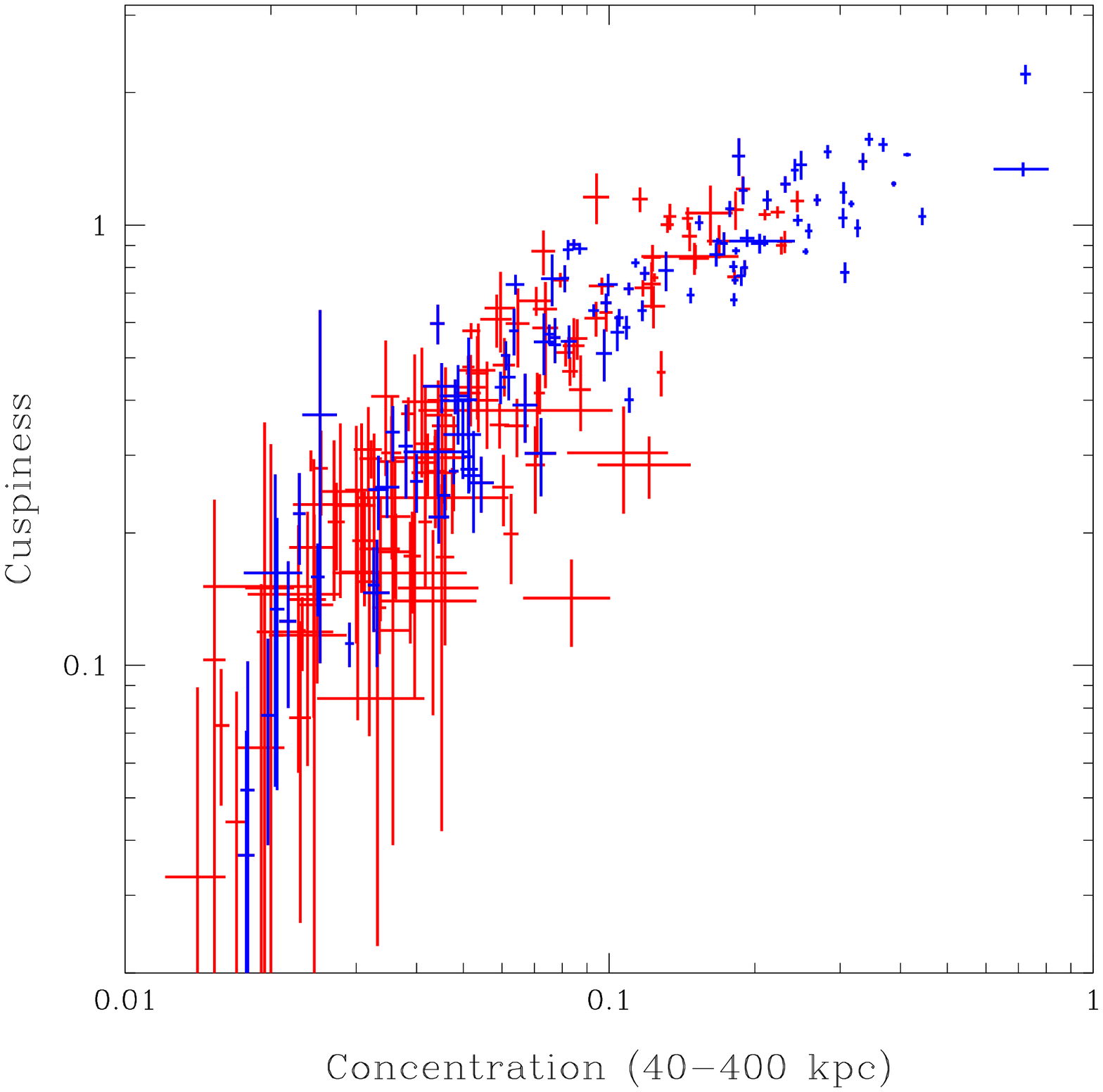}
\includegraphics[width=0.33\textwidth]{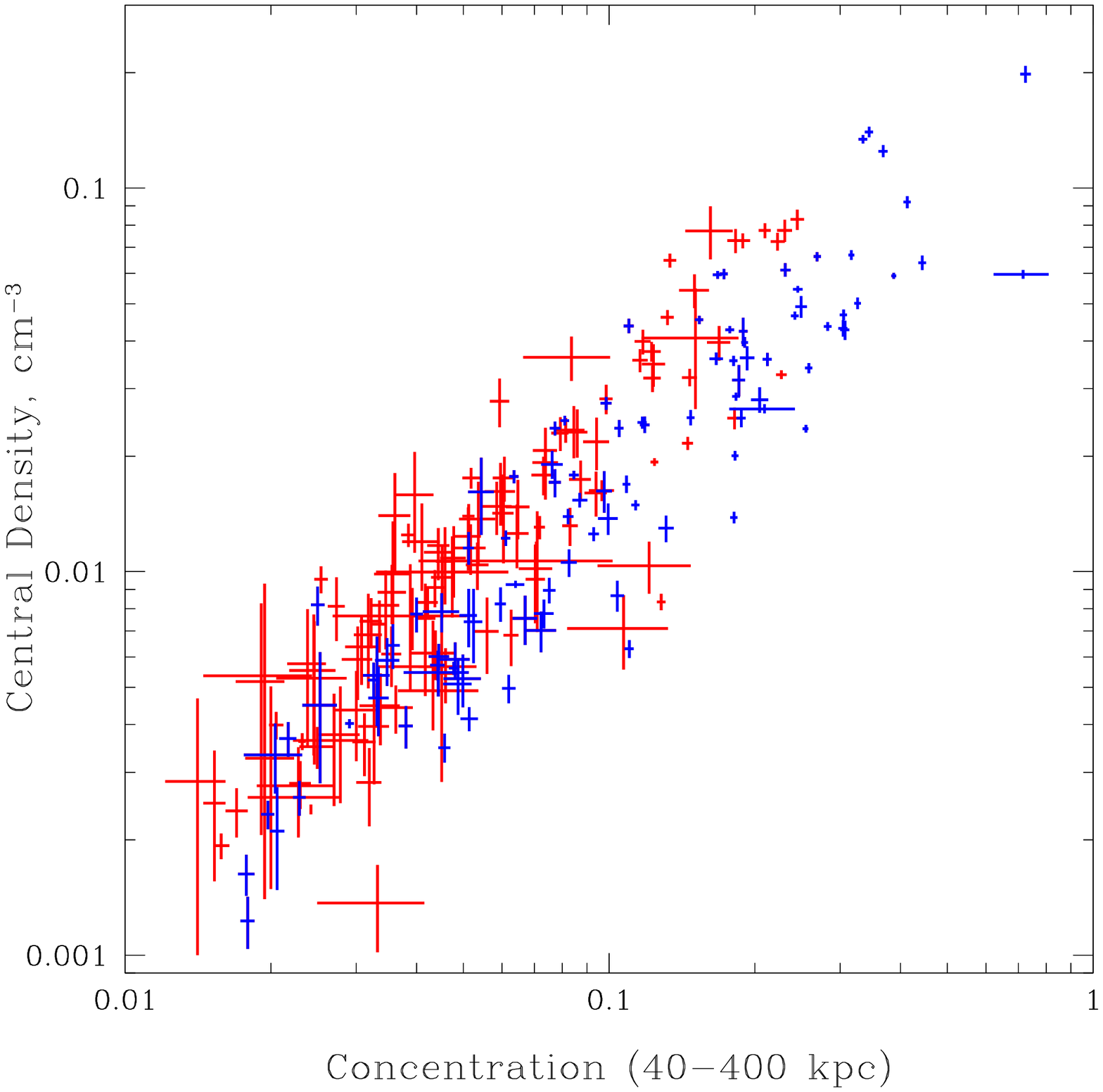}
\includegraphics[width=0.33\textwidth]{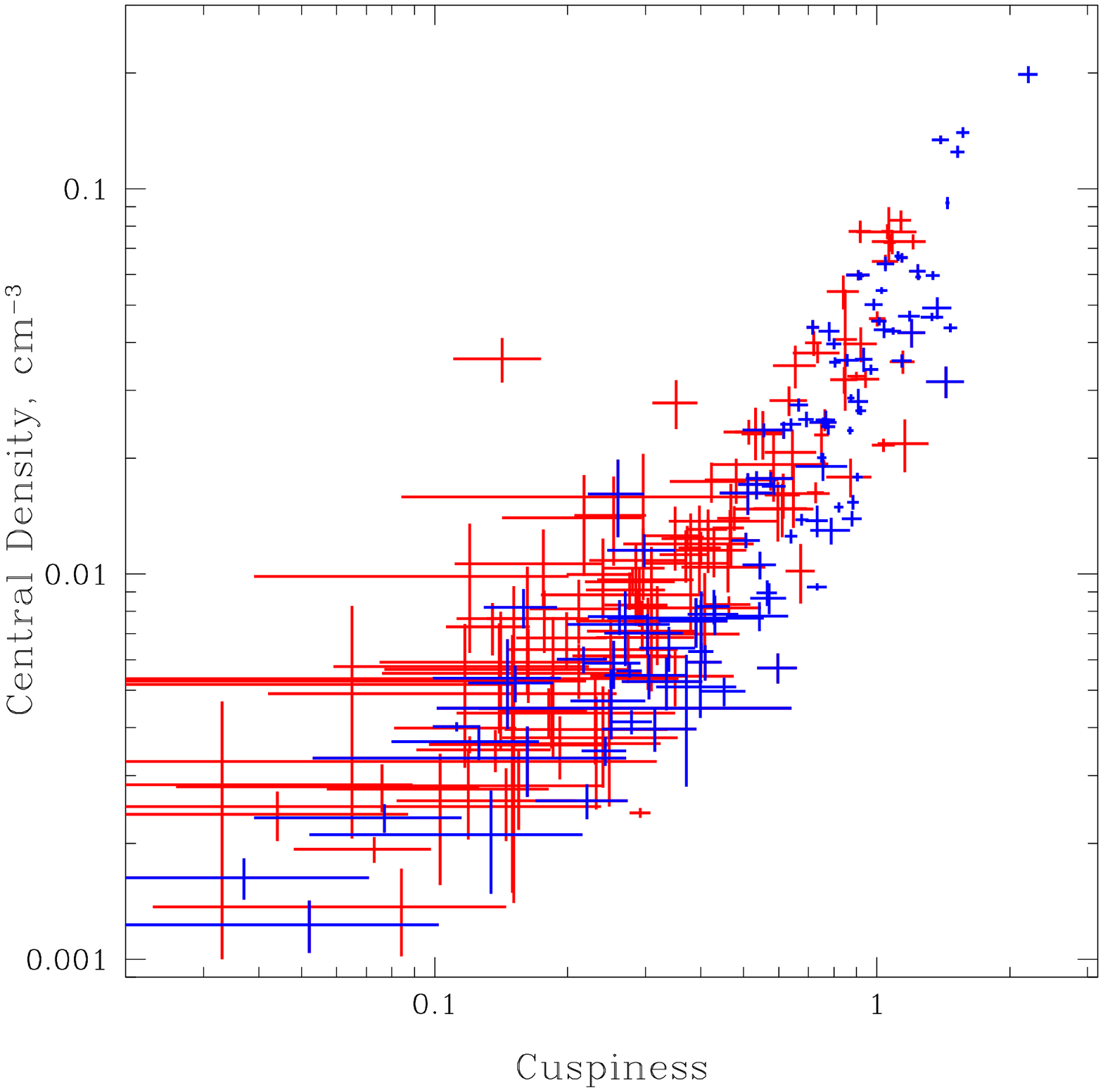}
}
\caption{\small{
Top Left: Concentration parameter in the 40--400 kpc range ($C_{\rm
  SB4}$)  vs. Concentration parameter in the 0.15--1.0 $r_{500}$ range
($C_{\rm SB}$). 
Top Center: Cuspiness ($\delta$) vs. Concentration parameter in the
0.15--1.0 $r_{500}$ range ($C_{\rm SB}$). 
Top Right: Central density ($n_{\rm core}$) vs. Concentration parameter in
the 0.15--1.0 $r_{500}$ range ($C_{\rm SB}$) . 
Bottom Left: Cuspiness ($\delta$) vs. Concentration parameter in the
40--400 kpc range ($C_{\rm SB4}$). 
Bottom Center: Central density ($n_{\rm core}$) vs. Concentration
parameter in the 40--400 kpc range ($C_{\rm SB4}$). 
Bottom Right: Central density ($n_{\rm core}$) vs. Cuspiness ($\delta$). 
In red, the ESZ sample, in blue,
the X-ray sample. This figure shows the strong correlation
between different metrics, whose correlations are quantified in Table \ref{table:corr}.
}}\label{fig:corr}
\end{figure*}

%%%%%%%%%%%%%%%%%%%%%%%%%%%%%%%%%%%%%%%%%%%%%%%%%%%%%%%%%%%%%%%%%%%%%%%%%%%%%%%%%%
%%%%%%%%%%%%%%%%%%%%%%%%%%%%%%%%%%%%%%%%%%%%%%%%%%%%%%%%%%%%%%%%%%%%%%%%%%%%%%%%%%
%%
%%                               RESULTS
%%
%%%%%%%%%%%%%%%%%%%%%%%%%%%%%%%%%%%%%%%%%%%%%%%%%%%%%%%%%%%%%%%%%%%%%%%%%%%%%%%%%%
%%%%%%%%%%%%%%%%%%%%%%%%%%%%%%%%%%%%%%%%%%%%%%%%%%%%%%%%%%%%%%%%%%%%%%%%%%%%%%%%%%

\section{Results}

With the cluster gas density and emission measure profiles, we are able to
compute the cuspiness of the gas density profile, concentration, and central gas density
for the X-ray and {\em Planck} ESZ cluster samples.
The uncertainties on the metrics for each cluster were computed
using 100 Monte Carlo realizations of the density profile, 
including also a Gaussian distribution for $r_{500}$ ($r_{500} \pm
\sigma_{r_{500}}$). The top left panel of Figure \ref{fig:hist_metrics}
presents the distribution of concentration parameters in the 0.15--1.0
$r_{500}$ range for both cluster samples. We used a Kolmogorov--Smirnov
(K-S) test for the SZ and X-ray samples to evaluate the probability that the two samples were drawn from the same distribution. 
We obtained $p$-values that are smaller than 3.1 $\times 10^{-2}$ for all metrics, which 
suggests that the fraction of cool cores in the sample of X-ray selected
clusters is different from that in the SZ sample. Defining a cool-core cluster as
one that presents a concentration parameter in the 0.15--1.0
$r_{500}$ range, $C_{\rm SB} >
0.4$, the fraction of cool-cores in
the X-ray sample is 44$\pm$7\%, whereas in the SZ sample, the fraction is
$28\pm4\%$. The uncertainties on the fraction of cool-core
clusters were computed using a Bootstrap re-sampling method, including
Poisson statistics on the number of clusters satisfying the cool-core
criterion: a metric value greater than the break value used to segregate clusters into CC and NCC.
With a break value of 0.075 \citep{2008Santos} for the
concentration parameter in the 40-400 kpc range (Figure
\ref{fig:hist_metrics}, top right panel), we have a CC fraction of 61$\pm$8\% in
the X-ray sample and $36\pm5\%$ in the SZ sample. The high fraction of CC clusters in the X-ray
selected sample compared to that in the SZ sample agrees quite well with the recent results
presented by \citet{2017Rossetti} ($59\pm5\%$ in their X-ray sample vs. $29\pm4\%$ in their SZ sample).
%despite the fact that the X-ray samples are very different (their
%X-ray sample contains only very massive clusters). 
With a break value of 0.5 \citep{2007Vik} for the cuspiness of the gas density
profile (Figure \ref{fig:hist_metrics}, bottom left panel), we have a
CC fraction of 64$\pm$8\% in the X-ray sample and $38\pm5\%$ in the
SZ sample. \citet{2012Maughan} used a value of 0.8, more
appropriate for moderate to strong CC clusters. They also use a break
value of 0.5 for the concentration parameter in the 0.15--1.0
$r_{500}$ range, which we chose to be 0.4 to also include weak CC
clusters. Finally, using a break value of $1.5 \times 10^{-2}
\rm cm^{-3}$ \citep{2010Hudson} for the central gas density to
distinguish cool and non-cool core clusters (Figure
\ref{fig:hist_metrics}, bottom right panel), 
we find that $53\pm7\%$  of the clusters in
the X-ray sample have cool cores, whereas the SZ sample shows a 
fraction of $39\pm5\%$. The fraction of CC clusters and K-S test
results are listed in Table \ref{table:metrics}. 
We also include in Table \ref{table:metrics} 
a systematic uncertainty on the fraction of CC clusters by varying 
the break value by $\pm$10\%. The magnitude of this systematic uncertainty is
comparable to the statistical uncertainty.  

Using all four comparisons of the X-ray and SZ cluster samples, 
we find that CC clusters are significantly more common in X-ray
selected cluster samples than in SZ selected samples.  

%From the fraction of CC clusters in X-ray vs. SZ samples, we see
%that the concentration parameter in the 40-400 kpc range and the
%cuspiness of the density profile are the most effective metrics to
%create a distinction on the fraction of the CC population between these samples.

Figure \ref{fig:corr} shows the correlations between all
metrics. Visually, we observe a strong correlation between all metrics,
which we verified numerically using a Spearman test. 
This provides a correlation coefficient ranging between 0 (no
correlation) and (-)+1, in the case of perfect (anti)correlation. 
Results for our metrics are listed in Table \ref{table:corr}.

%To examine a possible mass dependence in the comparisons between
%the X-ray and SZ samples, we excluded all 22 clusters with masses less
%than $10^{14}~M_{\odot}$ from the X-ray sample, so both the X-ray and
%SZ samples cover the same mass range (see Figure
%\ref{fig:histm}). We then re-analyzed all metrics. The results were
%very similar to the results when all clusters were
%included in the analysis.

Using a set of cosmological hydrodynamic simulations of galaxy
clusters, \citet{2015Rasia} found that 38\% (11/29) of their simulated
clusters at $z$ = 0 are classified as CC using the central entropy
\citep{2009Cavagnolo} and pseudo-entropy ratio \citep{2010Leccardi}
as metrics. This result agrees very well with our observed
fraction of CC clusters in the SZ sample (28 -- 39\% according
to the metric used, suggesting that the fraction of CC clusters
in SZ samples is representative of the fraction of CC clusters in
the universe). \citet{2015Lin} showed that constraints on the fraction of CC
clusters in SZ selected datasets are only subject to a systematic bias
of order one percent, a significant reduction compared to X-ray selected samples,
supporting that SZ selected samples of galaxy clusters are robust
cosmological probes.

\subsection{Numerical Simulations}

In this section, we apply the four metrics used in this paper to the set of numerical
simulations of galaxy clusters from \citet{2015Rasia}. We obtain the
following results: a) using the concentration parameter in the 0.15--1.0
$r_{500}$ range, we obtain a fraction of CC clusters of 33$\pm$11\%, b) using 
the concentration parameter in the 40--400 kpc range, the fraction of
CC clusters is 26$\pm$10\%, c) using the cuspiness of the gas density, the
fraction of CC clusters is 38$\pm$11\%, and d) using the central gas density, the
fraction of CC clusters is 48$\pm$13\%. 
These numbers are in good agreement
with the fraction of CC clusters in our ESZ sample.

\begin{table}[!b]
\caption{Spearman Rank Test}
\centering
\begin{tabular}{l c c}
\hline\hline
Relation & Correlation (X-ray) & Correlation (ESZ)\\ % inserts table heading
\hline
$C_{\rm SB}$ vs. $C_{\rm SB4}$ & 0.87 & 0.84 \\
$C_{\rm SB}$ vs. $\delta$ & 0.75 & 0.69 \\
$C_{\rm SB}$ vs. $n_{\rm core}$  & 0.90 & 0.87 \\
$C_{\rm SB4}$ vs. $\delta$ & 0.93 & 0.90 \\
$C_{\rm SB4}$ vs. $n_{\rm core}$ & 0.93 & 0.91 \\
$\delta$ vs. $n_{\rm core}$ & 0.92 & 0.88 \\
\hline
\end{tabular}
\label{table:corr}
\end{table}

\begin{table*}
\caption{Over-population of CC clusters}
\centering
\begin{tabular}{l c c c}
\hline\hline
Metric & $<L_{CC}>/<L_{NCC}>$ & Expected Over-population & Observed Over-population  \\ % inserts table heading
\hline
$C_{\rm SB}$   & 1.80 $\pm$ 0.16 & 2.7 $\pm$ 1.0 & 1.6 $\pm$ 0.3 \\
$C_{\rm SB4}$  & 1.63 $\pm$ 0.15 & 2.3 $\pm$ 0.7 & 1.7 $\pm$ 0.3 \\
$\delta$      & 1.57 $\pm$ 0.15 & 2.1 $\pm$ 0.7 & 1.7 $\pm$ 0.3 \\
$n_{\rm core}$ & 1.69 $\pm$ 0.15 & 2.4 $\pm$ 0.8 & 1.4 $\pm$ 0.3 \\ 
\hline
\end{tabular}
\label{table:overpop}
\end{table*}

%%%%%%%%%%%%%%%%%%%%%%%%%%%%%%%%%%%%%%%%%%%%%%%%%%%%%%%%%%%%%%%%%%%%%%%%%%%%%%%%%%
%%%%%%%%%%%%%%%%%%%%%%%%%%%%%%%%%%%%%%%%%%%%%%%%%%%%%%%%%%%%%%%%%%%%%%%%%%%%%%%%%%
%%
%%                           SELECTION EFFECTS
%%
%%%%%%%%%%%%%%%%%%%%%%%%%%%%%%%%%%%%%%%%%%%%%%%%%%%%%%%%%%%%%%%%%%%%%%%%%%%%%%%%%%
%%%%%%%%%%%%%%%%%%%%%%%%%%%%%%%%%%%%%%%%%%%%%%%%%%%%%%%%%%%%%%%%%%%%%%%%%%%%%%%%%%

\section{Selection Effects: Malmquist Bias in X-ray}

%\begin{figure}[!t]
%\centerline{
%\includegraphics[width=0.47\textwidth]{hist_l_over_lmean.eps}
%}
%\caption{\small{
%Black, blue, and red histograms show the complete X-ray sample, and
%the CC and NCC subsamples, respectively.
%To estimate the fractional increase in luminosity of the CC
%subsample, we compare the ratio of the observed luminosity to
%the mean luminosity -- $L/L_{\rm mean}$ (all clusters in the X-ray sample) for 
%CC and NCC clusters (those with the cuspiness parameter, $\delta$, 
%larger and smaller than 0.5, respectively). We find that the
%CC clusters are on average $\sim$ 1.36 times more luminous for the same
%mass than the NCC clusters.
%}}\label{fig:hist_l_lmean}
%\end{figure}

%The X-ray cluster sample used in this paper is derived from the ROSAT X-ray
%catalogs which formally used the total X-ray flux as the only
%selection criterion. 
%Our objects are nearly an order of magnitude
%brighter than the sensitivity limit of the parent ROSAT All-Sky
%Survey. Therefore, it is safe to assume that 
%there are no selection
%effects related to detectability of clusters with different X-ray
%morphologies 
%all clusters with fluxes above our limit, $f_{\rm X} > 7.5 \times 10^{-12} \rm
%~erg ~ s^{-1} ~ cm^{-2}$, were included in our selection.

The X-ray cluster sample used in this paper is derived from the ROSAT X-ray
catalogs which formally used the total X-ray flux as the only
selection criterion. 
Thus, it can be 
%However, the X-ray cluster sample can still be 
affected by the Malmquist bias
leading to over-representation of the CC clusters. {\it CC
clusters tend to be more X-ray luminous for the same mass and
thus they become over-represented in a purely X-ray flux-limited
survey}. To estimate the fractional increase in X-ray luminosity of the CC
subsample, we compare the ratio of the observed luminosity to
the expected luminosity for the measured mass (using the $L_{\rm X}$--$M$ relation given by Equation (22) from \citet{2009Vik}) for 
CC and NCC clusters for all four metrics used to identify CC clusters. We find that 
CC clusters are on average $\sim$ 1.6--1.8 times more X-ray luminous for the same
mass (Table \ref{table:overpop}) than are NCC clusters. These results are consistent with early studies based on
{\em Einstein} imaging data (central excesses over the $\beta$-model,
\citep{1999Jones}), and normalizations of the $L_{\rm X}$ -- $T$ relations for the CC
and NCC populations \citep{1998Allen}.

The impact of this difference in the total X-ray luminosity on
the fractions of CC and NCC clusters is substantial. In
a low-z flux-limited survey, the search volume is $\propto$ $L^{3/2}$ so a
subpopulation which is intrinsically more luminous by a factor of
$\sim$ 1.7 becomes over-represented by a factor of 2.2 above a fixed mass
threshold.

A similar bias is still present if we consider clusters in a narrow
redshift range, where there is no difference in the search volume. In
this case, a flux limit is equivalent to an X-ray luminosity
threshold. CC clusters are less massive than NCC clusters for a fixed $L_{\rm X}$ and
hence are more numerous than NCC clusters.

We can quantify the selection effects of CC clusters in X-ray
surveys. For simplicity, let us approximate 
the cluster mass function locally as a power law given by:
\begin{equation}
N(>M) \propto  M^{-\gamma},
\end{equation}
and assume a power law $M$--$L_{\rm X}$ relation as:
\begin{equation}
L_{\rm X} \propto M^\beta.
\end{equation}
Let $l$ be the ratio of average luminosities (0.5--2.0 keV band, in the 0--$r_{500}$ range)
at a fixed mass for the CC and NCC populations:
\begin{equation}
l \equiv \langle L_{\rm CC} \rangle/ \langle L_{\rm NCC} \rangle, 
\end{equation}
then we would expect the ratio of the number of CC to NCC clusters to be:
\begin{equation}
\Delta \equiv l^{\gamma/\beta}. 
\end{equation}
We computed the ratio of average luminosities 
at a fixed mass for the CC and NCC populations, $l$, to be in the
range 1.6--1.8 (Table \ref{table:overpop}). From \citet{2009Vik}, 
$\beta = 1.61 \pm 0.14$. To compute the slope of the halo mass
function, we averaged the slope of the mass function at
the location of each cluster mass in our X-ray sample, using the mass
function provided by \citet{2006Warren}. We obtained
$\gamma =  2.54 \pm 0.79$. With these numbers in hand, we estimate that the CC
clusters are over-represented in our X-ray sample
by a factor of $\Delta =$ 2.1 -- 2.7 (depending on the metric) because of the Malmquist bias.

%The estimated selection effect are big. The best way to address it
%would be to work with volume-limited samples (currently, such
%samples do not have a sufficient number of clusters) or select the
%objects by a technique which is not sensitive to the presence of a
%cooling flow (or its associated optical characteristics); a promising
%approach is an all-sky Sunyaev-Zel‘dovich effect survey with, e.g. Planck.

%%%%%%%%%%%%%%%%%%%%%%%%%%%%%%%%%%%%%%%%%%%%%%%%%%%%%%%%%%%%%%%%%%%%%%%%%%%%%%%%%%
%%%%%%%%%%%%%%%%%%%%%%%%%%%%%%%%%%%%%%%%%%%%%%%%%%%%%%%%%%%%%%%%%%%%%%%%%%%%%%%%%%
%%
%%                               CONCLUSIONS
%%
%%%%%%%%%%%%%%%%%%%%%%%%%%%%%%%%%%%%%%%%%%%%%%%%%%%%%%%%%%%%%%%%%%%%%%%%%%%%%%%%%%
%%%%%%%%%%%%%%%%%%%%%%%%%%%%%%%%%%%%%%%%%%%%%%%%%%%%%%%%%%%%%%%%%%%%%%%%%%%%%%%%%%

\section{Conclusions}

Using {\em Chandra} observations, we derived and compared the fraction of cool-core clusters in the {\em Planck}
Early Sunyaev-Zel'dovich (ESZ) sample of 164 detected clusters with $z \leq 0.35$  
and in a flux-limited X-ray sample of 100 clusters with $z \leq
0.30$. We use four metrics to identify the
presence of a cool-core: 1) the concentration
parameter: the ratio of the integrated surface brightness within 0.15 $r_{500}$ to
that within $r_{500}$, and 2) within 40 kpc to
that within 400 kpc, 3) the cuspiness of the gas density profile: the
negative of the logarithmic derivative of the gas density with respect to the
radius measured at 0.04 $r_{500}$, and 4) the central gas density,
measured at  0.01 $r_{500}$.
We find that:

\begin{itemize}

   \item In all four metrics that we used, the sample of X-ray selected clusters contains a 
significantly larger fraction of cool-core clusters compared to the
sample of SZ selected clusters (44$\pm$7\% vs. 28$\pm$4\% using the concentration
parameter in the 0.15--1.0 $r_{500}$ range
as a metric for cool-cores, 61$\pm$8\% vs. 36$\pm$5\% using the concentration
parameter in the 40--400 kpc range, 64$\pm$8\% vs. 38$\pm$5\% using the cuspiness,
and 53$\pm$7\% vs. 39$\pm$5\% using the central density). 
Our results for the concentration parameter in the 40--400 kpc range
agree well with the recent results by \citet{2017Rossetti}.

   \item Qualitatively,  
cool-core clusters are more X-ray luminous at fixed mass. Hence, our X-ray 
flux-limited sample, compared to the approximately mass-limited SZ
sample, is over-represented with cool-core clusters. 
We describe a simple quantitative model that successfully predicts the 
observed difference based on the selection bias. Our model predicts an
over-population of CC clusters in our X-ray selected sample compared
to SZ samples of 2.1 -- 2.7, depending on the metric used to
identify CC clusters, with a typical uncertainty of $\sim$ 0.8, 
which is consistent within the uncertainties with the observed values
in the range 1.4 -- 1.7 with a typical uncertainty of $\sim$ 0.3. 

   \item The results of the four metrics we used to measure the
     over-population of CC clusters in X-ray samples compared to that
     in SZ samples are all consistent within their uncertainties.   

   \item While differences in X-ray and SZ cluster selection are significant,
they can be quantitatively explained by the effect of cool-cores on
X-ray luminosities.

\end{itemize}

CC clusters are more X-ray luminous than
NCC clusters for a fixed cluster mass. Thus, an X-ray flux-limited
sample will select a larger fraction of CC clusters compared to an SZ
selected cluster sample. The determination of cosmological parameters
from an X-ray flux-limited sample
in the local Universe can be summarized by determining confidence
levels in the highly degenerate $\Omega_{\rm M}$--$\sigma_{8}$ plane. If cluster masses 
are determined using a proxy other than the X-ray luminosity (e.g.,
gas mass, $M$--$Y_{\rm X}$ scaling relation, $T_{\rm X}$,
weak-lensing, hydrostatic mass) there will be no Malmquist bias on the
determination of cosmological parameters, simply because when the mass
function is constructed, the CC clusters that were wrongly included in the
selection will now be excluded from the mass function because they do not
satisfy the criterion that their masses are above the mass limit given
their redshifts. On the other hand, if the cluster masses are determined by the 
$L_{\rm X}$--$M$ scaling relation, the masses of the CC clusters will be biased high, and
their inclusion in the mass function will
lead to a shift towards higher values of $\Omega_{\rm M}$ and $\sigma_{8}$.

%Hence, the effect of cool-cores on the X-ray selection are clearly
%understood and can be accounted for in cosmological studies. 

\acknowledgments

\noindent
F.A-S. acknowledges support from {\em Chandra} grant GO3-14131X. 
C.J., W.R.F., A.V. are supported by the Smithsonian Institution. 
R.J.W. is supported by a Clay Fellowship awarded by the 
Harvard-Smithsonian Center for Astrophysics.
Basic research in radio astronomy at the Naval Research Laboratory by
S.G. and T.E.C is supported by 6.1 base funding. 
S.B. acknowledges financial support from the PRIN-MIUR 201278X4FL
grant. M.A., G.W.P., and J.D. acknowledge support from the European
Research Council under the European Union's Seventh Framework 
Programme (FP7/2007-2013) / ERC grant agreement n$^\circ$ 340519.
L.L. acknowledges support from the {\em Chandra} X-ray Observatory grant 
GO3-14130B and by the {\em Chandra} X-ray Center through NASA contract NAS8-03060.
We thank the anonymous referee for their useful comments.

\newpage

\appendix

\setcounter{table}{0}
\renewcommand{\thetable}{A\arabic{table}}

We present in Table \ref{tab:ESZmetrics} the values of the metrics for
all clusters in the ESZ sample, including the secondary subclusters (on the lines following the primary subcluster, indicated by --). Columns list the
cluster name (the prefix PLCKESZ is omitted for simplicity), RA, DEC,
redshift, concentration parameter in the 0.15--1.0 $r_{500}$ range,
concentration parameter in the 40--400 kpc range, cuspiness of the gas density profile,
central gas density, and the maximum radius where the emission integral is computed. Each
metric value is followed by its uncertainty.

\begin{longtable}{lllcccccccccc}
\caption[Concentration parameter, Cuspiness, and Central Density for
  the SZ sample]{Concentration parameter, Cuspiness, and Central
  Density for the {\em Planck} ESZ sample.} \label{tab:ESZmetrics} \\

\hline \hline
\multicolumn{1}{c}{Cluster} & 
\multicolumn{1}{c}{RA} &
\multicolumn{1}{c}{DEC} &
\multicolumn{1}{c}{$z$} & 
\multicolumn{1}{c}{$C_{\rm SB}$} &
\multicolumn{1}{c}{$\sigma_{C_{\rm SB}}$} &
\multicolumn{1}{c}{$C_{\rm SB4}$} &
\multicolumn{1}{c}{$\sigma_{C_{\rm SB4}}$} &
\multicolumn{1}{c}{$\delta$} &
\multicolumn{1}{c}{$\sigma_{\delta}$} &
\multicolumn{1}{c}{$n_{\rm core}$} &
\multicolumn{1}{c}{$\sigma_{n_{\rm core}}$} &
\multicolumn{1}{c}{$r_{\rm max}$} \\ 
\multicolumn{1}{c}{} & 
\multicolumn{1}{c}{} &
\multicolumn{1}{c}{} &
\multicolumn{1}{c}{} & 
\multicolumn{1}{c}{} &
\multicolumn{1}{c}{} &
\multicolumn{1}{c}{} &
\multicolumn{1}{c}{} &
\multicolumn{1}{c}{} &
\multicolumn{1}{c}{} &
\multicolumn{1}{c}{($\rm cm^{-3}$)} &
\multicolumn{1}{c}{($\rm cm^{-3}$)} &
\multicolumn{1}{c}{($r_{500}$)} \\ 
\hline 

\endfirsthead

\multicolumn{13}{c}%
{\tablename\ \thetable{} -- continued from previous page} \\
\hline 
\multicolumn{1}{c}{Cluster} & 
\multicolumn{1}{c}{RA} &
\multicolumn{1}{c}{DEC} &
\multicolumn{1}{c}{$z$} & 
\multicolumn{1}{c}{$C_{\rm SB}$} &
\multicolumn{1}{c}{$\sigma_{C_{\rm SB}}$} &
\multicolumn{1}{c}{$C_{\rm SB4}$} &
\multicolumn{1}{c}{$\sigma_{C_{\rm SB4}}$} &
\multicolumn{1}{c}{$\delta$} &
\multicolumn{1}{c}{$\sigma_{\delta}$} &
\multicolumn{1}{c}{$n_{\rm core}$} &
\multicolumn{1}{c}{$\sigma_{n_{\rm core}}$} &
\multicolumn{1}{c}{$r_{\rm max}$} \\ 
\multicolumn{1}{c}{} & 
\multicolumn{1}{c}{} &
\multicolumn{1}{c}{} &
\multicolumn{1}{c}{} & 
\multicolumn{1}{c}{} &
\multicolumn{1}{c}{} &
\multicolumn{1}{c}{} &
\multicolumn{1}{c}{} &
\multicolumn{1}{c}{} &
\multicolumn{1}{c}{} &
\multicolumn{1}{c}{($\rm cm^{-3}$)} &
\multicolumn{1}{c}{($\rm cm^{-3}$)} &
\multicolumn{1}{c}{($r_{500}$)} \\ 
\hline 

\endhead

\hline 
\multicolumn{13}{c}{{Continued on next page}} \\ 
\hline
\endfoot

\hline
\endlastfoot
G000.44-41.83 & 21:04:18.603 & -41:20:39.36 & 0.165 & 0.275 & 0.008 & 0.0585 & 0.0044 & 0.611 & 0.084 & 0.01482 & 0.00233 & 1.68 \\
G002.74-56.18 & 22:18:39.822 & -38:53:58.47 & 0.141 & 0.393 & 0.007 & 0.0604 & 0.0030 & 0.254 & 0.047 & 0.01423 & 0.00371 & 1.55 \\
G003.90-59.41 & 22:34:27.334 & -37:44:07.88 & 0.151 & 0.391 & 0.007 & 0.0335 & 0.0011 & 0.135 & 0.029 & 0.00729 & 0.00113 & 1.37 \\
G006.47+50.54 & 15:10:56.117 & +5:44:40.38 & 0.077 & 0.591 & 0.005 & 0.1674 & 0.0035 & 0.920 & 0.015 & 0.05946 & 0.00131 & 1.43 \\
G006.70-35.54 & 20:34:46.912 & -35:49:24.54 & 0.089 & 0.188 & 0.006 & 0.0334 & 0.0016 & 0.251 & 0.048 & 0.00469 & 0.00096 & 1.66 \\
G006.78+30.46 & 16:15:46.073 & -6:08:54.61 & 0.203 & 0.304 & 0.009 & 0.0250 & 0.0008 & 0.159 & 0.030 & 0.00819 & 0.00096 & 1.08 \\
G008.30-64.75 & 22:58:48.095 & -34:48:04.62 & 0.312 & 0.218 & 0.005 & 0.0422 & 0.0020 & 0.289 & 0.047 & 0.00831 & 0.00070 & 1.50 \\
G008.44-56.35 & 22:17:45.701 & -35:43:32.55 & 0.149 & 0.370 & 0.009 & 0.0939 & 0.0049 & 0.614 & 0.056 & 0.01607 & 0.00215 & 1.82 \\
G008.93-81.23 & 0:14:19.305 & -30:23:29.33 & 0.307 & 0.249 & 0.004 & 0.0273 & 0.0011 & 0.212 & 0.048 & 0.00813 & 0.00153 & 1.37 \\
G018.53-25.72 & 20:03:30.848 & -23:23:37.54 & 0.317 & 0.110 & 0.003 & 0.0141 & 0.0020 & 0.033 & 0.056 & 0.00284 & 0.00184 & 1.42 \\
G021.09+33.25 & 16:32:46.854 & +5:34:31.61 & 0.151 & 0.638 & 0.005 & 0.3344 & 0.0069 & 1.395 & 0.061 & 0.13411 & 0.00361 & 1.42 \\
G029.00+44.56 & 16:02:14.068 & +15:58:16.23 & 0.035 & 0.183 & 0.005 & 0.0380 & 0.0013 & 0.315 & 0.076 & 0.00396 & 0.00050 & 1.01 \\
G033.46-48.43 & 21:52:21.245 & -19:32:54.19 & 0.094 & 0.282 & 0.005 & 0.1452 & 0.0037 & 1.036 & 0.061 & 0.02163 & 0.00083 & 1.68 \\
G033.78+77.16 & 13:48:52.710 & +26:35:31.20 & 0.062 & 0.592 & 0.004 & 0.1774 & 0.0036 & 1.090 & 0.046 & 0.04278 & 0.00092 & 1.53 \\
G036.72+14.92 & 18:04:31.215 & +10:03:24.21 & 0.152 & 0.415 & 0.009 & 0.1236 & 0.0068 & 0.655 & 0.073 & 0.03475 & 0.00440 & 1.62 \\
-- & 18:04:27.892 & +10:02:35.55 & 0.152 & 0.392 & 0.008 & 0.0773 & 0.0050 & 0.433 & 0.078 & 0.01403 & 0.00255 & 1.67 \\
G039.85-39.98 & 21:27:12.470 & -12:10:00.46 & 0.176 & 0.124 & 0.004 & 0.0228 & 0.0041 & 0.119 & 0.062 & 0.00277 & 0.00072 & 1.69 \\
-- & 21:26:37.301 & -12:06:49.76 & 0.176 & 0.075 & 0.006 & 0.0808 & 0.0158 & 0.081 & 0.074 & 0.00736 & 0.00148 & 2.03 \\
G042.82+56.61 & 15:22:29.473 & +27:42:18.76 & 0.072 & 0.341 & 0.005 & 0.0773 & 0.0019 & 0.556 & 0.059 & 0.02364 & 0.00097 & 1.47 \\
G044.22+48.68 & 15:58:21.100 & +27:13:47.87 & 0.089 & 0.493 & 0.005 & 0.0810 & 0.0017 & 0.758 & 0.053 & 0.02480 & 0.00073 & 1.25 \\
G046.50-49.43 & 22:10:19.489 & -12:10:10.03 & 0.085 & 0.277 & 0.006 & 0.0348 & 0.0020 & 0.254 & 0.038 & 0.00588 & 0.00084 & 1.60 \\
G046.88+56.49 & 15:24:11.019 & +29:52:45.70 & 0.115 & 0.144 & 0.014 & 0.0228 & 0.0049 & 0.145 & 0.063 & 0.00258 & 0.00055 & 1.53 \\
-- & 15:24:22.521 & +30:01:09.10 & 0.115 & 0.108 & 0.004 & 0.0371 & 0.0026 & 0.508 & 0.085 & 0.00468 & 0.00083 & 1.90 \\
G048.05+57.17 & 15:21:12.694 & +30:38:00.59 & 0.078 & 0.133 & 0.004 & 0.0178 & 0.0007 & 0.037 & 0.034 & 0.00163 & 0.00020 & 1.77 \\
G049.20+30.86 & 17:20:09.957 & +26:37:30.79 & 0.164 & 0.620 & 0.007 & 0.2313 & 0.0055 & 1.238 & 0.054 & 0.06126 & 0.00245 & 1.48 \\
G049.33+44.38 & 16:20:30.305 & +29:53:35.91 & 0.097 & 0.241 & 0.007 & 0.0459 & 0.0029 & 0.350 & 0.125 & 0.00543 & 0.00089 & 1.81 \\
G049.66-49.50 & 22:14:32.554 & -10:22:17.84 & 0.098 & 0.437 & 0.009 & 0.0977 & 0.0037 & 0.511 & 0.069 & 0.01625 & 0.00201 & 1.81 \\
G053.44-36.26 & 21:35:11.371 & -1:02:53.24 & 0.325 & 0.273 & 0.008 & 0.0536 & 0.0046 & 0.468 & 0.129 & 0.01374 & 0.00339 & 1.70 \\
-- & 21:35:25.878 & -0:57:44.27 & 0.325 & 0.350 & 0.094 & 0.4745 & 0.0922 & 0.115 & 0.113 & 0.16250 & 0.02681 & 2.39 \\
G053.52+59.54 & 15:10:12.700 & +33:30:34.01 & 0.113 & 0.293 & 0.004 & 0.0326 & 0.0009 & 0.152 & 0.033 & 0.00522 & 0.00057 & 1.44 \\
-- & 15:10:13.060 & +33:32:26.21 & 0.113 & 0.259 & 0.004 & 0.0298 & 0.0007 & 0.072 & 0.025 & 0.00312 & 0.00025 & 1.58 \\
G055.60+31.86 & 17:22:27.300 & +32:07:57.98 & 0.224 & 0.495 & 0.007 & 0.1173 & 0.0045 & 0.720 & 0.031 & 0.03980 & 0.00297 & 1.52 \\
G055.97-34.88 & 21:35:16.105 & +1:25:03.07 & 0.124 & 0.282 & 0.013 & 0.0451 & 0.0085 & 0.150 & 0.108 & 0.00489 & 0.00206 & 2.04 \\
G056.81+36.31 & 17:02:42.571 & +34:03:38.15 & 0.095 & 0.485 & 0.006 & 0.1049 & 0.0023 & 0.616 & 0.028 & 0.02366 & 0.00123 & 1.61 \\
G057.33+88.01 & 12:59:47.654 & +27:57:06.81 & 0.023 & 0.234 & 0.003 & 0.0232 & 0.0004 & 0.120 & 0.023 & 0.00361 & 0.00018 & 1.12 \\
G057.61+34.94 & 17:09:45.792 & +34:27:20.36 & 0.080 & 0.163 & 0.005 & 0.0253 & 0.0021 & 0.371 & 0.270 & 0.00449 & 0.00168 & 1.82 \\
G057.92+27.64 & 17:44:15.426 & +32:59:31.71 & 0.076 & 0.555 & 0.005 & 0.2121 & 0.0046 & 1.140 & 0.058 & 0.03579 & 0.00142 & 1.91 \\
G058.28+18.59 & 18:25:22.292 & +30:26:38.42 & 0.065 & 0.199 & 0.005 & 0.0312 & 0.0017 & 0.192 & 0.056 & 0.00360 & 0.00067 & 1.45 \\
G062.42-46.41 & 22:23:47.779 & -1:39:00.82 & 0.091 & 0.294 & 0.008 & 0.1466 & 0.0051 & 0.943 & 0.071 & 0.03207 & 0.00167 & 1.96 \\
-- & 22:23:56.949 & -1:35:01.68 & 0.091 & 0.256 & 0.007 & 0.1044 & 0.0040 & 0.840 & 0.037 & 0.02285 & 0.00159 & 1.90 \\
-- & 22:23:14.663 & -1:39:36.81 & 0.091 & 0.317 & 0.029 & 0.2421 & 0.0271 & 1.177 & 0.164 & 0.03242 & 0.00496 & 3.04 \\
G062.92+43.70 & 16:28:38.232 & +39:33:03.36 & 0.030 & 0.536 & 0.006 & 0.1906 & 0.0031 & 0.801 & 0.031 & 0.03963 & 0.00122 & 0.76 \\
G067.23+67.46 & 14:26:00.269 & +37:49:40.52 & 0.171 & 0.492 & 0.008 & 0.0544 & 0.0033 & 0.260 & 0.038 & 0.01615 & 0.00367 & 1.39 \\
-- & 14:26:03.448 & +37:49:28.87 & 0.171 & 0.442 & 0.007 & 0.0972 & 0.0036 & 0.784 & 0.030 & 0.02393 & 0.00121 & 1.41 \\
G071.61+29.79 & 17:47:12.236 & +45:12:29.69 & 0.157 & 0.141 & 0.005 & 0.0270 & 0.0048 & 0.232 & 0.092 & 0.00363 & 0.00118 & 1.83 \\
G072.63+41.46 & 16:40:19.460 & +46:42:45.63 & 0.228 & 0.347 & 0.005 & 0.0385 & 0.0013 & 0.373 & 0.033 & 0.01247 & 0.00087 & 1.18 \\
G072.80-18.72 & 21:22:27.115 & +23:11:50.12 & 0.143 & 0.330 & 0.007 & 0.0732 & 0.0041 & 0.872 & 0.102 & 0.01786 & 0.00206 & 1.47 \\
G073.96-27.82 & 21:53:36.797 & +17:41:43.53 & 0.233 & 0.502 & 0.006 & 0.1336 & 0.0041 & 1.046 & 0.071 & 0.06475 & 0.00273 & 1.19 \\
G077.90-26.64 & 22:00:53.012 & +20:58:43.93 & 0.147 & 0.368 & 0.008 & 0.0648 & 0.0037 & 0.597 & 0.122 & 0.01476 & 0.00259 & 1.64 \\
G080.38-33.20 & 22:26:02.754 & +17:22:34.09 & 0.107 & 0.316 & 0.005 & 0.0417 & 0.0013 & 0.212 & 0.062 & 0.00566 & 0.00092 & 1.86 \\
G080.99-50.90 & 23:11:33.144 & +3:38:08.17 & 0.300 & 0.375 & 0.007 & 0.0792 & 0.0033 & 0.751 & 0.028 & 0.02297 & 0.00232 & 1.46 \\
-- & 23:11:48.226 & +3:40:51.80 & 0.300 & 0.152 & 0.010 & 0.0618 & 0.0089 & 0.669 & 0.226 & 0.01322 & 0.00491 & 2.20 \\
G085.99+26.71 & 18:19:57.122 & +57:09:50.82 & 0.179 & 0.142 & 0.006 & 0.0278 & 0.0027 & 0.248 & 0.106 & 0.00376 & 0.00127 & 1.79 \\
G086.45+15.29 & 19:38:18.297 & +54:09:36.16 & 0.260 & 0.492 & 0.011 & 0.0860 & 0.0042 & 0.553 & 0.058 & 0.02315 & 0.00334 & 1.42 \\
G092.73+73.46 & 13:35:18.141 & +40:59:59.07 & 0.228 & 0.292 & 0.006 & 0.0444 & 0.0030 & 0.370 & 0.047 & 0.01125 & 0.00175 & 1.34 \\
G093.91+34.90 & 17:12:43.585 & +64:03:46.90 & 0.081 & 0.162 & 0.004 & 0.0197 & 0.0006 & 0.077 & 0.038 & 0.00233 & 0.00020 & 1.20 \\
G094.01+27.42 & 18:21:57.197 & +64:20:36.30 & 0.299 & 0.563 & 0.007 & 0.1890 & 0.0063 & 1.209 & 0.083 & 0.07306 & 0.00327 & 1.43 \\
G096.85+52.46 & 14:52:58.061 & +58:03:00.56 & 0.318 & 0.397 & 0.013 & 0.0835 & 0.0170 & 0.142 & 0.032 & 0.03621 & 0.00478 & 1.76 \\
G097.73+38.11 & 16:35:51.314 & +66:12:40.87 & 0.171 & 0.347 & 0.006 & 0.0392 & 0.0016 & 0.177 & 0.046 & 0.00766 & 0.00142 & 1.51 \\
G098.95+24.86 & 18:54:02.098 & +68:23:01.24 & 0.093 & 0.367 & 0.010 & 0.0966 & 0.0058 & 0.728 & 0.032 & 0.01629 & 0.00103 & 1.85 \\
G106.73-83.22 & 0:43:24.653 & -20:37:24.39 & 0.292 & 0.394 & 0.007 & 0.0458 & 0.0020 & 0.176 & 0.065 & 0.01064 & 0.00243 & 1.51 \\
G107.11+65.31 & 13:32:47.351 & +50:32:29.58 & 0.280 & 0.220 & 0.004 & 0.0322 & 0.0017 & 0.295 & 0.029 & 0.00742 & 0.00111 & 1.48 \\
-- & 13:32:38.660 & +50:33:45.07 & 0.280 & 0.206 & 0.005 & 0.0491 & 0.0022 & 0.600 & 0.055 & 0.01367 & 0.00123 & 1.56 \\
-- & 13:32:33.121 & +50:25:01.44 & 0.280 & 0.218 & 0.005 & 0.0435 & 0.0019 & 0.446 & 0.084 & 0.00880 & 0.00073 & 1.63 \\
G110.98+31.73 & 17:03:14.917 & +78:39:23.17 & 0.058 & 0.220 & 0.004 & 0.0217 & 0.0009 & 0.126 & 0.046 & 0.00367 & 0.00038 & 1.36 \\
G112.45+57.03 & 13:36:05.971 & +59:12:08.41 & 0.070 & 0.252 & 0.008 & 0.0444 & 0.0068 & 0.305 & 0.063 & 0.00546 & 0.00073 & 1.80 \\
G113.82+44.35 & 14:13:55.482 & +71:17:59.81 & 0.225 & 0.189 & 0.015 & 0.0710 & 0.0306 & 0.379 & 0.083 & 0.01066 & 0.00370 & 1.87 \\
-- & 14:14:14.032 & +71:17:19.62 & 0.225 & 0.138 & 0.008 & 0.0318 & 0.0058 & 0.232 & 0.075 & 0.00534 & 0.00306 & 1.97 \\
-- & 14:14:06.680 & +71:15:44.77 & 0.225 & 0.119 & 0.011 & 0.0308 & 0.0179 & 0.283 & 0.153 & 0.00498 & 0.00281 & 1.93 \\
G114.33+64.87 & 13:15:04.617 & +51:49:10.75 & 0.284 & 0.317 & 0.007 & 0.0444 & 0.0023 & 0.400 & 0.044 & 0.01172 & 0.00116 & 1.55 \\
G115.16-72.09 & 0:41:50.390 & -9:18:09.53 & 0.056 & 0.455 & 0.004 & 0.1535 & 0.0030 & 1.012 & 0.040 & 0.04534 & 0.00113 & 1.40 \\
G115.71+17.52 & 22:26:30.303 & +78:19:16.11 & 0.300 & 0.474 & 0.011 & 0.1686 & 0.0092 & 0.920 & 0.080 & 0.03959 & 0.00408 & 1.63 \\
G118.60+28.55 & 17:24:11.920 & +85:53:08.41 & 0.178 & 0.335 & 0.009 & 0.0534 & 0.0029 & 0.461 & 0.099 & 0.01043 & 0.00147 & 1.42 \\
G121.11+57.01 & 12:59:35.208 & +60:04:15.62 & 0.344 & 0.137 & 0.005 & 0.0194 & 0.0049 & 0.151 & 0.205 & 0.00536 & 0.00396 & 1.64 \\
G124.21-36.48 & 0:55:50.297 & +26:24:36.47 & 0.197 & 0.405 & 0.006 & 0.2269 & 0.0056 & 0.900 & 0.042 & 0.03266 & 0.00087 & 1.65 \\
G125.58-64.14 & 0:56:16.210 & -1:14:58.98 & 0.044 & 0.164 & 0.004 & 0.0229 & 0.0007 & 0.221 & 0.052 & 0.00258 & 0.00027 & 1.58 \\
G125.70+53.85 & 12:36:57.703 & +63:11:13.65 & 0.302 & 0.338 & 0.008 & 0.0607 & 0.0032 & 0.481 & 0.073 & 0.01758 & 0.00235 & 1.56 \\
G139.19+56.35 & 11:42:24.113 & +58:31:37.19 & 0.322 & 0.189 & 0.006 & 0.0191 & 0.0022 & 0.065 & 0.088 & 0.00517 & 0.00311 & 1.53 \\
G139.59+24.18 & 6:21:49.156 & +74:42:05.10 & 0.300 & 0.519 & 0.008 & 0.1826 & 0.0072 & 1.085 & 0.109 & 0.07301 & 0.00533 & 1.47 \\
G143.24+65.21 & 11:59:13.697 & +49:47:41.37 & 0.211 & 0.409 & 0.018 & 0.1210 & 0.0265 & 0.285 & 0.046 & 0.01037 & 0.00162 & 1.60 \\
-- & 11:59:32.831 & +49:47:07.28 & 0.211 & 0.167 & 0.013 & 0.0409 & 0.0075 & 0.360 & 0.221 & 0.00402 & 0.00182 & 1.83 \\
G146.33-15.59 & 2:54:27.386 & +41:34:46.82 & 0.017 & 0.476 & 0.012 & 0.1241 & 0.0023 & 0.760 & 0.018 & 0.01932 & 0.00045 & 0.45 \\
G149.24+54.18 & 10:58:26.061 & +56:47:35.78 & 0.137 & 0.285 & 0.009 & 0.0308 & 0.0023 & 0.250 & 0.104 & 0.00638 & 0.00131 & 1.51 \\
G149.73+34.69 & 8:30:59.262 & +65:50:22.84 & 0.182 & 0.325 & 0.005 & 0.0512 & 0.0015 & 0.476 & 0.040 & 0.01397 & 0.00103 & 1.32 \\
G159.85-73.47 & 1:31:53.170 & -13:36:43.29 & 0.206 & 0.311 & 0.007 & 0.0436 & 0.0022 & 0.276 & 0.056 & 0.00909 & 0.00100 & 1.39 \\
G161.44+26.23 & 7:21:30.328 & +55:45:41.56 & 0.038 & 0.313 & 0.008 & 0.0671 & 0.0039 & 0.390 & 0.070 & 0.00755 & 0.00111 & 1.10 \\
G163.72+53.53 & 10:22:28.815 & +50:06:24.68 & 0.158 & 0.342 & 0.009 & 0.0517 & 0.0039 & 0.428 & 0.079 & 0.01155 & 0.00173 & 1.64 \\
G164.18-38.89 & 2:58:57.595 & +13:34:44.19 & 0.074 & 0.308 & 0.005 & 0.0400 & 0.0012 & 0.262 & 0.040 & 0.00776 & 0.00081 & 1.34 \\
G164.61+46.38 & 9:38:19.645 & +52:02:58.03 & 0.342 & 0.357 & 0.009 & 0.0872 & 0.0045 & 0.423 & 0.083 & 0.01742 & 0.00209 & 1.70 \\
G165.08+54.11 & 10:23:39.906 & +49:08:32.59 & 0.144 & 0.337 & 0.009 & 0.0511 & 0.0032 & 0.415 & 0.089 & 0.01238 & 0.00232 & 1.56 \\
G166.13+43.39 & 9:17:53.077 & +51:43:41.38 & 0.217 & 0.326 & 0.006 & 0.0519 & 0.0022 & 0.575 & 0.024 & 0.01756 & 0.00111 & 1.51 \\
G167.65+17.64 & 6:38:04.950 & +47:47:50.92 & 0.174 & 0.260 & 0.009 & 0.0302 & 0.0022 & 0.163 & 0.088 & 0.00591 & 0.00128 & 1.61 \\
G171.94-40.65 & 3:12:57.303 & +8:22:08.53 & 0.270 & 0.235 & 0.009 & 0.0356 & 0.0024 & 0.290 & 0.116 & 0.00885 & 0.00156 & 1.59 \\
G172.88+65.32 & 11:11:40.472 & +40:50:28.25 & 0.079 & 0.256 & 0.007 & 0.0559 & 0.0032 & 0.400 & 0.090 & 0.00698 & 0.00157 & 2.11 \\
G176.28-35.05 & 3:38:40.698 & +9:58:03.07 & 0.035 & 0.718 & 0.004 & 0.3876 & 0.0047 & 1.242 & 0.017 & 0.05916 & 0.00104 & 1.09 \\
G180.62+76.65 & 11:57:17.375 & +33:36:39.11 & 0.213 & 0.378 & 0.007 & 0.1228 & 0.0050 & 0.844 & 0.059 & 0.03195 & 0.00255 & 1.64 \\
G182.44-28.29 & 4:13:25.196 & +10:27:53.66 & 0.088 & 0.622 & 0.005 & 0.1726 & 0.0033 & 0.908 & 0.056 & 0.05973 & 0.00179 & 1.38 \\
G182.63+55.82 & 10:17:03.531 & +39:02:53.32 & 0.206 & 0.467 & 0.006 & 0.0986 & 0.0031 & 0.634 & 0.062 & 0.02822 & 0.00252 & 1.63 \\
G186.39+37.25 & 8:42:57.490 & +36:21:59.64 & 0.282 & 0.338 & 0.007 & 0.0361 & 0.0027 & 0.218 & 0.076 & 0.01399 & 0.00410 & 1.32 \\
G195.62+44.05 & 9:20:23.927 & +30:30:37.51 & 0.2952 & 0.131 & 0.005 & 0.0170 & 0.0009 & 0.044 & 0.043 & 0.00238 & 0.00035 & 1.52 \\
-- & 9:19:34.939 & +30:31:54.67 & 0.427 & 0.326 & 0.026 & 0.1066 & 0.0131 & 0.392 & 0.207 & 0.01544 & 0.00611 & 2.56 \\
-- & 9:21:10.775 & +30:28:04.61 & 0.427 & 0.082 & 0.035 & 0.0399 & 0.0544 & 0.480 & 0.568 & 0.01593 & 0.02757 & 2.63 \\
G195.77-24.30 & 4:54:10.021 & +2:55:33.68 & 0.203 & 0.214 & 0.004 & 0.0205 & 0.0007 & 0.117 & 0.036 & 0.00398 & 0.00033 & 1.36 \\
-- & 4:54:25.560 & +2:59:08.98 & 0.203 & 0.090 & 0.004 & 0.0251 & 0.0062 & 0.095 & 0.073 & 0.00156 & 0.00090 & 2.18 \\
G209.56-36.49 & 4:33:37.913 & -13:15:42.03 & 0.033 & 0.438 & 0.005 & 0.1892 & 0.0039 & 1.200 & 0.086 & 0.04229 & 0.00359 & 0.79 \\
G218.85+35.50 & 9:09:12.462 & +10:58:29.96 & 0.175 & 0.385 & 0.006 & 0.0831 & 0.0031 & 0.466 & 0.035 & 0.01317 & 0.00150 & 1.65 \\
G226.17-21.91 & 5:52:50.842 & -21:03:15.06 & 0.099 & 0.291 & 0.008 & 0.0437 & 0.0046 & 0.274 & 0.069 & 0.00613 & 0.00091 & 1.75 \\
G226.24+76.76 & 11:55:17.943 & +23:24:20.25 & 0.143 & 0.488 & 0.005 & 0.0985 & 0.0026 & 0.666 & 0.033 & 0.02751 & 0.00107 & 1.39 \\
G228.49+53.12 & 10:25:58.011 & +12:41:08.71 & 0.143 & 0.603 & 0.009 & 0.2305 & 0.0082 & 0.917 & 0.052 & 0.07758 & 0.00512 & 1.72 \\
G229.21-17.24 & 6:16:24.830 & -21:56:19.24 & 0.171 & 0.199 & 0.006 & 0.0300 & 0.0029 & 0.231 & 0.119 & 0.00436 & 0.00115 & 1.60 \\
G229.64+77.96 & 12:01:13.211 & +23:06:29.50 & 0.269 & 0.218 & 0.008 & 0.0238 & 0.0022 & 0.141 & 0.082 & 0.00575 & 0.00223 & 1.43 \\
G229.94+15.29 & 8:17:25.557 & -7:30:30.54 & 0.070 & 0.483 & 0.005 & 0.0718 & 0.0019 & 0.415 & 0.043 & 0.01305 & 0.00090 & 1.48 \\
G234.59+73.01 & 11:44:44.158 & +19:42:15.97 & 0.021 & 0.095 & 0.001 & 0.0242 & 0.0001 & 0.292 & 0.016 & 0.00240 & 0.00007 & 0.95 \\
G236.95-26.67 & 5:47:37.339 & -31:52:09.32 & 0.148 & 0.340 & 0.006 & 0.0452 & 0.0024 & 0.309 & 0.076 & 0.00968 & 0.00210 & 1.59 \\
G239.28+24.76 & 9:09:20.461 & -9:41:03.53 & 0.054 & 0.333 & 0.003 & 0.0477 & 0.0011 & 0.276 & 0.018 & 0.00561 & 0.00022 & 1.35 \\
-- & 9:08:47.478 & -9:38:28.45 & 0.054 & 0.323 & 0.183 & 0.1382 & 0.2078 & 0.713 & 0.843 & 0.02846 & 0.05574 & 1.42 \\
G241.74-30.88 & 5:32:55.628 & -37:01:35.71 & 0.271 & 0.447 & 0.009 & 0.0846 & 0.0044 & 0.532 & 0.082 & 0.02341 & 0.00361 & 1.51 \\
G241.77-24.00 & 6:05:53.936 & -35:18:08.71 & 0.139 & 0.536 & 0.008 & 0.2448 & 0.0078 & 1.134 & 0.062 & 0.08282 & 0.00499 & 1.88 \\
G241.85+51.53 & 10:39:40.051 & +5:09:39.68 & 0.070 & 0.090 & 0.005 & 0.0332 & 0.0083 & 0.084 & 0.061 & 0.00137 & 0.00035 & 2.30 \\
G241.97+14.85 & 8:41:58.360 & -17:29:45.45 & 0.169 & 0.118 & 0.003 & 0.0158 & 0.0006 & 0.073 & 0.025 & 0.00193 & 0.00015 & 1.50 \\
-- & 8:41:51.490 & -17:27:46.97 & 0.169 & 0.183 & 0.004 & 0.0350 & 0.0013 & 0.302 & 0.055 & 0.00598 & 0.00092 & 1.51 \\
G243.57+67.76 & 11:32:51.155 & +14:29:32.08 & 0.083 & 0.246 & 0.006 & 0.0362 & 0.0031 & 0.181 & 0.040 & 0.00442 & 0.00064 & 1.70 \\
-- & 11:32:50.676 & +14:27:21.08 & 0.083 & 0.268 & 0.008 & 0.0696 & 0.0036 & 0.635 & 0.025 & 0.01276 & 0.00092 & 1.76 \\
G244.34-32.13 & 5:28:52.997 & -39:28:13.50 & 0.284 & 0.411 & 0.007 & 0.1159 & 0.0043 & 1.145 & 0.074 & 0.03555 & 0.00249 & 1.39 \\
G244.69+32.49 & 9:45:24.590 & -8:39:19.44 & 0.153 & 0.264 & 0.009 & 0.0432 & 0.0100 & 0.140 & 0.063 & 0.00566 & 0.00180 & 1.78 \\
G246.52-26.05 & 6:01:41.584 & -39:59:19.90 & 0.047 & 0.133 & 0.003 & 0.0179 & 0.0006 & 0.052 & 0.050 & 0.00123 & 0.00019 & 1.54 \\
-- & 6:02:11.776 & -39:57:26.35 & 0.047 & 0.184 & 0.005 & 0.0431 & 0.0015 & 0.287 & 0.050 & 0.00278 & 0.00034 & 1.31 \\
G247.17-23.32 & 6:16:31.784 & -39:47:46.01 & 0.152 & 0.268 & 0.014 & 0.1071 & 0.0252 & 0.304 & 0.083 & 0.00712 & 0.00156 & 1.88 \\
G249.87-39.86 & 4:49:56.204 & -44:40:20.25 & 0.150 & 0.321 & 0.007 & 0.0704 & 0.0032 & 0.285 & 0.064 & 0.00955 & 0.00221 & 2.05 \\
-- & 4:49:53.125 & -44:40:23.36 & 0.150 & 0.216 & 0.007 & 0.0763 & 0.0049 & 0.790 & 0.136 & 0.01097 & 0.00120 & 2.08 \\
G250.90-36.25 & 5:10:16.218 & -45:19:12.68 & 0.200 & 0.377 & 0.010 & 0.0645 & 0.0037 & 0.350 & 0.053 & 0.01260 & 0.00239 & 1.58 \\
G252.96-56.05 & 3:17:57.637 & -44:14:17.40 & 0.075 & 0.581 & 0.004 & 0.2692 & 0.0048 & 1.140 & 0.035 & 0.06626 & 0.00186 & 1.67 \\
G253.47-33.72 & 5:25:48.812 & -47:15:10.22 & 0.191 & 0.375 & 0.009 & 0.0739 & 0.0045 & 0.584 & 0.158 & 0.01927 & 0.00387 & 1.79 \\
G256.45-65.71 & 2:25:53.140 & -41:54:52.95 & 0.220 & 0.388 & 0.018 & 0.1507 & 0.0342 & 0.848 & 0.054 & 0.04070 & 0.01416 & 1.60 \\
-- & 2:25:25.618 & -42:00:50.57 & 0.220 & 0.197 & 0.017 & 0.1490 & 0.0277 & 1.047 & 0.058 & 0.03069 & 0.00388 & 2.46 \\
G257.34-22.18 & 6:37:14.638 & -48:28:18.15 & 0.203 & 0.282 & 0.009 & 0.0942 & 0.0058 & 1.158 & 0.152 & 0.02179 & 0.00342 & 1.78 \\
-- & 6:37:29.500 & -48:29:40.52 & 0.203 & 0.172 & 0.008 & 0.0299 & 0.0041 & 0.146 & 0.071 & 0.00405 & 0.00197 & 1.69 \\
G260.03-63.44 & 2:32:18.714 & -44:20:46.38 & 0.284 & 0.416 & 0.023 & 0.1618 & 0.0182 & 1.065 & 0.165 & 0.07744 & 0.01227 & 1.69 \\
G262.25-35.36 & 5:16:36.189 & -54:30:34.93 & 0.295 & 0.156 & 0.005 & 0.0245 & 0.0027 & 0.185 & 0.109 & 0.00553 & 0.00221 & 1.37 \\
G263.16-23.41 & 6:38:48.541 & -53:58:26.09 & 0.227 & 0.472 & 0.006 & 0.1319 & 0.0040 & 1.003 & 0.043 & 0.04609 & 0.00206 & 1.37 \\
-- & 6:38:44.287 & -53:58:26.85 & 0.227 & 0.453 & 0.007 & 0.1073 & 0.0034 & 0.743 & 0.030 & 0.02970 & 0.00158 & 1.41 \\
G263.20-25.21 & 6:26:48.188 & -54:32:57.24 & 0.051 & 0.117 & 0.003 & 0.0442 & 0.0015 & 0.598 & 0.062 & 0.00571 & 0.00051 & 1.73 \\
-- & 6:27:36.008 & -54:26:46.65 & 0.051 & 0.116 & 0.002 & 0.0279 & 0.0007 & 0.171 & 0.025 & 0.00216 & 0.00017 & 1.72 \\
G263.66-22.53 & 6:45:28.586 & -54:13:43.00 & 0.164 & 0.403 & 0.010 & 0.0738 & 0.0042 & 0.644 & 0.086 & 0.02070 & 0.00301 & 1.39 \\
G264.41+19.48 & 10:00:01.753 & -30:16:37.46 & 0.240 & 0.271 & 0.008 & 0.0596 & 0.0043 & 0.648 & 0.134 & 0.01621 & 0.00302 & 1.65 \\
G265.00-48.94 & 3:42:53.069 & -53:37:53.35 & 0.059 & 0.317 & 0.005 & 0.0445 & 0.0021 & 0.217 & 0.028 & 0.00601 & 0.00048 & 1.25 \\
G266.03-21.25 & 6:58:29.918 & -55:56:31.03 & 0.296 & 0.331 & 0.006 & 0.0254 & 0.0008 & 0.280 & 0.061 & 0.00955 & 0.00077 & 1.11 \\
-- & 6:58:20.053 & -55:56:28.77 & 0.296 & 0.299 & 0.005 & 0.0989 & 0.0033 & 0.962 & 0.051 & 0.02120 & 0.00112 & 1.21 \\
G266.84+25.07 & 10:23:50.137 & -27:15:22.24 & 0.254 & 0.598 & 0.007 & 0.2097 & 0.0059 & 1.056 & 0.031 & 0.07762 & 0.00345 & 1.59 \\
G269.31-49.87 & 3:28:36.732 & -55:42:41.79 & 0.085 & 0.396 & 0.009 & 0.0627 & 0.0023 & 0.199 & 0.046 & 0.00682 & 0.00114 & 1.83 \\
G269.51+26.42 & 10:36:40.918 & -27:31:35.50 & 0.013 & 0.426 & 0.010 & 0.1282 & 0.0027 & 0.463 & 0.055 & 0.00835 & 0.00041 & 0.46 \\
G271.50-56.55 & 2:45:28.461 & -53:02:03.51 & 0.300 & 0.354 & 0.008 & 0.0357 & 0.0030 & 0.120 & 0.081 & 0.00988 & 0.00364 & 1.55 \\
G272.10-40.15 & 4:31:30.202 & -61:24:48.44 & 0.059 & 0.205 & 0.088 & 0.0204 & 0.0028 & 0.162 & 0.109 & 0.00333 & 0.00069 & 0.98 \\
-- & 4:31:12.696 & -61:27:20.58 & 0.059 & 0.199 & 0.004 & 0.0419 & 0.0013 & 0.531 & 0.025 & 0.00796 & 0.00038 & 0.90 \\
G273.64+63.28 & 12:00:22.549 & +3:20:23.47 & 0.134 & 0.203 & 0.004 & 0.0355 & 0.0017 & 0.304 & 0.064 & 0.00610 & 0.00109 & 1.52 \\
G275.21+43.92 & 11:30:22.201 & -14:34:38.17 & 0.107 & 0.258 & 0.006 & 0.0417 & 0.0020 & 0.319 & 0.110 & 0.00756 & 0.00175 & 1.65 \\
G278.60+39.17 & 11:31:56.126 & -19:56:06.98 & 0.307 & 0.259 & 0.008 & 0.0410 & 0.0037 & 0.397 & 0.130 & 0.01199 & 0.00307 & 1.39 \\
-- & 11:31:54.302 & -19:55:42.44 & 0.307 & 0.317 & 0.016 & 0.1091 & 0.0105 & 0.780 & 0.087 & 0.03229 & 0.00395 & 1.53 \\
G280.19+47.81 & 11:49:46.508 & -12:18:50.35 & 0.156 & 0.224 & 0.006 & 0.0337 & 0.0032 & 0.184 & 0.058 & 0.00448 & 0.00095 & 1.65 \\
G282.49+65.17 & 12:17:41.196 & +3:39:22.03 & 0.077 & 0.286 & 0.006 & 0.0331 & 0.0021 & 0.146 & 0.047 & 0.00537 & 0.00141 & 1.51 \\
G286.58-31.25 & 5:31:28.179 & -75:10:37.86 & 0.210 & 0.248 & 0.009 & 0.0396 & 0.0037 & 0.296 & 0.212 & 0.01587 & 0.00465 & 1.67 \\
G288.61-37.65 & 3:52:30.180 & -74:01:56.35 & 0.127 & 0.225 & 0.006 & 0.0345 & 0.0023 & 0.408 & 0.139 & 0.00817 & 0.00189 & 1.43 \\
G292.51+21.98 & 12:01:04.779 & -39:51:52.89 & 0.300 & 0.181 & 0.004 & 0.0318 & 0.0021 & 0.309 & 0.077 & 0.00686 & 0.00190 & 1.45 \\
-- & 12:01:10.496 & -39:54:44.42 & 0.300 & 0.215 & 0.009 & 0.1284 & 0.0104 & 0.940 & 0.113 & 0.02524 & 0.00323 & 1.95 \\
G294.66-37.02 & 3:03:43.147 & -77:52:47.22 & 0.274 & 0.321 & 0.010 & 0.0474 & 0.0144 & 0.240 & 0.041 & 0.00998 & 0.00237 & 1.40 \\
-- & 2:59:20.807 & -77:52:10.09 & 0.274 & 0.258 & 0.017 & 0.0725 & 0.0091 & 0.399 & 0.203 & 0.01175 & 0.00627 & 2.20 \\
G295.33+23.33 & 12:15:27.399 & -39:01:58.34 & 0.119 & 0.169 & 0.005 & 0.0327 & 0.0024 & 0.240 & 0.096 & 0.00395 & 0.00116 & 1.66 \\
G296.41-32.48 & 3:51:31.364 & -82:13:27.86 & 0.061 & 0.238 & 0.007 & 0.0499 & 0.0044 & 0.334 & 0.069 & 0.00526 & 0.00082 & 1.81 \\
G303.75+33.65 & 12:54:40.703 & -29:13:40.55 & 0.054 & 0.330 & 0.010 & 0.1854 & 0.0057 & 1.435 & 0.141 & 0.03161 & 0.00295 & 1.94 \\
-- & 12:54:22.138 & -29:00:47.20 & 0.054 & 0.331 & 0.007 & 0.1652 & 0.0060 & 0.939 & 0.094 & 0.03347 & 0.00192 & 1.89 \\
G304.49+32.44 & 12:57:22.033 & -30:21:49.34 & 0.055 & 0.205 & 0.006 & 0.0481 & 0.0034 & 0.409 & 0.037 & 0.00591 & 0.00062 & 1.37 \\
G304.67-31.66 & 23:40:12.802 & -85:11:01.81 & 0.193 & 0.131 & 0.006 & 0.0200 & 0.0023 & 0.150 & 0.168 & 0.00326 & 0.00177 & 1.74 \\
G304.89+45.45 & 12:57:11.570 & -17:24:33.57 & 0.047 & 0.195 & 0.003 & 0.0846 & 0.0019 & 0.905 & 0.024 & 0.01788 & 0.00048 & 1.14 \\
G306.68+61.06 & 12:58:41.433 & -1:45:43.44 & 0.085 & 0.456 & 0.006 & 0.1186 & 0.0027 & 0.777 & 0.029 & 0.02416 & 0.00118 & 1.60 \\
G306.80+58.60 & 12:59:22.339 & -4:11:45.31 & 0.085 & 0.438 & 0.007 & 0.0773 & 0.0022 & 0.535 & 0.049 & 0.01708 & 0.00144 & 1.50 \\
G311.99+30.71 & 13:27:56.824 & -31:29:43.84 & 0.048 & 0.302 & 0.005 & 0.0611 & 0.0013 & 0.506 & 0.038 & 0.01224 & 0.00057 & 1.66 \\
-- & 13:29:47.769 & -31:36:26.05 & 0.048 & 0.149 & 0.004 & 0.0606 & 0.0048 & 0.701 & 0.056 & 0.00968 & 0.00094 & 2.08 \\
G313.36+61.11 & 13:11:29.460 & -1:20:27.68 & 0.183 & 0.590 & 0.006 & 0.1098 & 0.0027 & 0.717 & 0.024 & 0.04374 & 0.00190 & 1.32 \\
G313.87-17.10 & 16:01:48.426 & -75:45:15.56 & 0.153 & 0.540 & 0.007 & 0.0813 & 0.0037 & 0.513 & 0.036 & 0.02341 & 0.00170 & 1.39 \\
G315.70-18.04 & 16:31:25.411 & -75:06:38.34 & 0.105 & 0.201 & 0.004 & 0.0249 & 0.0020 & 0.137 & 0.046 & 0.00350 & 0.00044 & 1.42 \\
G316.34+28.54 & 13:47:28.110 & -32:51:57.98 & 0.039 & 0.394 & 0.006 & 0.0636 & 0.0014 & 0.576 & 0.070 & 0.01769 & 0.00069 & 0.80 \\
G318.13-29.57 & 19:47:14.773 & -76:23:44.99 & 0.217 & 0.441 & 0.010 & 0.1502 & 0.0107 & 0.841 & 0.070 & 0.05416 & 0.00547 & 1.54 \\
G321.96-47.97 & 22:49:58.191 & -64:25:46.79 & 0.094 & 0.313 & 0.006 & 0.0723 & 0.0054 & 0.303 & 0.061 & 0.00702 & 0.00084 & 1.61 \\
G324.49-44.97 & 22:18:00.518 & -65:10:52.42 & 0.095 & 0.342 & 0.009 & 0.1225 & 0.0054 & 0.735 & 0.088 & 0.03748 & 0.00221 & 1.87 \\
G332.23-46.36 & 22:01:53.194 & -59:56:43.51 & 0.098 & 0.388 & 0.005 & 0.0512 & 0.0013 & 0.298 & 0.052 & 0.01155 & 0.00114 & 1.53 \\
G332.88-19.28 & 18:13:15.663 & -61:26:54.65 & 0.147 & 0.347 & 0.008 & 0.0477 & 0.0029 & 0.296 & 0.072 & 0.01085 & 0.00228 & 1.58 \\
G335.59-46.46 & 21:54:05.793 & -57:51:41.87 & 0.076 & 0.230 & 0.006 & 0.0451 & 0.0039 & 0.430 & 0.056 & 0.00787 & 0.00093 & 1.72 \\
G336.59-55.44 & 22:46:21.491 & -52:44:17.54 & 0.097 & 0.181 & 0.006 & 0.0230 & 0.0012 & 0.076 & 0.050 & 0.00281 & 0.00040 & 1.68 \\
G337.09-25.97 & 19:14:37.070 & -59:28:17.20 & 0.120 & 0.501 & 0.009 & 0.1818 & 0.0063 & 0.764 & 0.027 & 0.02517 & 0.00163 & 2.02 \\
-- & 19:13:51.095 & -59:33:51.75 & 0.120 & 0.488 & 0.010 & 0.3395 & 0.0096 & 1.053 & 0.069 & 0.04536 & 0.00329 & 2.62 \\
G340.88-33.34 & 20:12:37.869 & -56:50:45.53 & 0.056 & 0.185 & 0.004 & 0.0291 & 0.0006 & 0.112 & 0.013 & 0.00402 & 0.00011 & 1.01 \\
G340.95+35.11 & 14:59:28.730 & -18:10:45.08 & 0.236 & 0.516 & 0.007 & 0.2230 & 0.0074 & 1.070 & 0.034 & 0.07261 & 0.00394 & 1.58 \\
G342.31-34.90 & 20:23:22.364 & -55:35:27.23 & 0.232 & 0.286 & 0.009 & 0.0388 & 0.0120 & 0.162 & 0.050 & 0.00767 & 0.00279 & 1.51 \\
G342.81-30.46 & 19:52:08.122 & -55:03:40.32 & 0.060 & 0.160 & 0.006 & 0.0319 & 0.0019 & 0.155 & 0.086 & 0.00282 & 0.00065 & 1.90 \\
G345.40-39.34 & 20:51:56.959 & -52:37:48.21 & 0.045 & 0.119 & 0.005 & 0.0707 & 0.0056 & 0.673 & 0.051 & 0.01019 & 0.00180 & 2.37 \\
-- & 20:51:59.518 & -52:47:09.10 & 0.045 & 0.076 & 0.003 & 0.0214 & 0.0012 & 0.114 & 0.131 & 0.00127 & 0.00053 & 1.78 \\
G346.59+35.04 & 15:15:02.301 & -15:22:53.25 & 0.223 & 0.118 & 0.004 & 0.0153 & 0.0008 & 0.103 & 0.135 & 0.00249 & 0.00093 & 1.48 \\
G347.18-27.35 & 19:34:51.447 & -50:52:16.67 & 0.237 & 0.181 & 0.006 & 0.0246 & 0.0041 & 0.117 & 0.103 & 0.00528 & 0.00214 & 1.56 \\
G349.46-59.94 & 22:48:44.562 & -44:31:48.94 & 0.347 & 0.521 & 0.008 & 0.0594 & 0.0027 & 0.352 & 0.041 & 0.02786 & 0.00403 & 1.21 \\
\end{longtable}

\clearpage

We present in Table \ref{tab:Xraymetrics} the values of the metrics for
all clusters in the X-ray sample, including the secondary subclusters
(on the lines following the primary subcluster, indicated by --). Columns list the
cluster name, {\em Planck} name (the prefix PLCKESZ is omitted for simplicity;
$^\dagger$ the prefix PSZ1 is omitted for simplicity), RA, DEC,
redshift, concentration parameter in the 0.15--1.0 $r_{500}$ range,
concentration parameter in the 40--400 kpc range, cuspiness of the gas density profile,
central gas density, and the maximum radius where the emission integral is computed. Each
metric value is followed by its uncertainty.

\begin{longtable}{l@{}l@{}l@{\hspace{1.0\tabcolsep}}c@{\hspace{1.0\tabcolsep}}l@{\hspace{1.0\tabcolsep}}c@{}c@{}c@{}c@{}c@{\hspace{1.0\tabcolsep}}cc@{}c@{}c@{}c}
\caption[Concentration parameter, Cuspiness, and Central Density for the X-ray sample]{Concentration
  parameter, Cuspiness, and Central Density for the X-ray sample.} \label{tab:Xraymetrics} \\

\hline \hline
\multicolumn{1}{c}{Cluster} &
\multicolumn{1}{c}{{\em Planck} name} & 
\multicolumn{1}{c}{RA} &
\multicolumn{1}{c}{DEC} &
\multicolumn{1}{c}{$z$} & 
\multicolumn{1}{c}{$C_{\rm SB}$} &
\multicolumn{1}{c}{$\sigma_{C_{\rm SB}}$} &
\multicolumn{1}{c}{$C_{\rm SB4}$} &
\multicolumn{1}{c}{$\sigma_{C_{\rm SB4}}$} &
\multicolumn{1}{c}{$\delta$} &
\multicolumn{1}{c}{$\sigma_{\delta}$} &
\multicolumn{1}{c}{$n_{\rm core}$} &
\multicolumn{1}{c}{$\sigma_{n_{\rm core}}$} &
\multicolumn{1}{c}{$r_{\rm max}$} \\ 
\multicolumn{1}{c}{} & 
\multicolumn{1}{c}{} & 
\multicolumn{1}{c}{} &
\multicolumn{1}{c}{} &
\multicolumn{1}{c}{} & 
\multicolumn{1}{c}{} &
\multicolumn{1}{c}{} &
\multicolumn{1}{c}{} &
\multicolumn{1}{c}{} &
\multicolumn{1}{c}{} &
\multicolumn{1}{c}{} &
\multicolumn{1}{c}{($\rm cm^{-3}$)} &
\multicolumn{1}{c}{($\rm cm^{-3}$)} &
\multicolumn{1}{c}{($r_{500}$)} \\  
\hline 

\endfirsthead

\multicolumn{14}{c}%
{\tablename\ \thetable{} -- continued from previous page} \\
\hline 
\multicolumn{1}{c}{Cluster} &
\multicolumn{1}{c}{{\em Planck} name} & 
\multicolumn{1}{c}{RA} &
\multicolumn{1}{c}{DEC} &
\multicolumn{1}{c}{$z$} & 
\multicolumn{1}{c}{$C_{\rm SB}$} &
\multicolumn{1}{c}{$\sigma_{C_{\rm SB}}$} &
\multicolumn{1}{c}{$C_{\rm SB4}$} &
\multicolumn{1}{c}{$\sigma_{C_{\rm SB4}}$} &
\multicolumn{1}{c}{$\delta$} &
\multicolumn{1}{c}{$\sigma_{\delta}$} &
\multicolumn{1}{c}{$n_{\rm core}$} &
\multicolumn{1}{c}{$\sigma_{n_{\rm core}}$} &
\multicolumn{1}{c}{$r_{\rm max}$} \\
\multicolumn{1}{c}{} & 
\multicolumn{1}{c}{} & 
\multicolumn{1}{c}{} &
\multicolumn{1}{c}{} &
\multicolumn{1}{c}{} & 
\multicolumn{1}{c}{} &
\multicolumn{1}{c}{} &
\multicolumn{1}{c}{} &
\multicolumn{1}{c}{} &
\multicolumn{1}{c}{} &
\multicolumn{1}{c}{} &
\multicolumn{1}{c}{($\rm cm^{-3}$)} &
\multicolumn{1}{c}{($\rm cm^{-3}$)} &
\multicolumn{1}{c}{($r_{500}$)} \\  
\hline 

\endhead

\hline 
\multicolumn{14}{c}{{Continued on next page}} \\ 
\hline
\endfoot

\hline
\endlastfoot
2A0335 & G176.28-35.05 & 03:38:40.698 & +09:58:03.07 & 0.035 & 0.718 & 0.004 & 0.3876 & 0.0047 & 1.242 & 0.017 & 0.05916 & 0.00104 & 1.09 \\
A85 & G115.16-72.09 & 00:41:50.390 & -09:18:09.53 & 0.056 & 0.455 & 0.004 & 0.1535 & 0.0030 & 1.012 & 0.040 & 0.04534 & 0.00113 & 1.40 \\
A119 & G125.58-64.14 & 00:56:16.210 & -01:14:58.98 & 0.044 & 0.164 & 0.004 & 0.0229 & 0.0007 & 0.221 & 0.052 & 0.00258 & 0.00027 & 1.58 \\
A133 & G149.55-84.16$^\dagger$ & 01:02:41.707 & -21:52:52.58 & 0.057 & 0.503 & 0.005 & 0.2421 & 0.0045 & 1.335 & 0.077 & 0.04650 & 0.00115 & 1.98 \\
A193 & & 01:25:07.559 & +08:41:59.95 & 0.049 & 0.295 & 0.006 & 0.0596 & 0.0016 & 0.428 & 0.036 & 0.00825 & 0.00085 & 1.44 \\
A376 & & 02:46:03.910 & +36:54:18.44 & 0.049 & 0.240 & 0.008 & 0.0870 & 0.0031 & 0.884 & 0.026 & 0.01538 & 0.00068 & 1.94 \\
A399 & & 02:57:53.422 & +13:01:57.47 & 0.071 & 0.251 & 0.004 & 0.0357 & 0.0012 & 0.339 & 0.049 & 0.00643 & 0.00086 & 1.52 \\
A401 & G164.18-38.89 & 02:58:57.595 & +13:34:44.19 & 0.074 & 0.308 & 0.005 & 0.0400 & 0.0012 & 0.262 & 0.040 & 0.00776 & 0.00081 & 1.34 \\
A478 & G182.44-28.29 & 04:13:25.196 & +10:27:53.66 & 0.088 & 0.622 & 0.005 & 0.1726 & 0.0033 & 0.908 & 0.056 & 0.05973 & 0.00179 & 1.38 \\
A496 & G209.56-36.49 & 04:33:37.913 & -13:15:42.03 & 0.033 & 0.438 & 0.005 & 0.1892 & 0.0039 & 1.200 & 0.086 & 0.04229 & 0.00359 & 0.79 \\
A548e & & 05:48:38.311 & -25:28:40.20 & 0.041 & 0.222 & 0.008 & 0.0995 & 0.0047 & 0.733 & 0.043 & 0.01378 & 0.00129 & 1.36 \\
A576 & G161.44+26.23 & 07:21:30.328 & +55:45:41.56 & 0.038 & 0.313 & 0.008 & 0.0671 & 0.0039 & 0.390 & 0.070 & 0.00755 & 0.00111 & 1.10 \\
A754 & G239.28+24.76 & 09:09:20.461 & -09:41:03.53 & 0.054 & 0.333 & 0.003 & 0.0477 & 0.0011 & 0.276 & 0.018 & 0.00561 & 0.00022 & 1.35 \\
-- & & 09:08:47.478 & -09:38:28.45 & 0.054 & 0.323 & 0.183 & 0.1382 & 0.2078 & 0.713 & 0.843 & 0.02846 & 0.05574 & 1.42 \\
A970 & & 10:17:23.807 & -10:41:07.78 & 0.059 & 0.325 & 0.007 & 0.0733 & 0.0034 & 0.543 & 0.087 & 0.00779 & 0.00068 & 1.73 \\
A1413 & G226.24+76.76 & 11:55:17.943 & +23:24:20.25 & 0.143 & 0.488 & 0.005 & 0.0985 & 0.0026 & 0.666 & 0.033 & 0.02751 & 0.00107 & 1.39 \\
A1644 & G304.89+45.45 & 12:57:11.570 & -17:24:33.57 & 0.047 & 0.195 & 0.003 & 0.0846 & 0.0019 & 0.905 & 0.024 & 0.01788 & 0.00048 & 1.14 \\
A1650 & G306.68+61.06 & 12:58:41.433 & -01:45:43.44 & 0.085 & 0.456 & 0.006 & 0.1186 & 0.0027 & 0.777 & 0.029 & 0.02416 & 0.00118 & 1.60 \\
A1651 & G306.80+58.60 & 12:59:22.339 & -04:11:45.31 & 0.085 & 0.438 & 0.007 & 0.0773 & 0.0022 & 0.535 & 0.049 & 0.01708 & 0.00144 & 1.50 \\
A1689 & G313.36+61.11 & 13:11:29.460 & -01:20:27.68 & 0.183 & 0.590 & 0.006 & 0.1098 & 0.0027 & 0.717 & 0.024 & 0.04374 & 0.00190 & 1.32 \\
A1736 & G312.64+35.09$^\dagger$ & 13:27:00.923 & -27:11:47.19 & 0.046 & 0.101 & 0.003 & 0.0206 & 0.0007 & 0.134 & 0.082 & 0.00211 & 0.00063 & 2.00 \\
A1767 & G112.45+57.03 & 13:36:05.971 & +59:12:08.41 & 0.070 & 0.252 & 0.008 & 0.0444 & 0.0068 & 0.305 & 0.063 & 0.00546 & 0.00073 & 1.80 \\
A1775 & & 13:41:48.863 & +26:22:19.89 & 0.076 & 0.279 & 0.005 & 0.0823 & 0.0021 & 0.880 & 0.046 & 0.01393 & 0.00065 & 1.93 \\
A1795 & G033.78+77.16 & 13:48:52.710 & +26:35:31.20 & 0.062 & 0.592 & 0.004 & 0.1774 & 0.0036 & 1.090 & 0.046 & 0.04278 & 0.00092 & 1.53 \\
A1831 & & 13:59:15.391 & +27:58:34.43 & 0.061 & 0.385 & 0.008 & 0.1312 & 0.0048 & 0.789 & 0.082 & 0.01298 & 0.00104 & 1.68 \\
A1914 & G067.23+67.46 & 14:26:00.269 & +37:49:40.52 & 0.171 & 0.492 & 0.008 & 0.0544 & 0.0033 & 0.260 & 0.038 & 0.01615 & 0.00367 & 1.39 \\
-- & & 14:26:03.448 & +37:49:28.87 & 0.171 & 0.442 & 0.007 & 0.0972 & 0.0036 & 0.784 & 0.030 & 0.02393 & 0.00121 & 1.41 \\
A2029 & G006.47+50.54 & 15:10:56.117 & +05:44:40.38 & 0.077 & 0.591 & 0.005 & 0.1674 & 0.0035 & 0.920 & 0.015 & 0.05946 & 0.00131 & 1.43 \\
A2034 & G053.52+59.54 & 15:10:12.700 & +33:30:34.01 & 0.113 & 0.293 & 0.004 & 0.0326 & 0.0009 & 0.152 & 0.033 & 0.00522 & 0.00057 & 1.44 \\
-- & & 15:10:13.060 & +33:32:26.21 & 0.113 & 0.259 & 0.004 & 0.0298 & 0.0007 & 0.072 & 0.025 & 0.00312 & 0.00025 & 1.58 \\
A2052 & & 15:16:44.489 & +07:01:17.84 & 0.035 & 0.532 & 0.005 & 0.2585 & 0.0043 & 0.970 & 0.037 & 0.03395 & 0.00100 & 1.05 \\
A2061 & G048.05+57.17 & 15:21:12.694 & +30:38:00.59 & 0.078 & 0.133 & 0.004 & 0.0178 & 0.0007 & 0.037 & 0.034 & 0.00163 & 0.00020 & 1.77 \\
A2063 & & 15:23:05.130 & +08:36:34.45 & 0.035 & 0.350 & 0.006 & 0.1133 & 0.0021 & 0.821 & 0.020 & 0.01493 & 0.00046 & 0.95 \\
A2065 & G042.82+56.61 & 15:22:29.473 & +27:42:18.76 & 0.072 & 0.341 & 0.005 & 0.0773 & 0.0019 & 0.556 & 0.059 & 0.02364 & 0.00097 & 1.47 \\
A2107 & & 15:39:39.091 & +21:46:58.25 & 0.041 & 0.366 & 0.006 & 0.1171 & 0.0025 & 0.640 & 0.035 & 0.02449 & 0.00091 & 1.19 \\
A2142 & G044.22+48.68 & 15:58:21.100 & +27:13:47.87 & 0.089 & 0.493 & 0.005 & 0.0810 & 0.0017 & 0.758 & 0.053 & 0.02480 & 0.00073 & 1.25 \\
A2147 & G029.00+44.56 & 16:02:14.068 & +15:58:16.23 & 0.035 & 0.183 & 0.005 & 0.0380 & 0.0013 & 0.315 & 0.076 & 0.00396 & 0.00050 & 1.01 \\
A2151 & & 16:04:35.792 & +17:43:17.15 & 0.037 & 0.412 & 0.016 & 0.2047 & 0.0084 & 0.908 & 0.047 & 0.02810 & 0.00214 & 1.02 \\
A2163 & G006.78+30.46 & 16:15:46.073 & -06:08:54.61 & 0.203 & 0.304 & 0.009 & 0.0250 & 0.0008 & 0.159 & 0.030 & 0.00819 & 0.00096 & 1.08 \\
A2199 & G062.92+43.70 & 16:28:38.232 & +39:33:03.36 & 0.030 & 0.536 & 0.006 & 0.1906 & 0.0031 & 0.801 & 0.031 & 0.03963 & 0.00122 & 0.76 \\
A2204 & G021.09+33.25 & 16:32:46.854 & +05:34:31.61 & 0.151 & 0.638 & 0.005 & 0.3344 & 0.0069 & 1.395 & 0.061 & 0.13411 & 0.00361 & 1.42 \\
A2244 & G056.81+36.31 & 17:02:42.571 & +34:03:38.15 & 0.095 & 0.485 & 0.006 & 0.1049 & 0.0023 & 0.616 & 0.028 & 0.02366 & 0.00123 & 1.61 \\
A2249 & G057.61+34.94 & 17:09:45.792 & +34:27:20.36 & 0.080 & 0.163 & 0.005 & 0.0253 & 0.0021 & 0.371 & 0.270 & 0.00449 & 0.00168 & 1.82 \\
A2255 & G093.91+34.90 & 17:12:43.585 & +64:03:46.90 & 0.081 & 0.162 & 0.004 & 0.0197 & 0.0006 & 0.077 & 0.038 & 0.00233 & 0.00020 & 1.20 \\
A2256 & G110.98+31.73 & 17:03:14.917 & +78:39:23.17 & 0.058 & 0.220 & 0.004 & 0.0217 & 0.0009 & 0.126 & 0.046 & 0.00367 & 0.00038 & 1.36 \\
A2415 & & 22:05:38.437 & -05:35:31.57 & 0.058 & 0.402 & 0.008 & 0.2494 & 0.0069 & 1.373 & 0.104 & 0.04912 & 0.00325 & 2.13 \\
A2420 & G046.50-49.43 & 22:10:19.489 & -12:10:10.03 & 0.085 & 0.277 & 0.006 & 0.0348 & 0.0020 & 0.254 & 0.038 & 0.00588 & 0.00084 & 1.60 \\
A2426 & G049.66-49.50 & 22:14:32.554 & -10:22:17.84 & 0.098 & 0.437 & 0.009 & 0.0977 & 0.0037 & 0.511 & 0.069 & 0.01625 & 0.00201 & 1.81 \\
A2457 & & 22:35:41.138 & +01:29:11.41 & 0.059 & 0.219 & 0.006 & 0.0763 & 0.0039 & 0.756 & 0.101 & 0.01903 & 0.00160 & 1.80 \\
A2572 & & 23:17:12.803 & +18:42:10.29 & 0.042 & 0.311 & 0.008 & 0.1040 & 0.0033 & 0.571 & 0.053 & 0.00867 & 0.00080 & 1.31 \\
A2589 & & 23:23:57.326 & +16:46:39.86 & 0.042 & 0.388 & 0.005 & 0.1086 & 0.0021 & 0.586 & 0.036 & 0.01691 & 0.00088 & 1.21 \\
A2593 & & 23:24:20.855 & +14:38:41.63 & 0.043 & 0.206 & 0.009 & 0.0487 & 0.0033 & 0.399 & 0.082 & 0.00509 & 0.00085 & 1.62 \\
A2597 & & 23:25:19.765 & -12:07:25.96 & 0.085 & 0.687 & 0.005 & 0.3163 & 0.0044 & 1.118 & 0.019 & 0.06686 & 0.00194 & 1.95 \\
A2626 & & 23:36:30.314 & +21:08:48.22 & 0.057 & 0.477 & 0.005 & 0.1830 & 0.0033 & 0.874 & 0.017 & 0.02861 & 0.00070 & 1.68 \\
A2634 & & 23:38:29.381 & +27:01:53.58 & 0.031 & 0.165 & 0.005 & 0.0640 & 0.0028 & 0.733 & 0.038 & 0.00926 & 0.00020 & 0.95 \\
A2657 & & 23:44:57.622 & +09:11:26.36 & 0.040 & 0.340 & 0.006 & 0.0929 & 0.0024 & 0.640 & 0.021 & 0.01253 & 0.00053 & 1.22 \\
A2665 & & 23:50:50.649 & +06:08:59.67 & 0.056 & 0.428 & 0.008 & 0.1928 & 0.0066 & 0.935 & 0.042 & 0.03611 & 0.00258 & 1.70 \\
A2734 & & 00:11:21.686 & -28:51:14.53 & 0.062 & 0.290 & 0.006 & 0.0826 & 0.0032 & 0.544 & 0.047 & 0.01058 & 0.00087 & 1.57 \\
A3112 & G252.96-56.05 & 03:17:57.637 & -44:14:17.40 & 0.075 & 0.581 & 0.004 & 0.2692 & 0.0048 & 1.140 & 0.035 & 0.06626 & 0.00186 & 1.67 \\
A3158 & G265.00-48.94 & 03:42:53.069 & -53:37:53.35 & 0.059 & 0.317 & 0.005 & 0.0445 & 0.0021 & 0.217 & 0.028 & 0.00601 & 0.00048 & 1.25 \\
A3266 & G272.10-40.15 & 04:31:30.202 & -61:24:48.44 & 0.059 & 0.205 & 0.088 & 0.0204 & 0.0028 & 0.162 & 0.109 & 0.00333 & 0.00069 & 0.98 \\
-- & & 04:31:12.696 & -61:27:20.58 & 0.059 & 0.199 & 0.004 & 0.0419 & 0.0013 & 0.531 & 0.025 & 0.00796 & 0.00038 & 0.90 \\
A3376 & G246.52-26.05 & 06:01:41.584 & -39:59:19.90 & 0.047 & 0.133 & 0.003 & 0.0179 & 0.0006 & 0.052 & 0.050 & 0.00123 & 0.00019 & 1.54 \\
-- & & 06:02:11.776 & -39:57:26.35 & 0.047 & 0.184 & 0.005 & 0.0431 & 0.0015 & 0.287 & 0.050 & 0.00278 & 0.00034 & 1.31 \\
A3391 & & 06:26:20.481 & -53:41:35.84 & 0.051 & 0.203 & 0.004 & 0.0514 & 0.0021 & 0.279 & 0.031 & 0.00414 & 0.00030 & 1.74 \\
A3395 & G263.20-25.21 & 06:26:48.188 & -54:32:57.24 & 0.051 & 0.117 & 0.003 & 0.0442 & 0.0015 & 0.598 & 0.062 & 0.00571 & 0.00051 & 1.73 \\
-- & & 06:27:36.008 & -54:26:46.65 & 0.051 & 0.116 & 0.002 & 0.0279 & 0.0007 & 0.171 & 0.025 & 0.00216 & 0.00017 & 1.72 \\
A3528n & & 12:54:22.206 & -29:00:46.76 & 0.054 & 0.332 & 0.005 & 0.1665 & 0.0055 & 0.858 & 0.053 & 0.03590 & 0.00137 & 1.90 \\
A3528s & G303.75+33.65 & 12:54:40.703 & -29:13:40.55 & 0.054 & 0.330 & 0.010 & 0.1854 & 0.0057 & 1.435 & 0.141 & 0.03161 & 0.00295 & 1.94 \\
-- & & 12:54:22.138 & -29:00:47.20 & 0.054 & 0.331 & 0.007 & 0.1652 & 0.0060 & 0.939 & 0.094 & 0.03347 & 0.00192 & 1.89 \\
A3532 & G304.49+32.44 & 12:57:22.033 & -30:21:49.34 & 0.055 & 0.205 & 0.006 & 0.0481 & 0.0034 & 0.409 & 0.037 & 0.00591 & 0.00062 & 1.37 \\
A3558 & G311.99+30.71 & 13:27:56.824 & -31:29:43.84 & 0.048 & 0.302 & 0.005 & 0.0611 & 0.0013 & 0.506 & 0.038 & 0.01224 & 0.00057 & 1.66 \\
-- & & 13:29:47.769 & -31:36:26.05 & 0.048 & 0.149 & 0.004 & 0.0606 & 0.0048 & 0.701 & 0.056 & 0.00968 & 0.00094 & 2.08 \\
A3560 & & 13:32:27.757 & -33:08:33.59 & 0.050 & 0.208 & 0.005 & 0.0457 & 0.0013 & 0.243 & 0.028 & 0.00348 & 0.00030 & 1.50 \\
A3562 & & 13:33:34.727 & -31:40:22.49 & 0.049 & 0.241 & 0.005 & 0.0752 & 0.0022 & 0.565 & 0.030 & 0.00895 & 0.00069 & 1.75 \\
-- & & 13:31:27.476 & -31:49:18.14 & 0.049 & 0.108 & 0.003 & 0.0400 & 0.0018 & 0.421 & 0.069 & 0.00483 & 0.00110 & 2.07 \\
A3571 & G316.34+28.54 & 13:47:28.110 & -32:51:57.98 & 0.039 & 0.394 & 0.006 & 0.0636 & 0.0014 & 0.576 & 0.070 & 0.01769 & 0.00069 & 0.80 \\
A3667 & G340.88-33.34 & 20:12:37.869 & -56:50:45.53 & 0.056 & 0.185 & 0.004 & 0.0291 & 0.0006 & 0.112 & 0.013 & 0.00402 & 0.00011 & 1.01 \\
A3695 & G006.70-35.54 & 20:34:46.912 & -35:49:24.54 & 0.089 & 0.188 & 0.006 & 0.0334 & 0.0016 & 0.251 & 0.048 & 0.00469 & 0.00096 & 1.66 \\
A3822 & G335.59-46.46 & 21:54:05.793 & -57:51:41.87 & 0.076 & 0.230 & 0.006 & 0.0451 & 0.0039 & 0.430 & 0.056 & 0.00787 & 0.00093 & 1.72 \\
A3827 & G332.23-46.36 & 22:01:53.194 & -59:56:43.51 & 0.098 & 0.388 & 0.005 & 0.0512 & 0.0013 & 0.298 & 0.052 & 0.01155 & 0.00114 & 1.53 \\
A3921 & G321.96-47.97 & 22:49:58.191 & -64:25:46.79 & 0.094 & 0.313 & 0.006 & 0.0723 & 0.0054 & 0.303 & 0.061 & 0.00702 & 0.00084 & 1.61 \\
A4038 & & 23:47:42.886 & -28:08:34.44 & 0.029 & 0.463 & 0.006 & 0.1821 & 0.0028 & 0.751 & 0.018 & 0.02007 & 0.00059 & 0.88 \\
A4059 & & 23:57:01.016 & -34:45:32.37 & 0.046 & 0.488 & 0.006 & 0.1474 & 0.0027 & 0.693 & 0.028 & 0.02525 & 0.00115 & 1.95 \\
AWM4 & & 16:04:56.644 & +23:55:57.63 & 0.033 & 0.436 & 0.006 & 0.1811 & 0.0032 & 0.677 & 0.023 & 0.01385 & 0.00049 & 1.23 \\
EXO0422 & & 04:25:51.164 & -08:33:34.34 & 0.039 & 0.568 & 0.007 & 0.3048 & 0.0054 & 1.186 & 0.067 & 0.04669 & 0.00149 & 1.38 \\
Hydra-A & & 09:18:05.641 & -12:05:43.98 & 0.052 & 0.616 & 0.006 & 0.2454 & 0.0056 & 1.026 & 0.030 & 0.05451 & 0.00115 & 1.37 \\
IC1262 & & 17:33:02.990 & +43:45:38.94 & 0.031 & 0.497 & 0.005 & 0.2548 & 0.0035 & 0.871 & 0.014 & 0.02359 & 0.00053 & 1.28 \\
IC1365 & & 21:13:55.893 & +02:33:50.66 & 0.049 & 0.243 & 0.007 & 0.0512 & 0.0021 & 0.401 & 0.155 & 0.00769 & 0.00134 & 1.33 \\
MKW3s & & 15:21:51.824 & +07:42:31.63 & 0.045 & 0.570 & 0.017 & 0.2093 & 0.0324 & 0.920 & 0.026 & 0.02659 & 0.00072 & 0.80 \\
MKW8 & & 14:40:39.469 & +03:28:13.13 & 0.027 & 0.191 & 0.003 & 0.0620 & 0.0020 & 0.452 & 0.052 & 0.00496 & 0.00042 & 2.37 \\
-- & & 14:38:21.874 & +03:40:11.93 & 0.027 & 0.512 & 0.010 & 0.0287 & 0.0011 & 0.084 & 0.002 & 0.12684 & 0.00011 & 0.43 \\
NGC6338 & & 17:15:22.990 & +57:24:40.27 & 0.028 & 0.415 & 0.013 & 0.2830 & 0.0053 & 1.469 & 0.053 & 0.04356 & 0.00112 & 0.67 \\
-- & & 17:15:23.167 & +57:26:04.65 & 0.028 & 0.290 & 0.009 & 0.1316 & 0.0033 & 0.874 & 0.070 & 0.01950 & 0.00083 & 0.69 \\
RXJ0341.3+1524 & & 03:41:16.625 & +15:24:01.81 & 0.029 & 0.347 & 0.007 & 0.1100 & 0.0024 & 0.401 & 0.026 & 0.00630 & 0.00034 & 2.73 \\
RXJ1252.5-3116 & & 12:52:34.721 & -31:15:59.24 & 0.053 & 0.714 & 0.007 & 0.4436 & 0.0079 & 1.047 & 0.047 & 0.06392 & 0.00279 & 1.90 \\
RXJ1504.1-0248 & & 15:04:07.427 & -02:48:16.47 & 0.215 & 0.778 & 0.005 & 0.3443 & 0.0067 & 1.567 & 0.054 & 0.14018 & 0.00441 & 1.39 \\
RXJ1524.2-3154 & & 15:24:12.912 & -31:54:22.66 & 0.103 & 0.597 & 0.006 & 0.4128 & 0.0069 & 1.447 & 0.014 & 0.09196 & 0.00335 & 1.95 \\
RXJ1539.5-8335 & & 15:39:34.529 & -83:35:23.27 & 0.073 & 0.715 & 0.007 & 0.3070 & 0.0069 & 0.781 & 0.043 & 0.04268 & 0.00246 & 2.15 \\
RXJ1558.3-1410 & & 15:58:21.651 & -14:09:59.28 & 0.097 & 0.548 & 0.006 & 0.1807 & 0.0036 & 0.805 & 0.025 & 0.03545 & 0.00112 & 1.84 \\
RXJ1720.1+2638~ & G049.20+30.86 & 17:20:09.957 & +26:37:30.79 & 0.164 & 0.620 & 0.007 & 0.2313 & 0.0055 & 1.238 & 0.054 & 0.06126 & 0.00245 & 1.48 \\
RXJ1958.2-3011 & & 19:58:14.918 & -30:11:11.53 & 0.117 & 0.694 & 0.019 & 0.7253 & 0.0185 & 2.203 & 0.112 & 0.19818 & 0.00995 & 3.45 \\
RXJ2014.8-2430 & & 20:14:51.619 & -24:30:22.88 & 0.161 & 0.647 & 0.007 & 0.3680 & 0.0076 & 1.523 & 0.056 & 0.12483 & 0.00452 & 1.53 \\
RXJ2344.3-042 & & 23:44:18.277 & -04:22:54.04 & 0.079 & 0.326 & 0.006 & 0.0524 & 0.0023 & 0.270 & 0.070 & 0.00740 & 0.00163 & 1.87 \\
S0405 & G296.41-32.48 & 03:51:31.364 & -82:13:27.86 & 0.061 & 0.238 & 0.007 & 0.0499 & 0.0044 & 0.334 & 0.069 & 0.00526 & 0.00082 & 1.81 \\
S0540 & & 05:40:06.677 & -40:50:11.66 & 0.036 & 0.398 & 0.006 & 0.1874 & 0.0040 & 0.767 & 0.039 & 0.02517 & 0.00137 & 2.66 \\
-- & & 05:42:49.463 & -40:59:57.11 & 0.036 & 0.517 & 0.040 & 0.0287 & 0.0000 & 0.085 & 0.008 & 0.12681 & 0.00043 & 1.18 \\
S1101 & & 23:13:58.595 & -42:43:31.06 & 0.058 & 0.681 & 0.005 & 0.3260 & 0.0053 & 0.985 & 0.045 & 0.05015 & 0.00175 & 2.40 \\
UGC03957 & & 07:40:58.208 & +55:25:38.21 & 0.034 & 0.554 & 0.009 & 0.3042 & 0.0066 & 1.039 & 0.054 & 0.04315 & 0.00213 & 1.35 \\
USGCS152 & & 10:50:26.109 & -12:50:42.05 & 0.015 & 0.791 & 0.070 & 0.7165 & 0.0936 & 1.339 & 0.049 & 0.05962 & 0.00160 & 1.13 \\
ZwCl1215 & G282.49+65.17 & 12:17:41.196 & +03:39:22.03 & 0.077 & 0.286 & 0.006 & 0.0331 & 0.0021 & 0.146 & 0.047 & 0.00537 & 0.00141 & 1.51 \\
ZwCl1742 & G057.92+27.64 & 17:44:15.426 & +32:59:31.71 & 0.076 & 0.555 & 0.005 & 0.2121 & 0.0046 & 1.140 & 0.058 & 0.03579 & 0.00142 & 1.91 \\
\end{longtable}


\newpage

\begin{thebibliography}{}

\bibitem[Allen \& Fabian(1998)]{1998Allen} Allen, S.~W., \& Fabian, A.~C.\ 1998, \mnras, 297, L57

\bibitem[Allen et al.(2011)]{2011Allen} Allen, S.~W., Evrard, A.~E., \& Mantz, A.~B.\ 2011, \araa, 49, 409

\bibitem[Anders \& Grevesse(1989)]{1989AndersGrevesse} Anders, E., \& Grevesse, N.\ 1989, \gca, 53, 197 

\bibitem[Andrade-Santos et al.(2012)]{2012Andrade-Santos} Andrade-Santos, F., Lima Neto, G.~B., \& Lagan{\'a}, T.~F.\ 2012, \apj, 746, 139 

\bibitem[Andrade-Santos et al.(2013)]{2013Andrade-Santos} Andrade-Santos, F., Nulsen, P.~E.~J., Kraft, R.~P., et al.\ 2013, \apj, 766, 107

\bibitem[Andrade-Santos et al.(2015)]{2015Andrade-Santos} Andrade-Santos, F., Jones, C., Forman, W.~R., et al.\ 2015, \apj, 803, 108 

\bibitem[Andrade-Santos et al.(2016)]{2016Andrade-Santos} Andrade-Santos, F., Bogd{\'a}n, {\'A}., Romani, R.~W., et al.\ 2016, \apj, 826, 91

\bibitem[Andrade-Santos et al.(2017)]{2017Andrade-Santos} Andrade-Santos, F., Jones, C., Forman, W.~R., Lovisari, L., \& Chandra-Planck Collaboration 2017, American Astronomical Society Meeting Abstracts, 229, 404.04

\bibitem[Andrade-Santos et al.(2017)]{2017Andrade-Santos}
  Andrade-Santos, F., Jones, C., Forman, W.~R., et al.\ 2017, \apj, in preparation

\bibitem[Angulo et al.(2012)]{2012Angulo} Angulo, R.~E., Springel, V., White, S.~D.~M., et al.\ 2012, \mnras, 426, 2046 

%\bibitem[Arnaud et al.(2010)]{2010Arnaud} Arnaud, M., Pratt, G.~W., Piffaretti, R., et al.\ 2010, \aap, 517, A92 

%\bibitem[Ascasibar \& Markevitch(2006)]{2006Ascasibar} Ascasibar, Y., \& Markevitch, M.\ 2006, \apj, 650, 102

%\bibitem[Beers et al.(1982)]{1982Beers} Beers, T.~C., Geller, %M.~J., \& Huchra, J.~P.\ 1982, \apj, 257, 23 

%\bibitem[Beers et al.(1990)]{1990Beers} Beers, T.~C., Flynn, K., \& Gebhardt, K.\ 1990, \aj, 100, 32 

%\bibitem[Beers et al.(1992)]{1992Beers} Beers, T.~C., Gebhardt, K., Huchra, J.~P., et al.\ 1992, \apj, 400, 410

%\bibitem[Begelman et al.(1980)]{1980Begelman} Begelman, M.~C., Blandford, R.~D., \& Rees, M.~J.\ 1980, \nat, 287, 307 

%\bibitem[Bell et al.(2003)]{2003Bell} Bell, E.~F., McIntosh, D.~H., Katz, N., \& Weinberg, M.~D.\ 2003, \apjs, 149, 289

%\bibitem[Bell et al.(2006)]{2006Bell} Bell, E.~F., Phleps, S., Somerville, R.~S., et al.\ 2006, \apj, 652, 270 

\bibitem[Benson et al.(2013)]{2013Benson} Benson, B.~A., de Haan, T., Dudley, J.~P., et al.\ 2013, \apj, 763, 147 

%\bibitem[B{\^i}rzan et al.(2008)]{2008Birzan} B{\^i}rzan, L., McNamara, B.~R., Nulsen, P.~E.~J., Carilli, C.~L., \& Wise, M.~W.\ 2008, \apj, 686, 859

%\bibitem[Bock et al.(1999)]{1999Bock} Bock, D.~C.-J., Large, M.~I., \& Sadler, E.~M.\ 1999, \aj, 117, 1578

\bibitem[B{\"o}hringer et al.(2007)]{2007Bohringer} B{\"o}hringer, H., Schuecker, P., Pratt, G.~W., et al.\ 2007, \aap, 469, 363

\bibitem[B{\"o}hringer et al.(2010)]{2010Bohringer} B{\"o}hringer, H., Pratt, G.~W., Arnaud, M., et al.\ 2010, \aap, 514, A32 

\bibitem[Buote \& Tsai(1996)]{1996Buote} Buote, D.~A., \& Tsai, J.~C.\ 1996, \apj, 458, 27 

%\bibitem[Burke-Spolaor(2011)]{2011Burke} Burke-Spolaor, S.\ 2011, \mnras, 410, 2113 

\bibitem[Cavagnolo et al.(2009)]{2009Cavagnolo} Cavagnolo, K.~W., Donahue, M., Voit, G.~M., \& Sun, M.\ 2009, \apjs, 182, 12 

%\bibitem[Cavagnolo et al.(2010)]{2010Cavagnolo} Cavagnolo, K.~W., McNamara, B.~R., Nulsen, P.~E.~J., et al.\ 2010, \apj, 720, 1066 

\bibitem[{{Cavaliere} \& {Fusco-Femiano}(1976)}]{1976Cavaliere} {Cavaliere}, A., \& {Fusco-Femiano}, R. 1976, \aap, 49, 137 

%\bibitem[Churazov et al.(2003)]{2003Churazov} Churazov, E., Forman, W., Jones, C., B{\"o}hringer, H.\ 2003, \apj, 590, 225 

\bibitem[David et al.(2001)]{2001David} David, L.~P., Nulsen, P.~E.~J., McNamara, B.~R., et al.\ 2001, \apj, 557, 546 

%\bibitem[Dawson(2013)]{2013Dawson} Dawson, W.~A.\ 2013, \apj, 772, 131

%\bibitem[Dawson(2014)]{2014Dawson} Dawson, W.~A.\ 2014, Astrophysics Source Code Library, 7004

%\bibitem[{{Demarco} {et~al.}(2003){Demarco}, {Magnard}, {Durret}, \& %{M{\'a}rquez}}]{2003Demarco} {Demarco}, R., {Magnard}, F., {Durret}, F., \& {M{\'a}rquez}, I. 2003, \aap, 407, 437

%\bibitem[Douglass et al.(2011)]{2011Douglas} Douglass, E.~M., Blanton, E.~L., Clarke, T.~E., Randall, S.~W., \& Wing, J.~D.\ 2011, \apj, 743, 199 

\bibitem[Ebeling et al.(2001)]{2001Ebeling} Ebeling, H., Edge, A.~C., \& Henry, J.~P.\ 2001, \apj, 553, 668

\bibitem[Eckert et al.(2011)]{2011Eckert} Eckert, D., Molendi, S., \& Paltani, S.\ 2011, \aap, 526, A79 

%\bibitem[Eracleous et al.(2012)]{2012Eracleous} Eracleous, M., Boroson, T.~A., Halpern, J.~P., \& Liu, J.\ 2012, \apjs, 201, 23

\bibitem[Fabian \& Nulsen(1977)]{1977Fabian} Fabian, A.~C., \& Nulsen, P.~E.~J.\ 1977, \mnras, 180, 479 

\bibitem[Fabian et al.(1984)]{1984Fabian} Fabian, A.~C., Nulsen, P.~E.~J., \& Canizares, C.~R.\ 1984, \nat, 310, 733 

\bibitem[Fabian(1994)]{1994Fabian} Fabian, A.~C.\ 1994,   \araa, 32, 277 

\bibitem[Fabian(2012)]{2012Fabian} Fabian, A.~C.\ 2012, \araa, 50, 455 

\bibitem[Forman \& Jones(1982)]{1982Forman} Forman, W., \& Jones, C.\ 1982, \araa, 20, 547 

%\bibitem[Green et al.(2015)]{2015Green} Green, G.~M., Schlafly, E.~F., Finkbeiner, D.~P., et al.\ 2015, \apj, 810, 25 

%\bibitem[Gilfanov(2004)]{2004Gilfanov} Gilfanov, M.\ 2004, \mnras, 349, 146

%\bibitem[Graham et al.(2015)]{2015Graham} Graham, M.~J., Djorgovski, S.~G., Stern, D., et al.\ 2015, \nat, 518, 74 

%\bibitem[Grange et al.(2010)]{2010Grange} Grange, Y.~G., Costantini, %E., de Plaa, J., et al.\ 2010, \aap, 513, A2  

\bibitem[Hudson et al.(2010)]{2010Hudson} Hudson, D.~S., Mittal, R., Reiprich, T.~H., et al.\ 2010, \aap, 513, A37 

\bibitem[Jeltema et al.(2005)]{2005Jeltema} Jeltema, T.~E., Canizares, C.~R., Bautz, M.~W., \& Buote, D.~A.\ 2005, \apj, 624, 606 

\bibitem[Jones et al.(1979)]{1979Jones} Jones, C., Mandel, E., Schwarz, J., et al.\ 1979, \apjl, 234, L21

\bibitem[Jones \& Forman(1984)]{1984Jones} Jones, C., \& Forman, W.\ 1984, \apj, 276, 38 

\bibitem[Jones \& Forman(1999)]{1999Jones} Jones, C., \& Forman, W.\ 1999, \apj, 511, 65

\bibitem[Jones et al.(2016)]{2016Jones} Jones, C., Santos, F.~A., Forman, W.~R., et al.\ 2016, American Astronomical Society Meeting Abstracts, 228, 110.04 

%\bibitem[Kahn \& Woltjer(1959)]{1959Kahn} Kahn, F.~D., \& Woltjer, L.\ 1959, \apj, 130, 705 

%\bibitem[Komossa et al.(2003)]{2003Komossa} Komossa, S., Burwitz, V., Hasinger, G., et al.\ 2003, \apjl, 582, L15

\bibitem[Kravtsov \& Borgani(2012)]{2012Kravtsov} Kravtsov, A.~V., \& Borgani, S.\ 2012, \araa, 50, 353 

\bibitem[Lagan{\'a} et al.(2010)]{2010Lagana} Lagan{\'a}, T.~F., Andrade-Santos, F., \& Lima Neto, G.~B.\ 2010, \aap, 511, A15 

\bibitem[Leccardi et al.(2010)]{2010Leccardi} Leccardi, A., Rossetti, M., \& Molendi, S.\ 2010, \aap, 510, A82

%\bibitem[Li \& White(2008)]{2008Li} Li, Y.-S., \& White,
%S.~D.~M.\ 2008, \mnras, 384, 1459 

\bibitem[Lin et al.(2015)]{2015Lin} Lin, H.~W., McDonald, M., Benson, B., \& Miller, E.\ 2015, \apj, 802, 34 

%\bibitem[{{Pislar} {et~al.}(1997){Pislar}, {Durret}, {Gerbal}, {Lima Neto}, \&  {Slezak}}]{1997Pislar} {Pislar}, V., {Durret}, F., {Gerbal}, D., {Lima Neto}, G.~B., \& {Slezak}, E.   1997, \aap, 322, 53

%\bibitem[Maness et al.(2004)]{2004Maness} Maness, H.~L., Taylor, G.~B., Zavala, R.~T., Peck, A.~B., \& Pollack, L.~K.\ 2004, \apj, 602, 123 

\bibitem[Mantz et al.(2010)]{2010Mantz} Mantz, A., Allen, S.~W., Rapetti, D., \& Ebeling, H.\ 2010, \mnras, 406, 1759

%\bibitem[Markevitch \& Vikhlinin (2007)]{2007Markevitch} Markevitch, M. \& Vikhlinin, A.\ 2007, Physics Reports, 443, 1

\bibitem[Maughan et al.(2012)]{2012Maughan} Maughan, B.~J., Giles, P.~A., Randall, S.~W., Jones, C., \& Forman, W.~R.\ 2012, \mnras, 421, 1583 

\bibitem[Mazzotta et al.(2004)]{2004Mazzotta} Mazzotta, P., Rasia, E., Moscardini, L., \& Tormen, G.\ 2004, \mnras, 354, 10 

\bibitem[McDonald et al.(2013)]{2013McDonald} McDonald, M., Benson, B.~A., Vikhlinin, A., et al.\ 2013, \apj, 774, 23

\bibitem[Mohr et al.(1995)]{1995Mohr} Mohr, J.~J., Evrard, A.~E., Fabricant, D.~G., \& Geller, M.~J.\ 1995, \apj, 447, 8 

\bibitem[Molendi \& Pizzolato(2001)]{2001Molendi} Molendi, S., \& Pizzolato, F.\ 2001, \apj, 560, 194 

%\bibitem[Nagai et al.(2007)]{2007Nagai} Nagai, D., Vikhlinin, A., \& Kravtsov, A.~V.\ 2007, \apj, 655, 98

%\bibitem[Navarro et al.(1996)]{1996Navarro} Navarro, J.~F., Frenk, C.~S., \& White, S.~D.~M.\ 1996, \apj, 462, 563

%\bibitem[O'Sullivan et al.(2011)]{2011OSul} O'Sullivan, E., Giacintucci, S., David, L.~P., et al.\ 2011, \apj, 735, 11 

\bibitem[Owen et al.(1985)]{1985Owen} Owen, F.~N., O'Dea, C.~P., Inoue, M., \& Eilek, J.~A.\ 1985, \apjl, 294, L85 

%\bibitem[Ota et al.(2013)]{2013Ota} Ota, N., Fujino, Y., Ibaraki, Y.,  B{\"o}hringer, H., \& Chon, G.\ 2013, \aap, 556, A21 

%\bibitem[Paterno-Mahler et al.(2013)]{2013Paterno-Mahler} Paterno-Mahler, R., Blanton, E.~L., Randall, S.~W., \& Clarke, T.~E.\ 2013, \apj, 773, 114

\bibitem[Peterson et al.(2003)]{2003Peterson} Peterson, J.~R., Kahn, S.~M., Paerels, F.~B.~S., et al.\ 2003, \apj, 590, 207 

\bibitem[Planck Collaboration et al.(2011)]{2011PlanckCol} Planck
  Collaboration, Ade, P.~A.~R., Aghanim, N., et al.\ 2011, \aap, 536,
  A8

\bibitem[Planck Collaboration et al.(2011)]{2011PlanckXMMvalid} Planck
  Collaboration, Aghanim, N., Arnaud, M., et al.\ 2011, \aap, 536, A9

\bibitem[Planck Collaboration et al.(2013)]{2013PlanckXMMvalid}
  Planck Collaboration, Ade, P.~A.~R., Aghanim, N., et al.\ 2013,
  \aap, 550, A130 


\bibitem[Planck Collaboration et al.(2014)]{2014Planck}
  Planck Collaboration, Ade, P.~A.~R., Aghanim, N., et al.\ 2014,
  \aap, 571, A20

\bibitem[Planck Collaboration et al.(2016)]{2016Planck}
  Planck Collaboration, Ade, P.~A.~R., Aghanim, N., et al.\ 2016,
  \aap, 594, A24

\bibitem[Piffaretti et al.(2003)]{2003Piffaretti} Piffaretti, R., Jetzer, P., \& Schindler, S.\ 2003, \aap, 398, 41 

%\bibitem[Planck Collaboration et al.(2013)]{2013PlanckIntermediateVI} Planck Collaboration, Ade, P.~A.~R., Aghanim, N., et al.\ 2013, \aap, 550, A132

%\bibitem[Pratt et al.(2010)]{2010Pratt} Pratt, G.~W., Arnaud, M., Piffaretti, R., et al.\ 2010, \aap, 511, A85 

\bibitem[Rasia et al.(2015)]{2015Rasia} Rasia, E., Borgani, S., Murante, G., et al.\ 2015, \apjl, 813, L17 

\bibitem[Reiprich \& B{\"o}hringer(2002)]{2002Reiprich} Reiprich, T.~H., \& B{\"o}hringer, H.\ 2002, \apj, 567, 716

%\bibitem[Revnivtsev et al.(2007)]{2007Revnivtsev} Revnivtsev, M., Churazov, E., Sazonov, S., Forman, W., \& Jones, C.\ 2007, \aap, 473, 783 

%\bibitem[Revnivtsev et al.(2008)]{2008Revnivtsev} Revnivtsev, M., Churazov, E., Sazonov, S., Forman, W., \& Jones, C.\ 2008, \aap, 490, 37 

%\bibitem[Richstone et al.(1992)]{1992Richstone} Richstone, D., Loeb, A., \& Turner, E.~L.\ 1992, \apj, 393, 477 

%\bibitem[Rodriguez et al.(2006)]{2006Rodriguez} Rodriguez, C., Taylor, G.~B., Zavala, R.~T., et al.\ 2006, \apj, 646, 49
 
%\bibitem[Roediger et al.(2012)]{2012Roediger} Roediger, E., Lovisari, L., Dupke, R., et al.\ 2012, \mnras, 420, 3632

%\bibitem[Romani et al.(2014)]{2014Romani} Romani, R.~W., Forman, %W.~R., Jones, C., et al.\ 2014, \apj, 780, 149

\bibitem[Rossetti et al.(2016)]{2016Rossetti} Rossetti, M., Gastaldello, F., Ferioli, G., et al.\ 2016, \mnras, 457, 4515 

\bibitem[Rossetti et al.(2017)]{2017Rossetti} Rossetti, M., Gastaldello, F., Eckert, D., et al.\ 2017, arXiv:1702.06961

\bibitem[Santos et al.(2008)]{2008Santos} Santos, J.~S., Rosati, P., Tozzi, P., et al.\ 2008, \aap, 483, 35 

%\bibitem[Schlegel et al.(1998)]{1998Schlegel} Schlegel, D.~J., Finkbeiner, D.~P., \& Davis, M.\ 1998, \apj, 500, 525

%\bibitem[Seigar et al.(2003)]{2003Seigar} Seigar, M.~S., Lynam, P.~D., \& Chorney, N.~E.\ 2003, \mnras, 344, 110 

%\bibitem[Smith et al.(2004)]{2004Smith} Smith, R.~J., Hudson, M.~J., Nelan, J.~E., et al.\ 2004, \aj, 128, 1558

\bibitem[Sommer \& Basu(2014)]{2014Sommer} Sommer, M.~W., \& Basu, K.\ 2014, \mnras, 437, 2163 

%\bibitem[Sun et al.(2007)]{2007Sun} Sun, M., Jones, C., Forman, W., et al.\ 2007, \apj, 657, 197 

%\bibitem[Vikhlinin et al.(2001)]{2001Vik} Vikhlinin, A., Markevitch, M., Forman, W., \& Jones, C.\ 2001, \apjl, 555, L87 

\bibitem[Vikhlinin et al.(2005)]{2005Vik} Vikhlinin, A., Markevitch, M., Murray, S.~S., et al.\ 2005, \apj, 628, 655

\bibitem[Vikhlinin et al.(2006)]{2006Vik} Vikhlinin, A., Kravtsov, A., Forman, W., et al.\ 2006, \apj, 640, 691

\bibitem[Vikhlinin et al.(2007)]{2007Vik} Vikhlinin, A., Burenin, R., Forman, W.~R., et al.\ 2007, Heating versus Cooling in Galaxies and Clusters of Galaxies, 48

\bibitem[Vikhlinin et al.(2009a)]{2009Vik} Vikhlinin, A., Burenin, R.~A., Ebeling, H., et al.\ 2009a, \apj, 692, 1033

\bibitem[Vikhlinin et al.(2009b)]{2009bVik} Vikhlinin, A., Kravtsov, A.~V., Burenin, R.~A., et al.\ 2009b, \apj, 692, 1060

\bibitem[Voevodkin \& Vikhlinin(2004)]{2004Voevodkin} Voevodkin, A., \& Vikhlinin, A.\ 2004, \apj, 601, 610

%\bibitem[Voit et al.(2005)]{2005Voit} Voit, G.~M., Kay, S.~T., \& Bryan, G.~L.\ 2005, \mnras, 364, 909 

%\bibitem[Wang et al.(2004)]{2004Wang} Wang, Q.~D., Owen, F., \& Ledlow, M.\ 2004, \apj, 611, 821 

\bibitem[Warren et al.(2006)]{2006Warren} Warren, M.~S., Abazajian,
  K., Holz, D.~E., \& Teodoro, L.\ 2006, \apj, 646, 881 

\bibitem[Wen \& Han(2013)]{2013Wen} Wen, Z.~L., \& Han, J.~L.\ 2013, \mnras, 436, 275 

%\bibitem[Yan et al.(2015)]{2015Yan} Yan, C.-S., Lu, Y., Dai, X., \& Yu, Q.\ 2015, \apj, 809, 117 

%\bibitem[ZuHone et al.(2010)]{2010ZuHone} ZuHone, J.~A., Markevitch, M., \& Johnson, R.~E.\ 2010, \apj, 717, 908

\end{thebibliography}
\end{document}